\documentclass[12pt]{iopart}

\usepackage{iopams}
\usepackage{graphicx}
\usepackage{mathrsfs}
\usepackage{color}
\usepackage[table]{xcolor}
\usepackage{hyperref}
\hypersetup{colorlinks, citecolor=black, filecolor=black, linkcolor=black, urlcolor=black}

\expandafter\let\csname equation*\endcsname\relax
\expandafter\let\csname endequation*\endcsname\relax
\usepackage{amsmath}

\usepackage{dcolumn}
\usepackage{bm}
\usepackage{setspace}

\usepackage[export]{adjustbox}

\definecolor{grayish}{RGB}{230,230,230}

\newcommand{\refEq}[1] {(\ref{#1})}

\newcommand{\Exp}[1]{\ensuremath{e^{#1}}}

\newcommand{\Nabla}{\ensuremath{\vec{\nabla}}}
\newcommand{\romanNum}[1]{\uppercase\expandafter{\romannumeral#1}}

\begin{document}

\title{A non-twisting flux tube for local gyrokinetic simulations}

\author{Justin Ball and Stephan Brunner}

\address{Ecole Polytechnique F\'{e}d\'{e}rale de Lausanne (EPFL), Swiss Plasma Center (SPC), CH-1015 Lausanne, Switzerland}

\ead{Justin.Ball@epfl.ch}

\begin{abstract}

Local gyrokinetic simulations use a field-aligned domain that twists due to the magnetic shear of the background magnetic equilibrium. However, if the magnetic shear is strong and/or the domain is long, the twist can become so extreme that it fails to properly resolve the turbulence. In this work, we derive and implement the ``non-twisting flux tube,'' a local simulation domain that remains rectangular at all parallel locations. Convergence and runtime tests indicate that it can calculate the heat flux more efficiently than the conventional flux tube. For one test case, it was 30 times less computationally expensive and we found no case for which it was more expensive. It is most advantageous when the magnetic shear is high and the domain includes at least two regions of turbulent drive (e.g. stellarator simulations, pedestal simulations, tokamak simulations with several poloidal turns). Additionally, it more accurately models the inboard midplane when the magnetic shear is large. Lastly, we show how the non-twisting flux tube can be generalized to allow further optimization and control of the simulation domain.

\end{abstract}



\section{Introduction}
\label{sec:intro}

In magnetic confinement fusion devices, plasma turbulence is usually an important, if not the dominant mechanism in determining the energy confinement time \cite{McKeeTurbulenceScale2001, GeigerW7XturbExp2019}. Turbulence ejects significant amounts of energy from the plasma, which then must be replaced using heating systems and large amounts of external electricity. Thus, it is vital to understand turbulent transport in order to enable a fusion power plant to generate net electricity at a competitive price.

The most successful approach to understand and predict turbulent transport is gyrokinetics, a high-fidelity kinetic model of the plasma \cite{CattoLinearizedGyrokinetics1978, FriemanNonlinearGyrokinetics1982, BrizardGKreview2007, AbelGyrokineticsDeriv2012}. Gyrokinetics has been rigorously derived from first principles using an asymptotic expansion of the Fokker-Planck and Maxwell's equations. This expansion makes the problem much more computationally tractable because it removes one velocity dimension from the model and, more importantly, eliminates a fast timescale --- the particle gyration around the magnetic field line. The crucial expansion parameter of gyrokinetics is $\rho_{\ast} \equiv \rho_{i} / a \ll 1$, the ratio of the ion gyroradius $\rho_{i}$ to the plasma minor radius $a$. Indeed, present-day large fusion devices \cite{LuxonDIIID2002} can have $\rho_{\ast} \approx 1 / 300$ and $\rho_{\ast}$ is anticipated to be even smaller in future high-performance devices \cite{NajmabadiARIES_AT2006}. Thus, we can be confident that our asymptotic expansion does not sacrifice much accuracy, making gyrokinetics one of the most reliable tools to simulate plasma turbulence in the core of fusion plasmas.

\begin{figure}
	\centering
	\hspace{-16em} (a) \hspace{15em} (b) \\
	\includegraphics[width=0.45\textwidth]{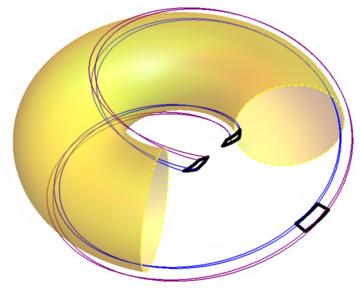}
	\includegraphics[width=0.45\textwidth]{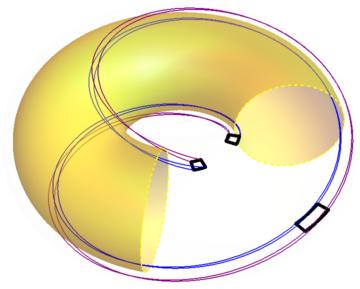}
	\caption{The boundaries of two local simulation domains (thin blue and purple lines): (a) a conventional flux tube and (b) a non-twisting flux tube. Both are one poloidal turn long. Note that at the outboard midplane, both flux tubes have a rectangular cross-section (thick black). However, away from the outboard midplane the cross-section of the conventional flux tube is twisted into a parallelogram, while the non-twisting flux tube remains rectangular. Also shown is the central flux surface of the flux tube (transparent yellow) with a toroidal wedge removed for visual clarity.}
	\label{fig:fluxTubeGeo}
\end{figure}

Despite its simplifications, gyrokinetic simulations remain near the limit of computational feasibility, with a single calculation typically requiring tens of thousands of CPU-hours and, hence, a supercomputer. To make simulations more accessible and practical, it is important to minimize the computational cost as much as possible. For grid-based codes, one straightforward way to do this is to minimize the number of grid points in the computational domain. This is a powerful motivation for so-called ``local'' simulations, which employ a reduced domain called a ``flux tube'' \cite{BeerBallooingCoordinates1995} (see figure \ref{fig:fluxTubeGeo}(a)). The flux tube is designed to exploit the physical properties of the turbulent eddies in order to minimize the number of grid points required. This is accomplished by carefully choosing both the {\it locations} of grid points within the flux tube as well as the overall {\it shape} of the flux tube.

The {\it locations} of the grid points are chosen to be field-aligned, meaning that two of the three spatial coordinates are constant along the magnetic field lines. This helps because it matches the underlying structure of the turbulence. Due to the strong magnetic field, plasma turbulence in fusion devices is very anisotropic --- individual turbulent eddies are very extended along the magnetic field lines, but are only a few gyroradii wide perpendicular to the field. Thus, the variation perpendicular to the magnetic field line has a small spatial scale, while the variation along the field line has a large spatial scale. By constructing a field-aligned grid, we take advantage of this and can get by with a much larger distance between grid points in the parallel direction. If our grid was not field-aligned we would require a grid spacing comparable to the particle gyroradius in all three spatial directions, instead of just two.

Additionally, it is helpful if the {\it shape} of the flux tube mimics the turbulent structures it seeks to model. This allows us to shrink our domain as much as possible and best utilize its volume. Thus, the flux tube is very elongated along a central field line and narrow across it. Specifically, as shown in figure \ref{fig:fluxTubeGeo}(a), it is typical for the domain to have a rectangular cross-section at the outboard midplane. Then, the four corners of the domain are fixed to four field lines, which determines how the cross-sectional shape changes along the parallel direction. This means that, due to magnetic shear, the domain twists into a parallelogram at other parallel locations. 

With such a narrow domain in the minor radial direction (i.e. across flux surfaces), the statistical properties of the turbulence can be assumed to be identical on both radial boundaries. For simplicity, instead of using {\it statistical} periodicity, one can apply {\it exact} periodicity between the boundaries as long as the domain is wider than a few turbulent correlation lengths. The same can be done with the toroidal boundaries. Because of this narrow periodic domain, flux tube codes can employ a Fourier representation of the turbulence in the perpendicular plane. Such a representation is advantageous because it respects the boundary conditions by construction and, through use of the three-halves rule \cite{Orszag3/2rule1971, ToldShiftedMetric2010}, can cleanly avoid numerical aliasing issues that arise from quadratic nonlinear coupling. The parallel boundaries can also use periodicity, but it is more complicated --- to ensure the parallel boundaries have statistically identical turbulence, the domain must be an integer number of poloidal turns long (i.e. the two ends must share the same poloidal location). This is because quantities like the strength and curvature of the magnetic field change the properties of turbulence and vary significantly with poloidal location. Additionally, implementing the parallel boundary condition while properly accounting for magnetic shear requires care \cite{BeerBallooingCoordinates1995, BallBoundaryCond2020}, as will be discussed. Regardless, the conventional flux tube is an elegant domain that appears to minimize the number of grid points, while still allowing for a physical treatment of turbulent eddies. However, the point of this paper is to demonstrate that further improvement is possible, thereby enabling more efficient numerical simulations. Specifically, we will formulate a simulation domain that does not twist due to the effect of magnetic shear, which we call the ``non-twisting flux tube'' and is shown in figure \ref{fig:fluxTubeGeo}(b).

The idea of minimizing the twist of the simulation domain has been investigated before. Most recently, Watanabe {\it et al.} \cite{WatanabeFluxTubeTrain2015} devised the ``flux tube train'' --- a domain formed by several conventional flux tubes coupled together into a chain. This enables a simulation domain to be several poloidal turns long without becoming extremely twisted. Normally, the twist would accrue along the entire length of the domain, but in the flux tube train it is reset between each individual flux tube (which can be just one poloidal turn long). However, the present paper is most closely related to the ``shifted metric'' procedure originally described in reference \cite{ScottShiftedMetric2001} and subsequently revisited in reference \cite{ToldShiftedMetric2010}. In the shifted metric procedure, one takes the discretized equations used by a turbulence code and performs a spatial shift of the metric coefficients (also known as the geometric coefficients). Specifically, each grid point is shifted in the toroidal direction by an amount that can depend on both the radial location and the location along the field line. Using such discrete shifts, one can cancel the twisting caused by the magnetic shear at each grid point. This enables the cross-section of a flux tube to stay approximately rectangular, while still maintaining a field-aligned grid. For reasons that will be explained later, this was theorized to provide a more efficient computational domain, especially when the magnetic shear is large. Such a procedure was first applied to a simplified fluid model of turbulence in a domain with Dirichlet radial boundary conditions \cite{ScottShiftedMetric2001}. Later, it was implemented in a gyrokinetic code, again with Dirichlet radial boundary conditions \cite{ToldShiftedMetric2010}. Both works investigated the applicability of the shifted metric procedure to a flux tube with the usual periodic radial boundary condition, but concluded that the procedure was inapplicable or infeasible when using a Fourier representation in the perpendicular plane.

In this paper, we will present a novel derivation that uses a coordinate system transformation of the gyrokinetic equations {\it before} they are discretized. This gives a different perspective on the non-twisting flux tube and its boundary conditions, which will reveal a straightforward way to implement it while maintaining radial periodicity and a Fourier representation in the perpendicular plane. Aspects of its implementation are conceptually similar to the improved ``wavevector-remap'' scheme used to model $E \times B$ flow shear \cite{HammettRemap2006, McMillanRemap2017} as magnetic shear twists the domain with parallel location analogously to how flow shear twists the domain with time. Flow shear will be omitted in this work for simplicity, but can be added to the non-twisting flux tube simply by including new terms analogously to the magnetic shear. The coordinate system discussed in this work may be useful for many purposes (e.g. global gyrokinetic simulations \cite{ToldShiftedMetric2010}, fluid simulations \cite{ScottShiftedMetric2001}), but this paper will focus exclusively on its application to local gyrokinetic simulations. It has the potential to be most helpful for simulations with high global magnetic shear (e.g. near the tokamak separatrix), high local magnetic shear (e.g. stellarators), or many poloidal turns (e.g. turbulence with long parallel correlation lengths).

In section \ref{sec:derivTwist}, we will review the analytic derivation of the conventional flux tube from reference \cite{BeerBallooingCoordinates1995} in order to provide appropriate context. Then, section \ref{sec:derivNonTwist} will present the novel derivation of the non-twisting flux tube, together with how it relates to both the conventional flux tube and the shifted metric approach. In section \ref{sec:benchmarking}, we will benchmark the implementation of the non-twisting flux tube in the gyrokinetic code GENE \cite{JenkoGENE2000, GoerlerGENE2011} and section \ref{sec:results} presents numerical results showing its computational performance relative to the conventional flux tube. Next, section \ref{sec:altTwist} details how the non-twisting flux tube can be generalized to give even more flexibility in structuring the spatial grid. Lastly, section \ref{sec:conclusions} offers some concluding thoughts.

\section{Analytic derivation of the conventional flux tube}
\label{sec:derivTwist}

A conventional flux tube \cite{BeerBallooingCoordinates1995} uses a field-aligned coordinate system given by $\left( x, y, \chi \right)$. The flux surface label $x = r - r_{0}$ is the minor radial coordinate $r$ relative to the center of the flux tube $r_{0}$, 
\begin{align}
  y \left( x, \zeta, \chi \right) \equiv \pm C_{y} \left( \zeta - q(x) \chi \right) \label{eq:yDef}
\end{align}
is the binormal coordinate (which labels the different field lines within the flux surfaces), and $\chi$ is the straight-field line poloidal angle (see \ref{app:straightCoord}). Here $\zeta$ is the toroidal angle and $q(x)$ is the safety factor. The normalization constant $C_{y}$ will be left arbitrary, though it is often taken to be either $C_{y} = r_{0}/q_{0}$ or $C_{y} = (1/B_{r}) d\psi/d r$ (where $q_{0} \equiv q(0)$, $\psi$ is the poloidal flux, and $B_{r}$ is a reference value of the magnetic field). Note that the sign in the definition of $y$ depends on what coordinate system convention is used. For the default coordinate system used by the GENE code, the lower sign (e.g. the minus sign in equation \refEq{eq:yDef}) should be taken in all formulas in this paper. For the opposite convention (e.g. that used by the GS2 gyrokinetic code \cite{KotschenreutherGS21995, DorlandETGturb2000}), the upper sign should be taken in all formulas.

In local simulations, due to the narrow domain, the safety factor profile is linearized about the center of the flux tube $x = 0$, giving
\begin{align}
  q(x) = q_{0} + \left. \frac{d q}{d x} \right|_{x=0} x . \label{eq:safetyFactor}
\end{align}
This enables us to rewrite equation \refEq{eq:yDef} as
\begin{align}
  y(x,\zeta,\chi) = \pm C_{y} \zeta \mp C_{y} q_{0} \chi \mp \hat{s} \chi x . \label{eq:yForm}
\end{align}
by introducing the global magnetic shear, $\hat{s} \equiv C_{y} \left. dq/dx \right|_{x=0}$. Note that this is a non-standard definition of $\hat{s}$ if $C_{y} \neq r_{0}/q_{0}$. It is in this $\left( x, y, \chi \right)$ coordinate system that we apply the perpendicular periodic boundary conditions \cite{BeerBallooingCoordinates1995}. In the radial direction, the real space electrostatic potential $\bar{\phi}$ must follow
\begin{align}
  \bar{\phi} ( x + L_{x}, y, \chi ) = \bar{\phi} ( x, y, \chi ) , \label{eq:radialBCy}
\end{align}
while the binormal boundary condition is
\begin{align}
  \bar{\phi} ( x, y + L_{y}, \chi ) = \bar{\phi} ( x, y, \chi ) , \label{eq:binormalBCy}
\end{align}
where $L_{x}$ and $L_{y}$ are the widths of the domain in the radial and binormal directions respectively. The parallel boundary condition is called the ``twist-and-shift'' condition \cite{BeerBallooingCoordinates1995}, which is more complicated as it must account for the effect of magnetic shear. To appropriately maintain the twisting effect of magnetic shear across the boundary, poloidal periodicity should be applied at constant toroidal angle $\zeta$, according to
\begin{align}
  \bar{\phi} ( x, y ( x, \zeta, \chi + 2 \pi N_{\text{pol}} ), \chi + 2 \pi N_{\text{pol}} ) = \bar{\phi} ( x, y ( x, \zeta, \chi ), \chi ) , \label{eq:parallelBCgen}
\end{align}
where $N_{\text{pol}} \in \mathbb{Z}_{+}$ must be a positive integer to ensure that the parallel boundaries share the same poloidal location and have statistically identical turbulence. Using equation \refEq{eq:yForm}, we can see that $y ( x, \zeta, \chi + 2 \pi N_{\text{pol}} ) = y ( x, \zeta, \chi ) \mp 2 \pi N_{\text{pol}} C_{y} q_{0} \mp 2 \pi N_{\text{pol}} \hat{s} x$. Thus, equation \refEq{eq:parallelBCgen} becomes
\begin{align}
  \bar{\phi} ( x, y \mp 2 \pi N_{\text{pol}} \hat{s} x, \chi + 2 \pi N_{\text{pol}} ) = \bar{\phi} ( x, y, \chi ) , \label{eq:parallelBCy}
\end{align}
where we have assumed that $\mp ( 2 \pi N_{\text{pol}} C_{y} / L_{y}) q_{0}$ is very close to some integer $N_{q} \in \mathbb{Z}$ and applied binormal periodicity $N_{q}$ times using equation \refEq{eq:binormalBCy}. This assumption is acceptable because equation \refEq{eq:yDef} shows that $C_{y}/L_{y} = 1/L_{\zeta}$ at constant $x$ and $\chi$, where $L_{\zeta}$ is the width of the flux tube in toroidal angle such that $L_{\zeta} = 2 \pi$ corresponds to the full toroidal domain. Thus, so long as the flux tube is very small compared to the full flux surface, $2 \pi N_{\text{pol}} C_{y} / L_{y}$ will be a very big number and a negligibly small change in $q_{0}$ will ensure that our assumption holds.

Importantly, by evaluating the parallel boundary condition of equation \refEq{eq:parallelBCy} at $x \rightarrow x + L_{x}$, applying the radial boundary condition of equation \refEq{eq:radialBCy} to both sides, and finally applying the parallel boundary condition to the left side we find
\begin{align}
  \bar{\phi} ( x, y \mp 2 \pi N_{\text{pol}} \hat{s} L_{x}, \chi ) = \bar{\phi} ( x, y, \chi ) . \label{eq:artificalCorry}
\end{align}
This shows that the combination of parallel and radial boundary conditions will introduce artificial correlations between different $y$ locations unless
\begin{align}
  L_{x} = \frac{N_{\text{asp}} L_{y}}{2 \pi N_{\text{pol}} \left| \hat{s} \right|} \label{eq:aspRatioDiscrete}
\end{align}
holds for a positive integer $N_{\text{asp}} \in \mathbb{Z}_{+}$, where the subscript ``$\text{asp}$'' indicates that it controls the domain aspect ratio. Fulfilling this constraint resolves the problem because the offending term in equation \refEq{eq:artificalCorry} can be eliminated by applying binormal periodicity $N_{\text{asp}}$ times using equation \refEq{eq:binormalBCy}. This constraint discretizes the aspect ratio of the domain. Note that it is a general consequence of the flux tube boundary conditions and must be satisfied regardless of whether one uses a real space or Fourier-space representation.

For the reasons outlined in the introduction, local gyrokinetic codes usually employ a Fourier representation in the perpendicular plane. Thus, we will use the Fourier-space electrostatic potential $\phi$, defined by
\begin{align}
  \bar{\phi} \left( x, y, \chi \right) = \sum_{k_{x}, k_{y}} \phi \left( k_{x}, k_{y}, \chi \right) \Exp{i k_{x} x + i k_{y} y} . \label{eq:FourierSeriesy}
\end{align}
According to Fourier's theorem, we see that the binormal and radial boundary conditions of equations \refEq{eq:radialBCy} and \refEq{eq:binormalBCy} require
\begin{align}
  k_{x} &= \frac{2 \pi}{L_{x}} m \label{eq:radialWavenumberDiscrete} \\
  k_{y} &= \frac{2 \pi}{L_{y}} n , \label{eq:binormalWavenumberDiscrete}
\end{align}
where $m \in \mathbb{Z}$ and $n \in \mathbb{Z}$ are integers. Note $n$ is analogous to the toroidal mode number in the restricted domain of the flux tube. Thus, the perpendicular boundary conditions can be satisfied in a simulation simply by constructing the Fourier coordinate grids appropriately. Typically they are constructed according to
\begin{align}
  k_{x} &\in \frac{2 \pi}{L_{x}} m ~~~~~~~ \text{for all} ~ m \in \left[ - \frac{N_{x} - 1}{2}, \frac{N_{x} - 1}{2} \right] \label{eq:kxGridy} \\
  k_{y} &\in \frac{2 \pi}{L_{y}} n ~~~~~~~~ \text{for all} ~ n \in \left[ 0, N_{y} - 1 \right] , \label{eq:kyGridy}
\end{align}
where $N_{x}$ and $N_{y}$ are the number of considered radial and binormal modes respectively and we have assumed $N_{x}$ is odd for simplicity. Additionally, we have omitted the negative $k_{y}$ values as they can be recovered using the reality condition $\phi \left( k_{x}, k_{y}, \chi \right) = \phi^{\ast} \left( -k_{x}, -k_{y}, \chi \right)$, where $^{\ast}$ indicates the complex conjugate.

The parallel boundary condition is more complicated. After substituting the Fourier representation of equation \refEq{eq:FourierSeriesy}, equation \refEq{eq:parallelBCy} becomes
\begin{align}
  \phi ( k_{x} \pm 2 \pi N_{\text{pol}} k_{y} \hat{s}, k_{y}, \chi + 2 \pi N_{\text{pol}} ) = \phi ( k_{x}, k_{y}, \chi ) . \label{eq:parallelBCfourierkx}
\end{align}
This is enforced by appropriately coupling different $k_{x}$ modes across the ends of the flux tube. For these modes to always line up properly, the smallest possible jump in the radial wavenumber (i.e. $2 \pi N_{\text{pol}} k_{y, \text{min}} \left| \hat{s} \right|$) must be equal to a multiple of the radial grid spacing (e.g. $N_{\text{asp}} k_{x, \text{min}}$). This condition turns out to be identical to the domain aspect ratio condition of equation \refEq{eq:aspRatioDiscrete} (since the minimum radial and binormal wavenumbers are $k_{x, \text{min}} = 2 \pi / L_{x}$ and $k_{y, \text{min}} = 2 \pi / L_{y}$ respectively). The only exceptions are the largest values of $k_{x}$, which are coupled to $\phi = 0$ because the corresponding modes are outside of the considered $k_{x}$ grid.

In the $\left( x, y, \chi \right)$ coordinate system, the Fourier-space electrostatic gyrokinetic model (neglecting collisions and background flow) is given by the gyrokinetic equation \cite{ParraUpDownSym2011}
\begin{align}
  \frac{\partial h_{s}}{\partial t} &+ v_{||} \hat{b} \cdot \vec{\nabla} \chi \left. \frac{\partial h_{s}}{\partial \chi} \right|_{k_{x}, k_{y}} + i \vec{v}_{d s} \cdot \left( k_{x} \Nabla x + k_{y} \Nabla y \right) h_{s} + a_{s ||} \frac{\partial h_{s}}{\partial v_{||}} \label{eq:GKeqy} \\
  &\mp \frac{1}{J B} \left\{ h_{s}, \phi J_{0} \left( k_{\perp} \rho_{s} \right) \right\} = \frac{Z_{s} e F_{M s}}{T_{s}} \frac{\partial \phi}{\partial t} J_{0} \left( k_{\perp} \rho_{s} \right) \mp i \frac{k_{y}}{J B} \phi J_{0} \left( k_{\perp} \rho_{s} \right) \frac{\partial F_{Ms}}{\partial x} \nonumber
\end{align}
and the quasineutrality equation
\begin{align}
  \phi = 2 \pi \left( \sum_{s} \frac{Z_{s}^{2} e^{2} n_{s}}{T_{s}} \right)^{-1} \sum_{s} \frac{Z_{s} e B}{m_{s}} \int dv_{||} \int d \mu J_{0} \left( k_{\perp} \rho_{s} \right) h_{s} , \label{eq:quasineuty}
\end{align}
where the perpendicular wavenumber is
\begin{align}
  k_{\perp} = \sqrt{k_{x}^{2} \left| \Nabla x \right|^{2} + 2 k_{x} k_{y} \Nabla x \cdot \Nabla y + k_{y}^{2} \left| \Nabla y \right|^{2}} \label{eq:kperpDefy}
\end{align}
and the nonlinear term in Fourier-space is
\begin{align}
  \left\{ h_{s}, \phi J_{0} \left( k_{\perp} \rho_{s} \right) \right\} &= \sum_{k_{x}', k_{y}'} \left( k_{x}' k_{y}'' - k_{x}'' k_{y}' \right) h_{s}' \phi'' J_{0} \left( k_{\perp}'' \rho_{s} \right) \label{eq:nonlinearFouriery}
\end{align}
with $k_{x}'' = k_{x} - k_{x}'$ and $k_{y}'' = k_{y} - k_{y}'$. The unknowns are the Fourier-analyzed electrostatic potential $\phi$ and the Fourier-analyzed non-adiabatic portion of the distribution function $h_{s} \equiv \delta f + Z_{s} e \phi F_{Ms} / T_{s}$. The coordinates are the time $t$, the radial wavenumber $k_{x}$, the binormal wavenumber $k_{y}$, the straight-field line poloidal angle $\chi$, the parallel velocity $v_{||}$, and the magnetic moment $\mu \equiv m_{s} v_{\perp}^{2} / ( 2 B )$. The species subscript $s \in \{i,e\}$ indicates ions or electrons respectively. Here $\hat{b}$ is the magnetic field unit vector, $B$ is the strength of the magnetic field, $\vec{v}_{ds}$ is the magnetic drift velocity, $a_{s||}$ is the parallel acceleration from the magnetic mirror force, $J_{0} ( \ldots )$ is the 0\textsuperscript{th} order Bessel function of the first kind, $Z_{s}$ is the species charge number, $e$ is the proton electric charge, $F_{Ms}$ is the Maxwellian velocity distribution, $\delta f$ is the turbulent portion of the Fourier-analyzed perturbed distribution function, $T_{s}$ is the species temperature, $n_{s}$ is the species density, $m_{s}$ is the species mass, and $J^{-1} \equiv \mp (\Nabla x \times \Nabla y) \cdot \hat{b}$ is closely related to the coordinate system Jacobian and is defined such that it is positive for both sign conventions. Lastly, to reduce the computational cost, most codes compute the nonlinear term in real space according to
\begin{align}
  \left\{ h_{s}, \phi J_{0} \left( k_{\perp} \rho_{s} \right) \right\} &= \frac{1}{N_{x} \left( 2 N_{y} - 1 \right)} \sum_{x, y} \left[ \left( \sum_{k'_{x}, k'_{y}} k_{x}' h_{s}' \Exp{i k'_{x} x + i k'_{y} y} \right) \left( \sum_{k''_{x}, k''_{y}} k_{y}'' \phi'' J_{0} \left( k_{\perp}'' \rho_{s} \right) \Exp{i k''_{x} x + i k''_{y} y} \right) \right. \label{eq:nonlinearRealy} \\
  &- \left. \left( \sum_{k'_{x}, k'_{y}} k_{y}' h_{s}' \Exp{i k'_{x} x + i k'_{y} y} \right) \left( \sum_{k''_{x}, k''_{y}} k_{x}'' \phi'' J_{0} \left( k_{\perp}'' \rho_{s} \right) \Exp{i k''_{x} x + i k''_{y} y} \right) \right] \Exp{-i k_{x} x - i k_{y} y} \nonumber
\end{align}
by using
\begin{align}
\phi \left( k_{x}, k_{y}, \chi \right) = \frac{1}{N_{x} \left( 2 N_{y} - 1 \right)} \sum_{x, y} \bar{\phi} \left( x, y, \chi \right) \Exp{- i k_{x} x - i k_{y} y} , \label{eq:invFourierSeriesy}
\end{align}
the inverse discrete Fourier transform of equation \refEq{eq:FourierSeriesy}. Note the (double) primed quantities indicate they are evaluated at the (double) primed wavenumbers. In the calculation of the nonlinear term in real space, high wavenumber perturbations may be generated that then get fictitiously mapped to lower wavenumbers in the return Fourier transform. In the code, this is avoided by using the three-halves rule \cite{Orszag3/2rule1971, ToldShiftedMetric2010}. First, before performing the Fourier transform to real space, the Fourier-space quantities are copied to an expanded wavenumber grid that is at least $3/2$ times bigger in both perpendicular directions. These new high wavenumber modes are given zero amplitude. Then, the full calculation of the nonlinear term is performed according to equation \refEq{eq:nonlinearRealy} on this expanded grid. At the end, the final Fourier-space result is taken and the high wavenumber modes are discarded to return to the original wavenumber grid. Thus, in GENE the summations in equation \refEq{eq:invFourierSeriesy} are really taken over an expanded wavenumber grid, $x \in 2 m L_{x} / \left( 3 ( N_{x} + 1 ) \right)$ for all $m \in \left[ 0, 3 ( N_{x} + 1 ) / 2 - 1 \right]$, and $y \in n L_{y} / \left( 3 N_{y} \right)$ for all $n \in \left[ 0, 3 N_{y} - 1 \right]$.

\begin{figure}
	\centering
	\includegraphics[width=1.0\textwidth]{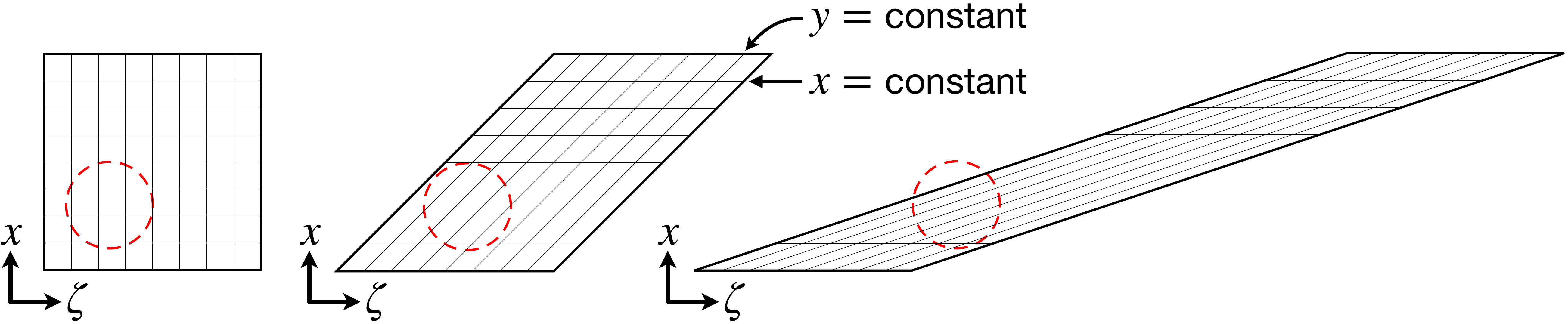}
	\caption{The $\left( x, y \right)$ coordinate system grid at three $\chi$ locations along a conventional flux tube. We see from the example ion gyroradius (red dashed circle) that, as the domain becomes increasingly sheared, ions will more effectively average over any of the turbulent perturbations allowed in the domain.}
	\label{fig:crossSec}
\end{figure}

Now let us consider the implications of this $\left( x, y, \chi \right)$ coordinate system (or equivalently $\left( k_{x}, k_{y}, \chi \right)$). A regularly-spaced rectangular grid in these coordinates produces a simulation domain that becomes sheared into a parallelogram as it extends along the field lines due to the effect of magnetic shear. This has important consequences for what type of turbulence can be modeled in the domain. For certain conditions, the twist of the coordinate system is appropriate. If the three-dimensional shape of a turbulent eddy is primarily an extrusion along field lines of its two-dimensional cross-section at the outboard midplane, the coordinate system will model it efficiently. However, as an eddy extends along the field lines, magnetic shear twists it and decreases the spatial scales of the structure. If the magnetic shear is large or the eddy is very extended along the magnetic field, the spatial scales of the eddy will become very small as is reflected in the extreme tilt of the parallelogram (see figure \ref{fig:crossSec}). Such fine structures are strongly damped due to finite gyroradius effects, so it would be surprising if they were important. Yet, at these parallel locations the conventional flux tube is optimized to model them.

To see this mathematically, one can look at the Bessel functions in the gyrokinetic model, which represent the finite gyroradius effects. In the conventional flux tube, the radial grid is centered around $k_{x} = 0$ at all parallel locations, for which we see that
\begin{align}
  J_{0} \left( k_{\perp} \rho_{s} \right) = J_{0} \left( \left| k_{y}\right| \rho_{s} \left| \Nabla y \right| \right) \label{eq:bessely}
\end{align}
using equation \refEq{eq:kperpDefy}. When one calculates the geometric coefficient
\begin{align}
  \Nabla y = \pm C_{y} \Nabla \zeta \mp C_{y} q_{0} \Nabla \chi \mp \hat{s} \chi \Nabla x \label{eq:grady}
\end{align}
at $x=0$, we see that the third term has a secular dependence on the parallel location $\chi$. Thus, as we move along the field line, $J_{0} \left( k_{\perp} \rho_{s} \right) \rightarrow 0$ for the central radial mode in the grid, indicating that it is damped. This means that, for large values of $\hat{s}$ or long flux-tube domains, the radial wavenumber grid is centered around a mode that is strongly damped at most values of $\chi$.

To summarize, the perpendicular grid of a conventional flux tube is optimized to model perturbations that are field line following. Away from $\chi = 0$, this is a distinct goal from efficiently modeling the perturbations that are least damped by finite gyroradius effects. The question then becomes empirical --- does turbulence prioritize following field lines from the outboard midplane or does it prioritize minimizing the amount of finite gyroradius damping? The short answer is both, to some degree, and the balance depends on the specifics of a particular simulation. We will delay a more in depth answer until section \ref{sec:results} and instead define a grid that is optimized for turbulence that minimizes finite gyroradius damping --- the non-twisting flux tube.

\section{Analytic derivation of the non-twisting flux tube}
\label{sec:derivNonTwist}

In order to produce a flux tube that does not twist (e.g. figure \ref{fig:fluxTubeGeo}(b)), we want to perform a coordinate system transformation that removes the effect of the magnetic shear --- both global shear and local shear. This can be accomplished by defining a new ``non-twisting'' binormal coordinate $Y$ such that $\Nabla x \cdot \Nabla Y = 0$ at all parallel locations. In other words, we want $(x, Y)$ to be an orthogonal coordinate system. The effect of global shear is clearly contained in the third term of equation \refEq{eq:yForm}, but the local correction to the shear is more subtle. It manifests through the definition of the straight-field line poloidal angle $\chi$ (see \ref{app:straightCoord}) in the second term of equation \refEq{eq:yForm} because $\Nabla x \cdot \Nabla \chi \neq 0$. To remove all magnetic shear, we make the coordinate transformation \cite{ScottShiftedMetric2001,  MartinParallelBCstell2018}
\begin{align}
  Y \left( x, y, \chi \right) &\equiv y - \frac{\Nabla x \cdot \Nabla y}{\left| \Nabla x \right|^{2}} x . \label{eq:Ydef}
\end{align}
Using equation \refEq{eq:grady} and $\Nabla x \cdot \Nabla \zeta = 0$, this geometric factor can be calculated to be 
\begin{align}
   \frac{\Nabla x \cdot \Nabla y}{\left| \Nabla x \right|^{2}} = \mp \hat{s} \chi \mp C_{y} q_{0} \frac{\Nabla x \cdot \Nabla \chi}{\left| \Nabla x \right|^{2}} , \label{eq:Yfactor}
\end{align}
where the first term contains the global magnetic shear and the second term contains the local correction to the magnetic shear through the definition of $\chi$ (see equations \refEq{eq:chiDef} and \refEq{eq:localShearDef}). Thus, we can rewrite equation \refEq{eq:Ydef} as
\begin{align}
  Y \left( x, \zeta, \chi \right) &= \pm C_{y} \zeta \mp C_{y} q_{0} \left( \chi - \frac{\Nabla x \cdot \Nabla \chi}{\left| \Nabla x \right|^{2}} x \right) \label{eq:Yform}
\end{align}
using equation \refEq{eq:yForm}.
Importantly, we can directly substitute this expression to show that $\Nabla x \cdot \Nabla Y = 0$, remembering that the geometric coefficients are evaluated at $x = 0$ in the {\it local} gyrokinetic model.

To accomplish this coordinate system transformation, we will first transform the boundary conditions. The binormal boundary condition is applied at constant $x$ and $\chi$, so it is written as
\begin{align}
  \bar{\Phi} ( x, Y \left( x, y + L_{y}, \chi \right), \chi ) = \bar{\Phi} ( x, Y \left( x, y, \chi \right), \chi ) .
\end{align} 
From equation \refEq{eq:Ydef} we see that $Y \left( x, y + L_{y}, \chi \right) = Y \left( x, y, \chi \right) + L_{y}$, demonstrating that the transformed binormal boundary condition is simply
\begin{align}
  \bar{\Phi} ( x, Y + L_{y}, \chi ) = \bar{\Phi} ( x, Y, \chi ) . \label{eq:binormalBCY}
\end{align}
Note that we have introduced a new symbol for the electrostatic potential $\bar{\Phi}$. This is because it has a different functional dependence on its arguments according to
\begin{align}
  \bar{\Phi} \left( x, Y, \chi \right) = \bar{\phi} \left( x, Y + \frac{\Nabla x \cdot \Nabla y}{| \Nabla x |^{2}} x, \chi \right) = \bar{\phi} \left( x, y \left( x, Y, \chi \right), \chi \right) . \label{eq:PhiRealFormY}
\end{align}
We also define a new distribution function $\bar{H}_{s}$ from $\bar{h}_{s}$ in an analogous fashion. More broadly, throughout the paper we will match the style of these symbols with the associated coordinate (i.e. lowercase $\bar{\phi}$ and $\phi$ are functions of lowercase $y$ and $k_{x}$ respectively, capital $\bar{\Phi}$ and $\Phi$ are functions of capital $Y$ and $K_{x}$, script $\varphi$ will be a function of script $\mathscr{K}_{x}$, etc.).

The parallel boundary condition must still be taken at constant toroidal angle $\zeta$ and $x$, according to
\begin{align}
  \bar{\Phi} ( x, Y ( x, \zeta, \chi + 2 \pi N_{\text{pol}} ), \chi + 2 \pi N_{\text{pol}} ) = \bar{\Phi} ( x, Y ( x, \zeta, \chi ), \chi ) . \label{eq:parallelBCgenY}
\end{align}
Equation \refEq{eq:Yform} shows that $Y ( x, \zeta, \chi + 2 \pi N_{\text{pol}} ) = Y ( x, \zeta, \chi ) \mp 2 \pi N_{\text{pol}} C_{y} q_{0}$ because the geometric coefficients $\Nabla x \cdot \Nabla \chi$ and $| \Nabla x |^{2}$ are both $2 \pi$-periodic in $\chi$.
This implies that
\begin{align}
  \bar{\Phi} ( x, Y, \chi + 2 \pi N_{\text{pol}} ) = \bar{\Phi} ( x, Y, \chi ) , \label{eq:parallelBCY}
\end{align}
again assuming that $( 2 \pi N_{\text{pol}} C_{y} / L_{y} ) q_{0}$ is very close to an integer and applying binormal periodicity as we did for the conventional flux tube. This parallel boundary condition is simpler than equation \refEq{eq:parallelBCy}, which is intuitive as a non-twisting flux tube has an identical rectangular cross-section at both parallel boundaries making it straightforward to copy turbulent structures across it.

Instead the complications from magnetic shear appear in the radial boundary condition. It is applied at constant $y$ and $\chi$, according to
\begin{align}
  \bar{\Phi} \left( x + L_{x}, Y \left( x + L_{x}, y, \chi \right), \chi \right) = \bar{\Phi} \left( x, Y \left( x, y, \chi \right), \chi \right) .
\end{align} 
Using equation \refEq{eq:Ydef} we see that $Y \left( x + L_{x}, y, \chi \right) = Y \left( x, y, \chi \right) - (\Nabla x \cdot \Nabla y ) L_{x} / | \Nabla x |^{2}$, so the radial boundary condition in the new coordinates is
\begin{align}
  \bar{\Phi} \left( x + L_{x}, Y - \frac{\Nabla x \cdot \Nabla y}{\left| \Nabla x \right|^{2}} L_{x}, \chi \right) = \bar{\Phi} \left( x, Y, \chi \right) . \label{eq:radialBCY}
\end{align}
We see that a new term, $(\Nabla x \cdot \Nabla y ) L_{x} / | \Nabla x |^{2}$, appears, which is needed to properly pair up field lines across the radial boundary. In the conventional flux tube, as one moved along the domain in $\chi$, the domain cross-section twisted with the field lines, so the field lines always stayed at the same position relative to the boundaries. In the new coordinates, all the field lines not at $x=0$ will shift relative to the binormal boundaries as you move in $\chi$. At some parallel location(s) they may exit through the binormal boundary of the domain and immediately re-enter through the opposite binormal boundary. Note that this means the ``non-twisting flux tube'' is technically not a true ``flux tube'' according to some definitions. However, we think the terminology is appropriate as the domain still changes area to retain a constant magnetic flux through its cross-section. Since the field lines move relative to the domain boundaries, the new term in equation \refEq{eq:radialBCY} is needed to properly maintain field line identity across the radial boundary, accounting for the local magnetic shear. If radial periodicity identifies two field lines on opposite radial boundaries to be the same at a particular parallel location, the two field lines must remain matched up at all other parallel locations. Otherwise a particle could move to a different field line by moving purely in the parallel direction, which is unphysical.

Field line identity must also be preserved across the parallel boundary condition, which can be checked by combining the parallel and radial boundary conditions of equations \refEq{eq:parallelBCY} and \refEq{eq:radialBCY}. Specifically, we evaluate the radial boundary condition at $\chi \rightarrow \chi + 2 \pi N_{\text{pol}}$, apply the parallel boundary condition to both sides, and then apply the radial boundary condition to the left side. In doing this, it is important to note that equation \refEq{eq:Yfactor} shows
\begin{align}
  \left. \frac{\Nabla x \cdot \Nabla y}{\left| \Nabla x \right|^{2}} \right|_{\chi + 2 \pi N_{\text{pol}}} - \left. \frac{\Nabla x \cdot \Nabla y}{\left| \Nabla x \right|^{2}} \right|_{\chi} = \mp 2 \pi N_{\text{pol}} \hat{s} , \label{eq:geoCoeffPeriodicity}
\end{align}
since $\Nabla x \cdot \Nabla \chi$ and $| \Nabla x |^{2}$ are both $2 \pi$-periodic in $\chi$. Equation \refEq{eq:geoCoeffPeriodicity} states that after an integer number of poloidal turns the local correction to the shear integrates to zero and only the twisting from global shear remains. Using this fact, we find the combination of parallel and radial boundary conditions in the new coordinate system gives
\begin{align}
  \bar{\Phi} ( x, Y \pm 2 \pi N_{\text{pol}} \hat{s} L_{x}, \chi ) = \bar{\Phi} ( x, Y, \chi ) . \label{eq:artificalCorrY}
\end{align}
This implies that the domain will have fictitious correlations between different $Y$ positions unless equation \refEq{eq:binormalBCY} can be used to eliminate the offending term. Enforcing this discretizes the aspect ratio of the domain according to equation \refEq{eq:aspRatioDiscrete} just like the conventional flux tube, as is expected.

We emphasize that the boundary conditions of the non-twisting flux tube (i.e. equations \refEq{eq:binormalBCY}, \refEq{eq:parallelBCY}, and \refEq{eq:radialBCY}) are consistent with those of the conventional flux tube (i.e. equations \refEq{eq:radialBCy}, \refEq{eq:binormalBCy}, and \refEq{eq:parallelBCy}). This can be checked by taking a boundary condition from the non-twisting flux tube, applying equation \refEq{eq:PhiRealFormY} to the left and right sides, and then substituting equation \refEq{eq:Ydef} to replace $Y$ with $y$.

As with the conventional flux tube, we would like to use a Fourier representation in the perpendicular plane. Because the binormal boundary condition in real space (i.e. equation \refEq{eq:binormalBCY}) has the same form as in the conventional flux tube (i.e. equation \refEq{eq:binormalBCy}), it will be satisfied using the standard Fourier representation
\begin{align}
  \bar{\Phi} \left( x, Y, \chi \right) = \sum_{k_{y}} \hat{\Phi} \left( x, k_{y}, \chi \right) \Exp{i k_{y} Y} \label{eq:halfFourierSeriesY}
\end{align}
with the wavenumbers discretized according to equation \refEq{eq:binormalWavenumberDiscrete}. However, the radial boundary condition of equation \refEq{eq:radialBCY} is more complicated than in the conventional flux tube. To find the allowed radial wavenumbers, we first Fourier analyze using equation \refEq{eq:halfFourierSeriesY} to get the radial boundary condition into the form
\begin{align}
  \hat{\Phi} \left( x + L_{x}, k_{y}, \chi \right) \Exp{- i k_{y} \frac{\Nabla x \cdot \Nabla y}{\left| \Nabla x \right|^{2}} L_{x}} = \hat{\Phi} \left( x, k_{y}, \chi \right) . \label{eq:radialBCYfloquet}
\end{align}
Then, using Floquet's theorem we make the substitution
\begin{align}
  \hat{\Phi} \left( x, k_{y}, \chi \right) = P \left( x, k_{y}, \chi \right) \Exp{i k_{y} \frac{\Nabla x \cdot \Nabla y}{\left| \Nabla x \right|^{2}} x} \label{eq:FloquetSubstitution}
\end{align}
and see that the radial boundary condition becomes simple periodicity, $P \left( x + L_{x}, k_{y}, \chi \right) =  P \left( x, k_{y}, \chi \right)$. This is satisfied by writing $P \left( x, k_{y}, \chi \right)$ as a standard Fourier series in $x$ with wavenumber discretized according to equation \refEq{eq:radialWavenumberDiscrete}. Thus, by substituting this result into equation \refEq{eq:FloquetSubstitution} and then equation \refEq{eq:halfFourierSeriesY}, we find the final form of 
\begin{align}
  \bar{\Phi} \left( x, Y, \chi \right) = \sum_{K_{x}, k_{y}} \Phi \left( K_{x}, k_{y}, \chi \right) \Exp{i K_{x} x + i k_{y} Y} , \label{eq:FourierSeriesY}
\end{align}
where the allowed radial wavenumbers are given by 
\begin{align}
  K_{x} = \frac{2 \pi}{L_{x}} m + k_{y} \frac{\Nabla x \cdot \Nabla y}{\left| \Nabla x \right|^{2}} \label{eq:KxAllowed}
\end{align}
for any integer $m \in \mathbb{Z}$. Comparing this with equation \refEq{eq:radialWavenumberDiscrete}, we see that the new radial wavenumber is defined by
\begin{align}
  K_{x} \equiv k_{x} + k_{y} \frac{\Nabla x \cdot \Nabla y}{\left| \Nabla x \right|^{2}} , \label{eq:KxDef}
\end{align}
which means that $K_{x} x + k_{y} Y = k_{x} x + k_{y} y$. Equation \refEq{eq:KxAllowed} indicates that the allowed radial wavenumbers now depend on the poloidal angle and the binormal wavenumber, which might seem strange. However, this is {\it not} because something physical has changed about the situation. We will see that it is purely a semantic consequence of our coordinate system transform.

\begin{figure}
	\centering
	\hspace{-16em} (a) \hspace{19em} (b) \\
	\includegraphics[width=0.45\textwidth]{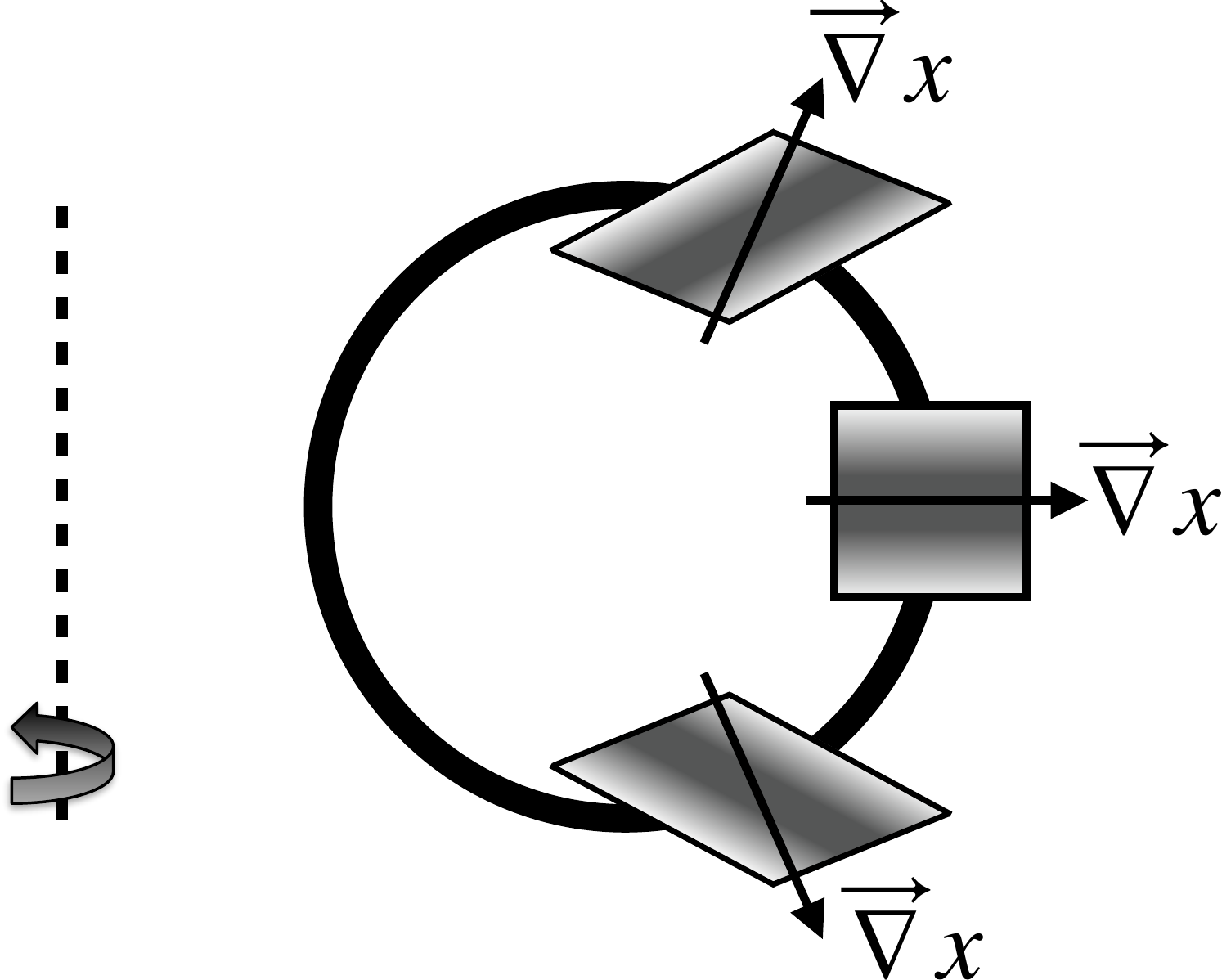}
	\hspace{3em}
	\includegraphics[width=0.45\textwidth]{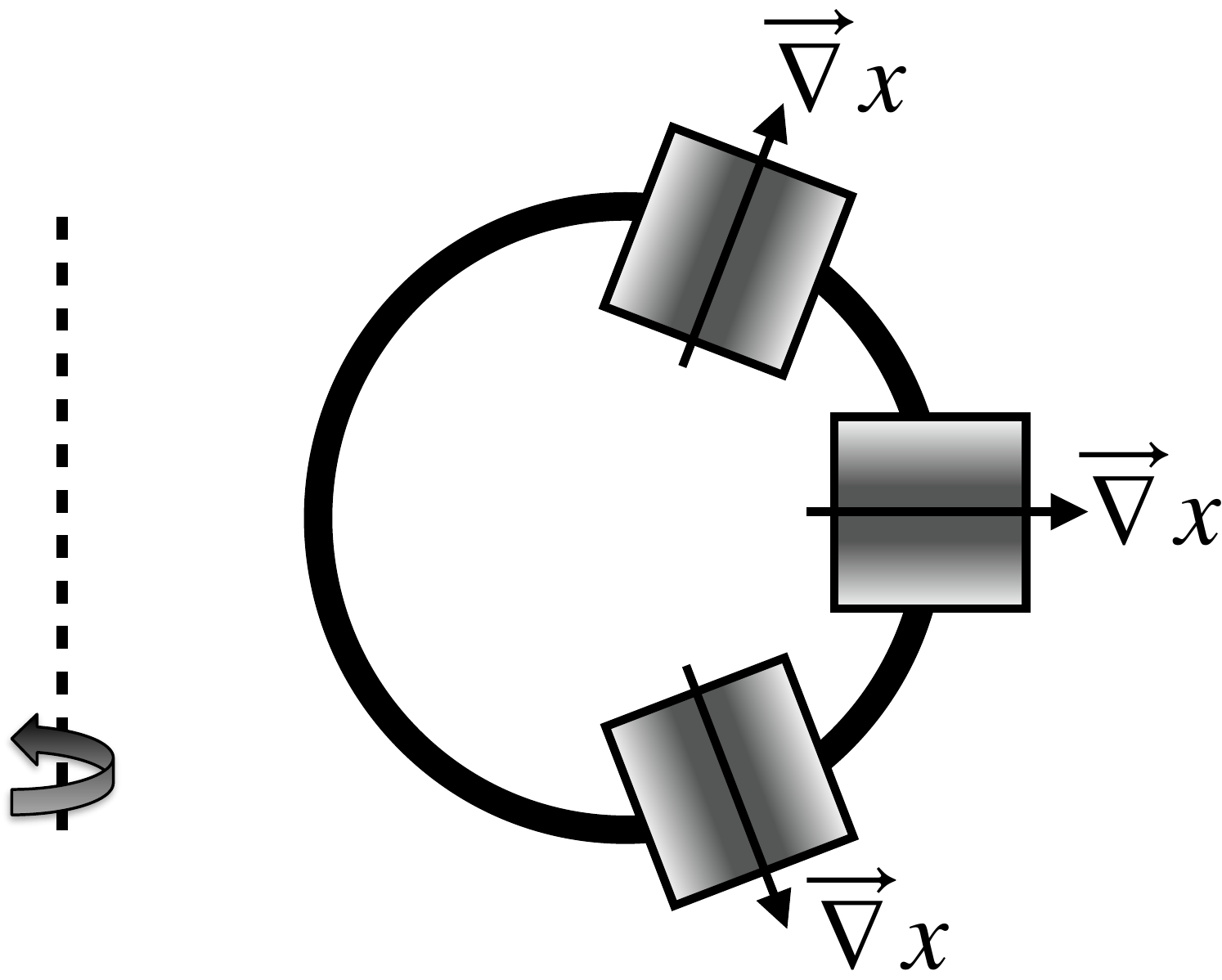}
	\caption{Three perpendicular cross-sections at different poloidal angles for (a) a cartoon conventional flux tube with a $(k_{x} = 0, k_{y} = 2\pi/L_{y})$ perturbation and (b) a cartoon non-twisting flux tube with a $(K_{x} = 0, k_{y} = 2\pi/L_{y})$ perturbation. Note that the three perturbations depicted in (a) are linearly coupled, unlike those in (b). Also shown is the central $x=0$ flux surface of the flux tube (thick black), which is circular, and the axis of toroidal symmetry (dotted black).}
	\label{fig:kxDef}
\end{figure}

Equation \refEq{eq:KxDef} defines our coordinate system transformation in Fourier-space just as equation \refEq{eq:Ydef} defined our coordinate system transform in real space, so it is important to understand the physical meaning of $K_{x}$ relative to $k_{x}$. We will call $K_{x}$ the {\it local} radial wavenumber and $k_{x}$ the {\it ballooning} radial wavenumber. We see from equation \refEq{eq:KxDef} that the two wavenumbers are identical wherever $\Nabla x \cdot \Nabla y = 0$, such as at $\chi = 0$. The ballooning radial wavenumber $k_{x}$ is defined by performing the Fourier analysis while holding $y$ and $\chi$ constant, so it is the radial wavenumber of the turbulence {\it along lines of constant $y$ and $\chi$}. However, lines of constant $y$ and $\chi$ are {\it not} perpendicular to the flux surfaces (i.e. lines of constant $x$) because the $( x, y )$ coordinate system is not orthogonal (see equation \refEq{eq:Yfactor} or figure \ref{fig:crossSec}). Thus, as shown in figure \ref{fig:kxDef}(a), $\left( k_{x} = 0, k_{y} \neq 0 \right)$ does not everywhere refer to a Fourier mode that is constant in the $\Nabla x$ direction, even for circular flux surfaces. Lines of constant $y$ and $\chi$ only run in the the $\Nabla x$ direction at the parallel location where the flux tube is rectangular, typically at the outboard midplane. Thus, away from the midplane the Fourier mode with zero variation in the $\Nabla x$ direction actually has $k_{x} \neq 0$. Instead $k_{x} = 0$ refers to the Fourier mode that, if you traced it along field lines back to the outboard midplane, would have zero variation in the $\Nabla x$ direction. Hence, its name --- the ballooning radial wavenumber. In contrast, the local radial wavenumber $K_{x}$ is defined by performing the Fourier analysis while holding $Y$ constant. Since we defined $Y$ such that $\Nabla x \cdot \Nabla Y = 0$, lines of constant $Y$ run in the $\Nabla x$ direction at all parallel locations. Thus, the $K_{x} = 0$ Fourier mode never varies in the $\Nabla x$ direction, as is shown in figure \ref{fig:kxDef}(b). Note that the $k_{x} = 0$ Fourier modes at each parallel location are linearly coupled, while the $K_{x} = 0$ Fourier modes are not.

The final Fourier-space boundary condition is in the parallel direction. By using equation \refEq{eq:FourierSeriesY} to perform a Fourier analysis of equation \refEq{eq:parallelBCY}, we find that it is
\begin{align}
  \Phi ( K_{x}, k_{y}, \chi + 2 \pi N_{\text{pol}} ) = \Phi ( K_{x}, k_{y}, \chi ) . \label{eq:parallelBCfourierKx}
\end{align}
Despite this simple form, it is not appropriate to use a Fourier representation in $\chi$ because the $K_{x}$ grid varies with $\chi$ as do the geometric coefficients in the gyrokinetic equations. Like before, we can combine the parallel and radial boundary conditions of equations \refEq{eq:KxAllowed} and \refEq{eq:parallelBCfourierKx} to derive a constraint on the aspect ratio of the domain. Using equations \refEq{eq:binormalWavenumberDiscrete} and \refEq{eq:geoCoeffPeriodicity}, we find that the usual condition of equation \refEq{eq:aspRatioDiscrete} still holds and ensures that the same values of $K_{x}$ exist at both $\chi$ and $\chi + 2 \pi N_{\text{pol}}$. Before we move on, we note that equation \refEq{eq:parallelBCfourierKx} is consistent with the parallel boundary condition in the conventional flux tube (i.e. equation \refEq{eq:parallelBCfourierkx}). To prove this we must first relate the functional forms of $\Phi$ and $\phi$ by substituting equations \refEq{eq:FourierSeriesy} and \refEq{eq:FourierSeriesY} into equation \refEq{eq:PhiRealFormY} to see
\begin{align}
  \Phi \left( K_{x}, k_{y}, \chi \right) = \phi \left( K_{x} - k_{y} \frac{\Nabla x \cdot \Nabla y}{| \Nabla x |^{2}}, k_{y}, \chi \right) = \phi \left( k_{x} ( K_{x}, k_{y}, \chi ), k_{y}, \chi \right) . \label{eq:PhiFourierFormY}
\end{align}
Then, we apply this to the left and right sides of equation \refEq{eq:parallelBCfourierKx} and substitute equation \refEq{eq:KxDef} to arrive at equation \refEq{eq:parallelBCfourierkx}.

Now that we know the boundary conditions and allowed modes, we will construct the coordinate system grids. We could determine our $K_{x}$ grid by taking equation \refEq{eq:KxAllowed} and choosing $m \in \left[ - \left( N_{x} - 1 \right) / 2, \left( N_{x} - 1 \right) / 2 \right]$ at all parallel locations. However, this will accomplish nothing new. We would have created the exact same grid for the exact same physical situation as with the conventional flux tube, just with different labels on the grid points. In other words, the new $K_{x}$ grid would still be centered around $k_{x} = 0$. Instead, we would like to lay down a grid centered around $K_{x} = 0$. However, in general it is not possible to do this exactly. In order to respect the radial boundary condition, we must adhere to equation \refEq{eq:KxAllowed}, which prohibits a grid point at exactly $K_{x} = 0$ at most parallel locations. Thus, we must be content to center the local radial wavenumber grid around $K_{x} \approx 0$, instead of exactly $0$. This can be accomplished by first finding the allowed mode number that comes closest to $K_{x} = 0$, given by
\begin{align}
  m_{0} \left( k_{y}, \chi \right) = - \text{NINT} \left[ \frac{L_{x}}{2 \pi} k_{y} \frac{\Nabla x \cdot \Nabla y}{\left| \Nabla x \right|^{2}} \right] , \label{eq:selectedModes}
\end{align}
where $\text{NINT} [\ldots]$ is a function that takes a real number and returns the nearest integer. Using equation \refEq{eq:selectedModes}, we can construct the $K_{x}$ grid around this mode number according to
\begin{align}
  K_{x} \in \frac{2 \pi}{L_{x}} \left( m + m_{0} \left( k_{y}, \chi \right) \right) + k_{y} \frac{\Nabla x \cdot \Nabla y}{\left| \Nabla x \right|^{2}} ~~~~~~~ \text{for all} ~ m \in \left[ - \frac{N_{x} - 1}{2}, \frac{N_{x} - 1}{2} \right] . \label{eq:KxGrid}
\end{align}
Thus, we see that the $K_{x}$ grid changes for different binormal wavenumbers and at different parallel locations, but always stays as close as possible to $K_{x} = 0$. The $k_{y}$ grid remains the same as in the conventional flux tube, specified according to equation \refEq{eq:kyGridy}.

We can now see the impact of this new $K_{x}$ grid on how a gyrokinetic code models turbulence. In our previous discussion concerning figure \ref{fig:crossSec}, we saw that the conventional flux tube prioritized turbulence that follows field lines rather than turbulence with minimal finite gyroradius damping. This could be seen by noting the secular dependence on $\hat{s} \chi$ in the argument of the Bessel function for $k_{x} = 0$, the central radial mode in the simulation domain (see equations \refEq{eq:bessely} and \refEq{eq:grady}). Now we can repeat the exercise for $K_{x} = 0$, the central radial mode in the new coordinate system. By substituting the definitions of $Y$ and $K_{x}$ (i.e. equations \refEq{eq:Ydef} and \refEq{eq:KxDef}) into equation \refEq{eq:kperpDefy}, we find that the perpendicular wavenumber becomes
\begin{align}
  K_{\perp} = \sqrt{K_{x}^{2} \left| \Nabla x \right|^{2} + k_{y}^{2} \left| \Nabla Y \right|^{2}} , \label{eq:kperpDefY}
\end{align}
which is similar to equation \refEq{eq:kperpDefy} except the cross term vanishes because $\Nabla x \cdot \Nabla Y = 0$. Thus, when we calculate the Bessel function for the central radial mode in the grid, $K_{x} \approx 0$, we find
\begin{align}
J_{0} \left( K_{\perp} \rho_{s} \right) = J_{0} \left( \left| k_{y}\right| \rho_{s} \left| \Nabla Y \right| \right) . \label{eq:besselY}
\end{align}
Using equation \refEq{eq:Yform}, this new geometric coefficient can be found to be
\begin{align}
\Nabla Y = \pm C_{y} \Nabla \zeta \mp C_{y} q_{0} \Nabla \chi \pm C_{y} q_{0} \frac{\Nabla x \cdot \Nabla \chi}{\left| \Nabla x \right|^{2}} \Nabla x , \label{eq:gradY}
\end{align}
which has no secular dependence with $\chi$. Thus, while the flux tube cross-section may expand and contract periodically due to the effect of flux surface shaping and toroidicity, it will not twist at all. This means the domain will include the modes that are least damped by finite gyroradius effects, regardless of the strength of the magnetic shear or the length of the simulation domain.

Lastly, the final step in changing our coordinate system is to transform the gyrokinetic model. Substituting equations \refEq{eq:Ydef}, \refEq{eq:KxDef}, and \refEq{eq:PhiFourierFormY}, we find that the gyrokinetic equation (i.e. equation \refEq{eq:GKeqy}) becomes
\begin{align}
\frac{\partial H_{s}}{\partial t} &+ v_{||} \hat{b} \cdot \vec{\nabla} \chi \left. \frac{\partial H_{s}}{\partial \chi} \right|_{K_{x} - k_{y} \frac{\Nabla x \cdot \Nabla y}{\left| \Nabla x \right|^{2}}, k_{y}} + i \vec{v}_{d s} \cdot \left( K_{x} \Nabla x + k_{y} \Nabla Y \right) H_{s} + a_{s ||} \frac{\partial H_{s}}{\partial v_{||}} \nonumber \\
&\mp \frac{1}{J B} \left\{ H_{s}, \Phi J_{0} \left( K_{\perp} \rho_{s} \right) \right\} = \frac{Z_{s} e F_{M s}}{T_{s}} \frac{\partial \Phi}{\partial t} J_{0} \left( K_{\perp} \rho_{s} \right) \mp i \frac{k_{y}}{J B} \Phi J_{0} \left( K_{\perp} \rho_{s} \right) \frac{\partial F_{Ms}}{\partial x} . \label{eq:GKeqY}
\end{align}
where $H_{s}$ is defined analogously to the definition of $\Phi$ in equation \refEq{eq:PhiFourierFormY}. The nonlinear term is either
\begin{align}
  \left\{ H_{s}, \Phi J_{0} \left( K_{\perp} \rho_{s} \right) \right\} &= \sum_{K_{x}', k_{y}'} \left( K_{x}' k_{y}'' - K_{x}'' k_{y}' \right) H_{s}' \Phi'' J_{0} \left( K_{\perp}'' \rho_{s} \right) \label{eq:nonlinearFourierY}
\end{align}
using $K_{x}'' = K_{x} - K_{x}'$ and $k_{y}'' = k_{y} - k_{y}'$ or 
\begin{align}
  \left\{ H_{s}, \Phi J_{0} \left( K_{\perp} \rho_{s} \right) \right\} &= \frac{1}{N_{x} \left( 2 N_{y} - 1 \right)} \sum_{x, Y} \left[ \left( \sum_{K'_{x}, k'_{y}} K_{x}' H_{s}' \Exp{i K'_{x} x + i k'_{y} Y} \right) \left( \sum_{K''_{x}, k''_{y}} k_{y}'' \Phi'' J_{0} \left( K_{\perp}'' \rho_{s} \right) \Exp{i K''_{x} x + i k''_{y} Y} \right) \right. \label{eq:nonlinearRealY} \\
  &- \left. \left( \sum_{K'_{x}, k'_{y}} k_{y}' H_{s}' \Exp{i K'_{x} x + i k'_{y} Y} \right) \left( \sum_{K''_{x}, k''_{y}} K_{x}'' \Phi'' J_{0} \left( K_{\perp}'' \rho_{s} \right) \Exp{i K''_{x} x + i k''_{y} Y} \right) \right] \Exp{-i K_{x} x - i k_{y} Y} \nonumber
\end{align}
using the new Fourier transform of
\begin{align}
  \Phi \left( K_{x}, k_{y}, \chi \right) = \frac{1}{N_{x} \left( 2 N_{y} - 1 \right)} \sum_{x, Y} \bar{\Phi} \left( x, Y, \chi \right) \Exp{- i K_{x} x - i k_{y} Y} . \label{eq:invFourierSeriesY}
\end{align}
Importantly, in order to retain a field-aligned grid, we see that the parallel derivative in equation \refEq{eq:GKeqY} must still be taken at constant $k_{x} = K_{x} - k_{y} \Nabla x \cdot \Nabla y/ | \Nabla x |^{2}$. If we applied the chain rule in order to take the parallel derivative at constant $K_{x}$, we would no longer be able to use a coarse grid in $\chi$. Quasineutrality remains the same as equation \refEq{eq:quasineuty}, except $H_{s}$ and $\Phi$ are functions of $K_{x}$ instead of $k_{x}$ and the perpendicular wavenumber is given by equation \refEq{eq:kperpDefY}. Thus, the complete non-twisting flux tube in Fourier-space is defined by equations \refEq{eq:kyGridy}, \refEq{eq:quasineuty} (using $\Phi$, $H_{s}$, and $K_{\perp}$), \refEq{eq:parallelBCfourierKx}, \refEq{eq:KxGrid}, \refEq{eq:kperpDefY}, \refEq{eq:GKeqY}, and \refEq{eq:nonlinearRealY}. The new geometric coefficients and equilibrium quantities that appear in these equations are simple to relate to those already calculated for the conventional flux tube. Using equation \refEq{eq:Ydef}, we find
\begin{align}
  \Nabla x \cdot \Nabla Y &= 0 \label{eq:gradXdotGradY} \\
  \left| \Nabla Y \right|^{2} &= \left| \Nabla y \right|^{2} - \frac{\left( \Nabla x \cdot \Nabla y \right)^{2}}{\left| \Nabla x \right|^{2}} \label{eq:gradYSq} \\
   \Nabla Y \cdot \Nabla \chi &=  \Nabla y \cdot \Nabla \chi - \frac{\Nabla x \cdot \Nabla y}{\left| \Nabla x \right|^{2}} \Nabla x \cdot \Nabla \chi \label{eq:gradYGradChi} \\
   J^{-1} &\equiv \mp \left( \Nabla x \times \Nabla Y \right) \cdot \hat{b} = \mp \left( \Nabla x \times \Nabla y \right) \cdot \hat{b} \label{eq:jacob} \\
   \vec{B} &= \frac{1}{C_{y}} \frac{d \psi}{d x} \Nabla x \times \Nabla y = \frac{1}{C_{y}} \frac{d \psi}{d x} \Nabla x \times \Nabla Y \label{eq:magField} \\
   \Nabla B &= \left( \frac{\partial B}{\partial x} + \frac{\Nabla x \cdot \Nabla y}{\left| \Nabla x \right|^{2}} \frac{\partial B}{\partial y} \right) \Nabla x + \frac{\partial B}{\partial y} \Nabla Y + \frac{\partial B}{\partial \chi} \Nabla \chi . \label{eq:gradB}
\end{align}

Before we move on, we have two comments that are important to provide perspective on the non-twisting flux tube. First, we believe that the crucial novel insight of this paper is specifying the radial wavenumber grid according to equations \refEq{eq:selectedModes} and \refEq{eq:KxGrid}. It reveals that not every parallel location permits Fourier modes with a radial wavenumber of zero. Prior works \cite{ToldShiftedMetric2010, ScottShiftedMetric2001} appear to have presupposed the existence of a local radial wavenumber of $K_{x} = 0$ at all parallel locations. This created a number of problems, which made a practical implementation appear impossible. Notably, if a mode exists at $K_{x} = 0$, this can violate the radial boundary condition, necessitate a prohibitively small $K_{x}$ grid spacing, and/or require arbitrary non-uniform $K_{x}$ grid spacing when using general geometry. All of these difficulties are resolved by constructing the grid according to equation \refEq{eq:KxGrid}, which prohibits the $K_{x} = 0$ mode at most parallel locations.

Second, the analytic derivation presented in this section actually accomplishes remarkably little. We have simply rewritten the Fourier-analyzed gyrokinetic model in terms of a new radial wavenumber $K_{x}$, which absorbs some of the information that was previously contained in the geometric coefficient $\Nabla y$. As noted above, if we constructed a $K_{x}$ grid using equation \refEq{eq:KxAllowed} and $m \in \left[ - \left( N_{x} - 1 \right) / 2, \left( N_{x} - 1 \right) / 2 \right]$ at all parallel locations, nothing consequential would change about the representation. Our change of coordinate system would be purely semantics and running a simulation would produce the exact same results as the conventional coordinate system. This again highlights the importance of equations \refEq{eq:selectedModes} and \refEq{eq:KxGrid}. Performing the change of coordinates to $K_{x}$ is not what creates a non-twisting flux tube, rather the essential step is {\it constructing a near-rectangular grid} in the new $\left( K_{x}, k_{y} \right)$ coordinates. Alternatively, one could have created a non-twisting flux tube without changing coordinates by constructing a non-rectangular grid in $\left( k_{x}, k_{y} \right)$. Specifically one would use a grid centered around $k_{x} \approx - k_{y} \Nabla x \cdot \Nabla y / | \Nabla x |^{2}$, which can be found by expressing the grid of equation \refEq{eq:KxGrid} in terms of $k_{x}$ by substituting equation \refEq{eq:KxDef}. In fact, this is essentially what is done in the shifted metric approach. Nevertheless, we believe that using a coordinate system transformation, as presented in this paper, is more intuitive and demonstrates the constraints on the system (e.g. boundary conditions) well.

\section{Code benchmarking}
\label{sec:benchmarking}

\begin{table}
	\centering
	\begin{tabular}{c|c||c|c}
		Parameter & Value & Parameter & Value \\
		\hline
		Minor radius of flux tube, $x_{0}/a$ & 0.54 & Major radius, $R_{0}/a$ & $3.0$ \\
		\hline
		Safety factor, $q_{0}$ & $1.4$ & Magnetic shear, $\hat{s}$ & $0.8$ \\
		\hline
		Temperature gradient, $a/L_{Ts}$ & $2.3$ & Density gradient, $a/L_{n}$ & $0.733$ \\
		\hline
		Ion-e\textsuperscript{-} mass ratio, $m_{i}/m_{e}$ & $4590$ & Ion-e\textsuperscript{-} temperature ratio, $T_{i}/T_{e}$ & $1.0$ \\
		\hline
		Effective ion charge, $Z_{eff}$ & $1.0$ & 	4\textsuperscript{th} order $y$ hyperdiffusion \cite{PueschelHyperDiff2010}, $\epsilon_{y}$ & $0.0$ \\
		\hline
		4\textsuperscript{th} order $\chi$ hyperdiffusion \cite{PueschelHyperDiff2010}, $\epsilon_{\chi}$ & $0.2$ & 4\textsuperscript{th} order $v_{||}$ hyperdiffusion \cite{PueschelHyperDiff2010}, $\epsilon_{v||}$ & $0.2$ \\
	\end{tabular}
	\caption{The nominal Cyclone Base Case (CBC) parameters \cite{DimitsCBC2000} used in this work. Unless otherwise noted the geometry is specified using the s-$\alpha$ model \cite{GreeneS_Alpha1981} with $\alpha = 0$ and the simulations are electrostatic and collisionless.}
	\label{tab:CBC}
\end{table}

\begin{table}
	\centering
	\begin{tabular}{c||c|c||c|c }
		Coordinate    & \multicolumn{2}{c||}{Grid range}
		& \multicolumn{2}{c}{Number of grid points}                \\
		\hline
		& Adiabatic & Kinetic & Adiabatic & Kinetic  \\
		\hline
		$x / \rho_{i}$ & $[ - 100, 100 )$ & $[ - 92, 92 )$ & $N_{x}$ & $N_{x}$ \\
		\hline
		$y / \rho_{i}$ & $[ - 63, 63 )$ & $[ - 63, 63 )$ & $64$ & $64$ \\
		\hline
		$\chi$ & $[ - N_{\text{pol}} \pi, N_{\text{pol}} \pi )$ & $[ - N_{\text{pol}} \pi, N_{\text{pol}} \pi )$ & $20 N_{\text{pol}}$ & $48 N_{\text{pol}}$ \\
		\hline
		$v_{||} / v_{th,s}$ & $[ - 3, 3 ]$ & $[ - 3, 3 ]$ & $32$ & $64$ \\
		\hline
		$\sqrt{\mu / (T_{s} / B_{r})}$ & $( 0, 2.75 )$ & $( 0, 2.75 )$ & $10$ & $10$ \\
		\hline
		$t/(a/c_{S})$ & $[1000 N_{\text{pol}}, 6000]$ & $[30, 200]$ & \multicolumn{2}{c}{time step $<$ CFL limit \cite{CourantCFL1967}} \\
	\end{tabular}
	\caption{The nominal GENE coordinate grids used in this work. Note that all grids are equally spaced, $v_{th,s} = \sqrt{2 T_{s} / m_{s}}$ is the thermal velocity of species $s$, and $c_{S} = \sqrt{T_{e} / m_{i}}$ is the sound speed.}
	\label{tab:defaultRes}
\end{table}

In this section, we will benchmark our implementation of the non-twisting flux tube in GENE. The practical details involved in implementing the results of section \ref{sec:derivNonTwist} are fairly technical and are therefore relegated to \ref{app:implementation}. However, this section on benchmarking is important because it presents a concrete illustration of the relationship between the grids in the non-twisting and conventional flux tubes. Unless otherwise noted, we will use standard Cyclone Base Case (CBC) parameters \cite{DimitsCBC2000} given in table \ref{tab:CBC} using the resolutions given in table \ref{tab:defaultRes}. Our first test is to compare linear growth rates against the conventional flux tube when the two coordinate systems are identical. To see when this occurs, we must return to the discussion surrounding equation \refEq{eq:selectedModes}. Remember, if $m_{0} \left( k_{y}, \chi \right) = 0$ for all binormal modes and all parallel locations, our modified radial grid becomes functionally identical to that of the conventional flux-tube. In other words, the non-twisting and conventional flux tubes model the exact same physical situation using the exact same grid, just with different labels on the grid points.

\begin{figure}
	(a) \hspace{18em} (b)~\\
	\includegraphics[width=0.49\textwidth]{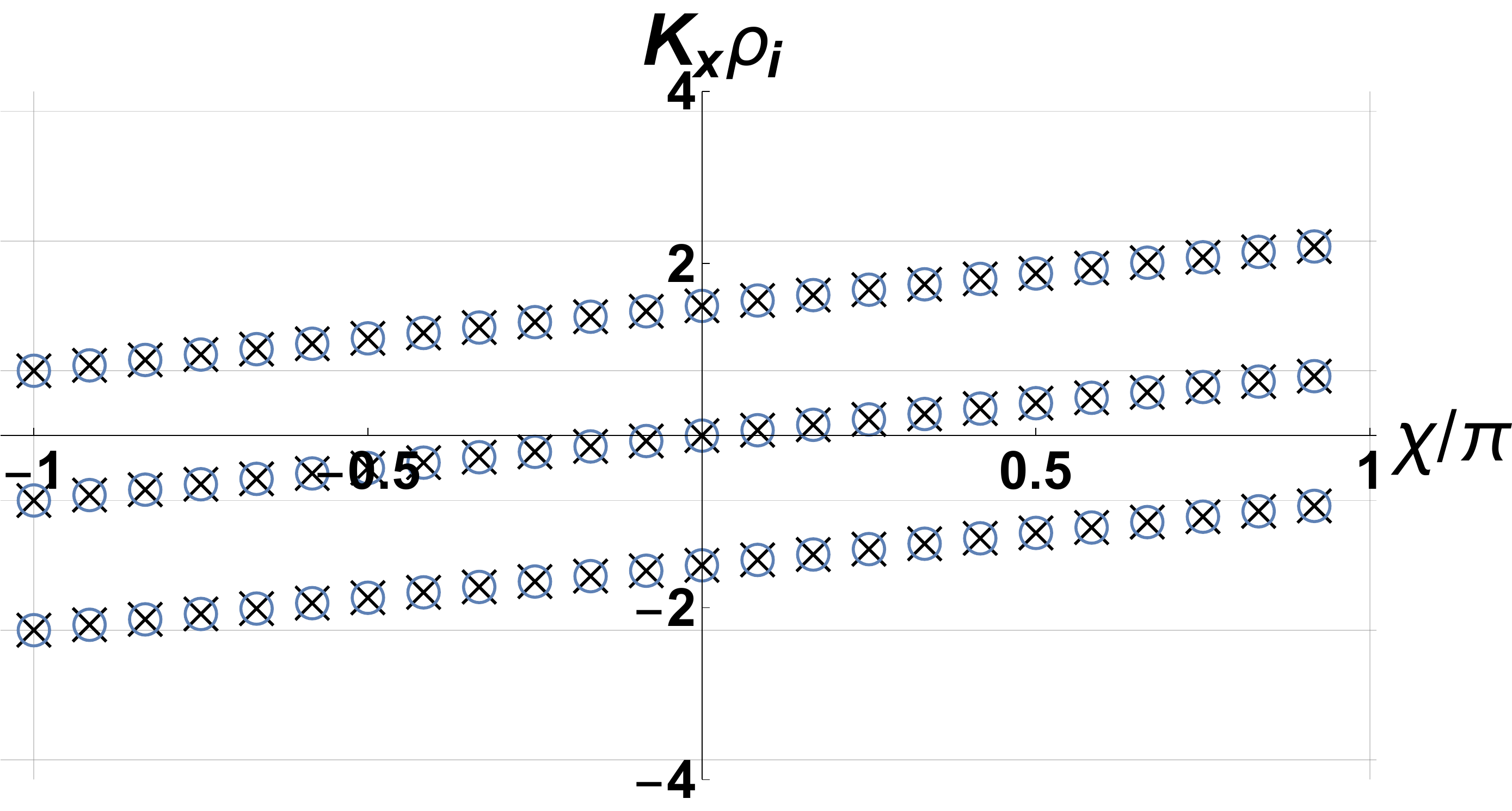} \hspace{1em}
	\includegraphics[width=0.49\textwidth]{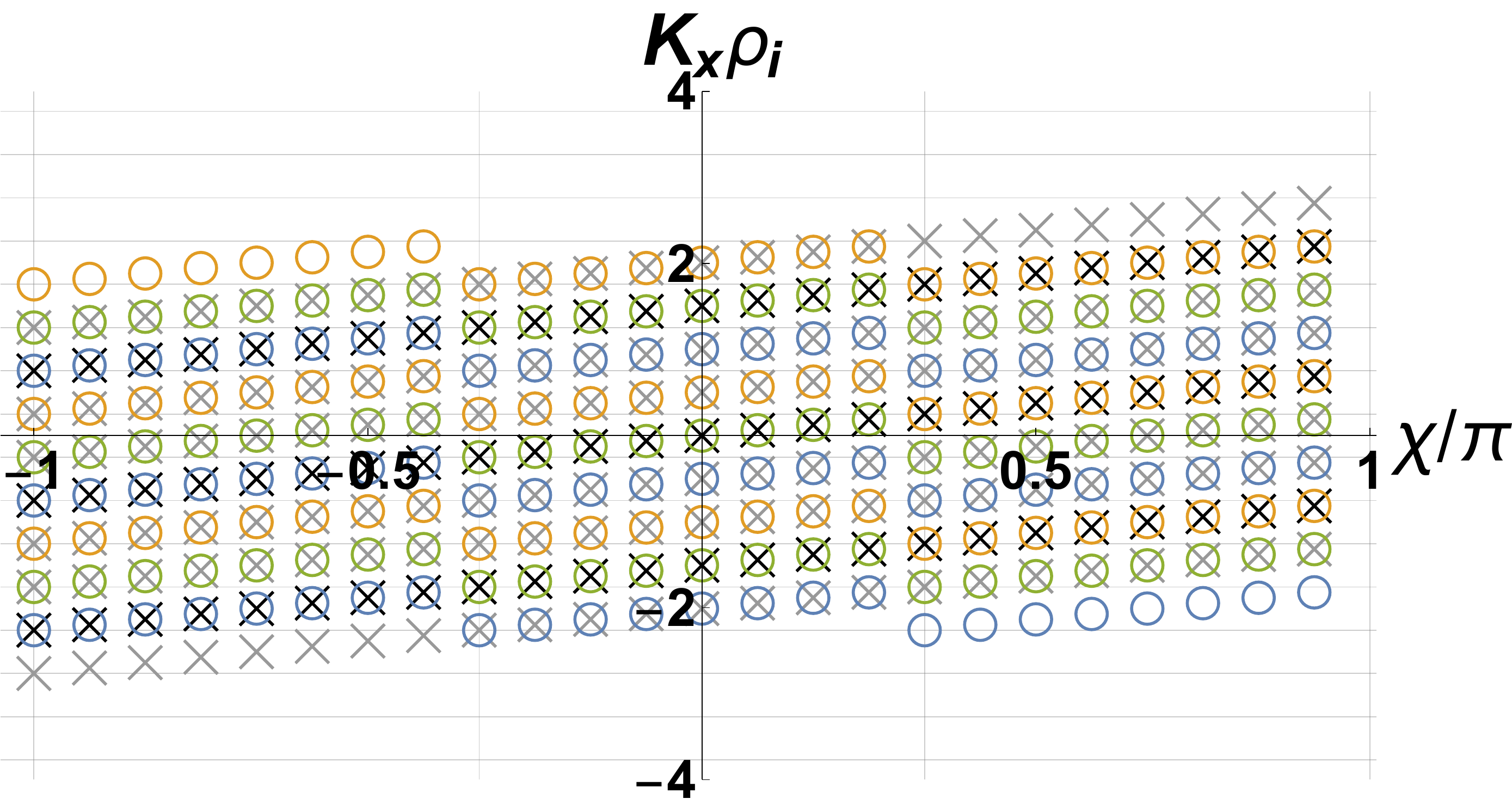}
	(c) ~\\
	\includegraphics[width=1.0\textwidth]{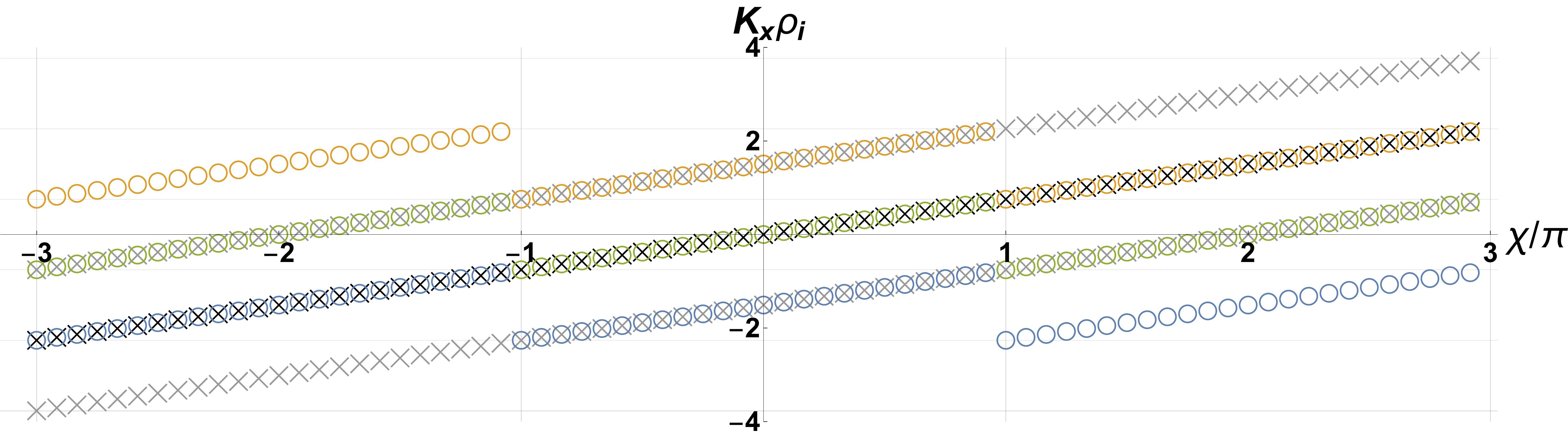}
	(d) ~\\
	\includegraphics[width=1.0\textwidth]{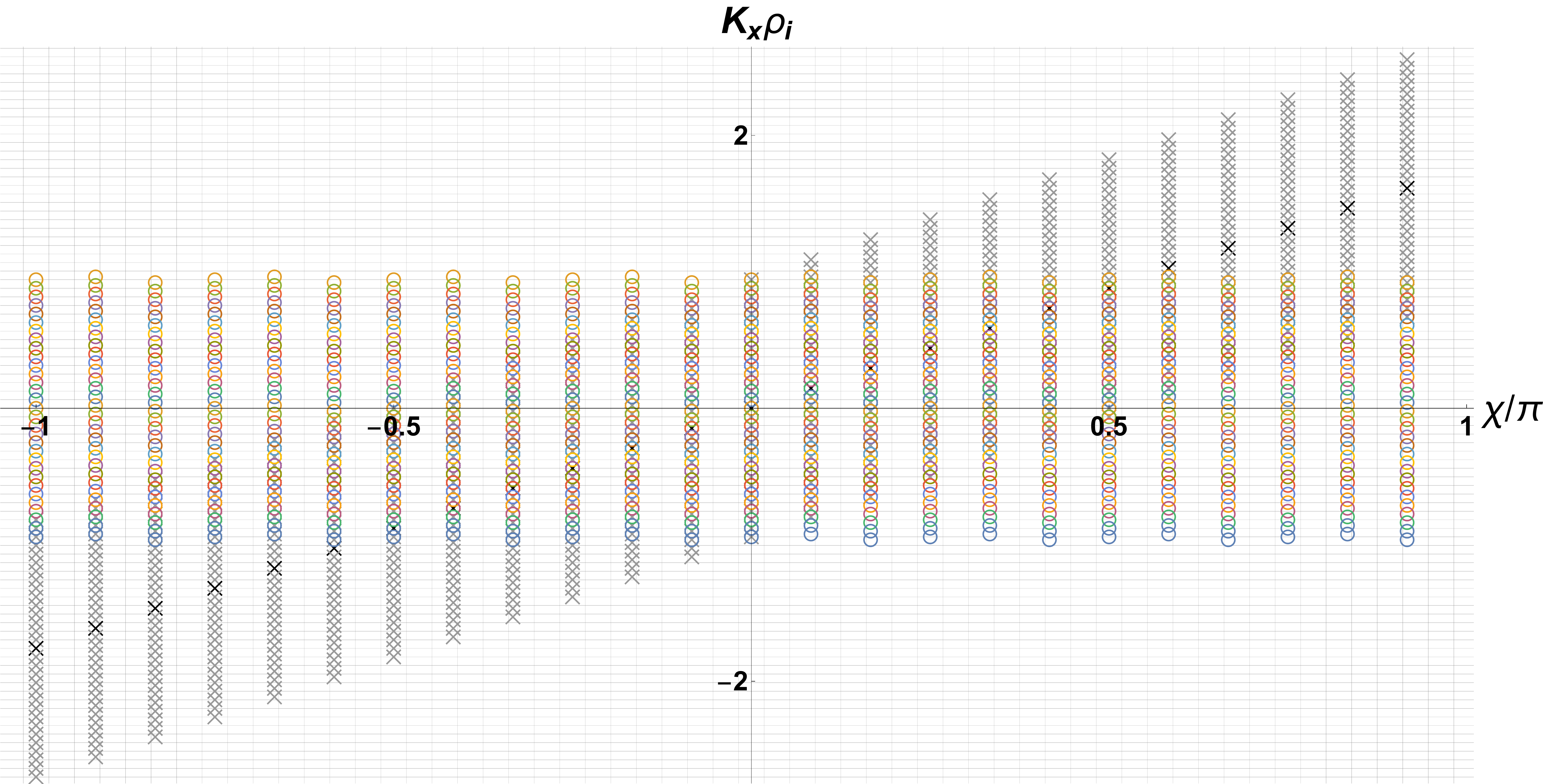}
	\caption{The parallel-radial spatial grid expressed in $\left( \chi, K_{x} \right)$ from the non-twisting (colored circles) and conventional (black and gray crosses) flux tubes for (a) $N_{\text{asp}} = 1$, $N_{\text{pol}} = 1$, $N_{x} = 3$; (b) $N_{\text{asp}} = 3$, $N_{\text{pol}} = 1$, $N_{x} = 9$; (c) $N_{\text{asp}} = 3$, $N_{\text{pol}} = 3$, $N_{x} = 3$; and (d) $N_{\text{asp}} = 4$, $N_{\text{pol}} = 1$, $N_{x} = 31$. The first three plots are grids for linear simulations and use a single binormal mode $k_{y} \rho_{i} = 0.3$, while the last plot is a nonlinear grid with $k_{y} \rho_{i} = 0.7$ and a minimum wavenumber of $k_{y,\text{min}} \rho_{i} = 0.05$. All use $N_{\chi} = 24 N_{\text{pol}}$ parallel grid points and have only global magnetic shear. The vertical grid lines and changes in color indicate where linear modes couple to a new radial index in the non-twisting flux tube (i.e. the argument to the $\text{NINT}$ function is half-integer), the horizontal grid lines correspond to the connections across the parallel boundary, and the black crosses indicate the linear mode with zero ballooning angle.}
	\label{fig:gridLayouts}
\end{figure}

\begin{figure}
	\centering
	\includegraphics[width=1.0\textwidth]{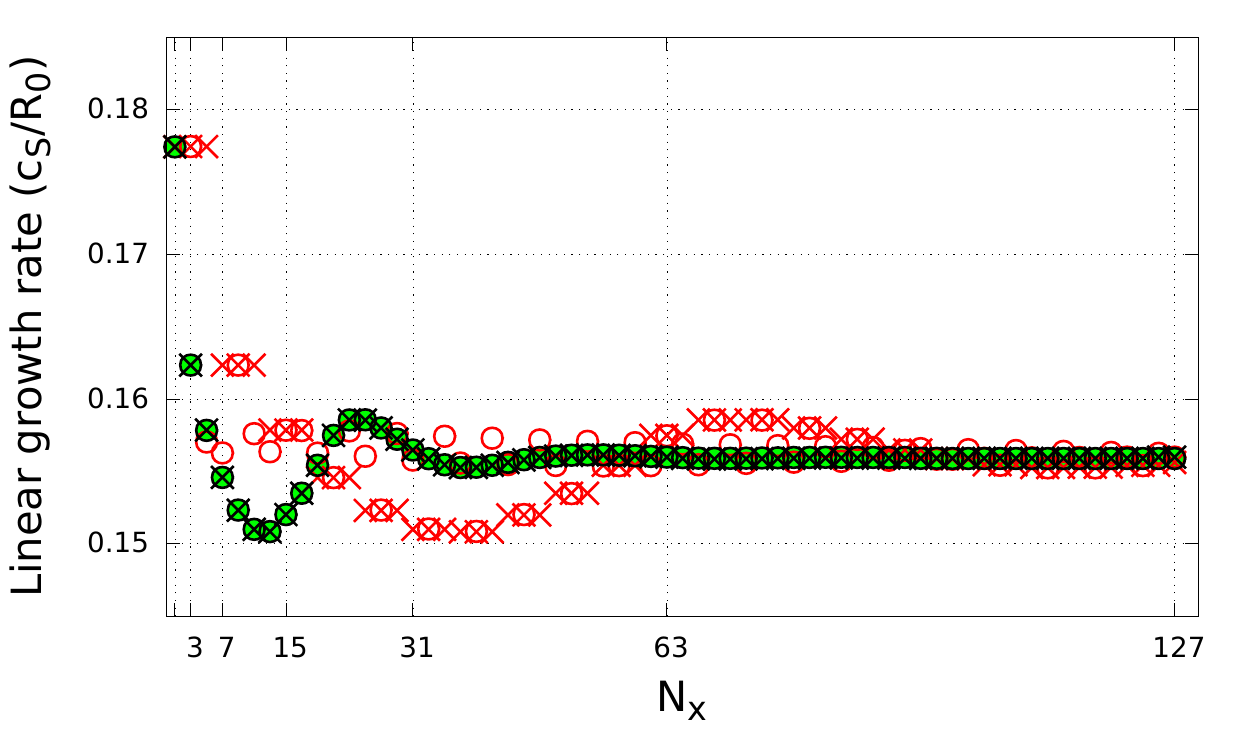}
	\caption{A resolution study of the linear growth rate of the most unstable mode using the non-twisting (empty or filled circles) or conventional (crosses) flux tube with $N_{\text{asp}} = 1$ and $N_{\text{pol}} = 1$ (black), $N_{\text{asp}} = 3$ and $N_{\text{pol}} = 1$ (red), or $N_{\text{asp}} = 3$ and $N_{\text{pol}} = 3$ (filled green circles). These correspond to the grids from figure \ref{fig:gridLayouts}(a)-(c).}
	\label{fig:linearResults}
\end{figure}

To simulate a single linear mode, one sets $N_{\text{asp}} = 1$ and $k_{y} = 2 \pi / L_{y}$. Thus, we see from equation \refEq{eq:selectedModes} that $m_{0} \left( k_{y}, \chi \right)$ becomes
\begin{align}
  m_{0} \left( \frac{2 \pi}{L_{y}}, \chi \right) = - \text{NINT} \left[ \mp \frac{1}{2} \text{sign} \left( \hat{s} \right) \frac{\chi}{\pi N_{\text{pol}}} \mp \frac{C_{y} q_{0}}{2 \pi N_{\text{pol}} \left| \hat{s} \right|} \frac{\Nabla x \cdot \Nabla \chi}{\left| \Nabla x \right|^{2}} \right] , \label{eq:linearProof}
\end{align}
where we have used equations \refEq{eq:aspRatioDiscrete} and \refEq{eq:Yfactor}. In general this expression is {\it not} necessarily zero because of the local correction to the magnetic shear, which is represented by the second term. For tokamaks with extreme shaping or a tight aspect ratio, the second term can be made arbitrarily large, meaning the expression will round to a non-zero value at some parallel locations. However, for typical tokamak parameters the local correction to the shear is often small. In fact, for circular flux surfaces in the large aspect ratio limit, this local correction vanishes entirely. This can be accomplished in GENE by specifying the magnetic geometry using the s-$\alpha$ representation \cite{GreeneS_Alpha1981} and setting $\alpha = 0$. Thus, with only global shear equation \refEq{eq:linearProof} becomes
\begin{align}
  m_{0} \left( \frac{2 \pi}{L_{y}}, \chi \right) = - \text{NINT} \left[ \mp \frac{1}{2} \text{sign} \left( \hat{s} \right) \frac{\chi}{\pi N_{\text{pol}}} \right] . \label{eq:linearProofNoLocalShear}
\end{align}
Given the range of $\chi \in \left[ - \pi N_{\text{pol}}, \pi N_{\text{pol}} \right)$, this expression will necessarily round to zero at all parallel locations as long as we are careful to properly round at $\chi = - \pi N_{\text{pol}}$ (see \ref{app:implementation} for more details). This ensures that the two grids are identical, as is shown in figure \ref{fig:gridLayouts}(a). Therefore, we have derived a simple linear test case for which the non-twisting and conventional flux tubes should produce identical results --- using a single $k_{y}$ mode with $N_{\text{asp}} = 1$ in the s-$\alpha$ geometry model with $\alpha = 0$. The linear growth rates for this test are shown in figure \ref{fig:linearResults} by the black markers. We see perfect agreement as expected, regardless of the number of radial grid points $N_{x}$ (which determines how far the linear mode is allowed to extend in ballooning space). This test gives confidence that the modifications to the geometric coefficients was done properly (e.g. replacing $|\Nabla y|^{2}$ with $|\Nabla Y|^{2}$).

In the next test case, we change $N_{\text{asp}}$ from $1$ to $3$. While $N_{x}$ determines how far the linear modes extend in ballooning space, the value of $N_{\text{asp}}$ determines the number of independent linear modes in the system. This can be seen in figure \ref{fig:gridLayouts}(b). Note that {\it both} the non-twisting and conventional flux tubes now contain three linear modes, whereas they only contained one linear mode in figure \ref{fig:gridLayouts}(a). We also see that when $N_{\text{asp}} \neq 1$ the non-twisting and conventional grids can differ. In particular, we see that linear modes in the non-twisting flux tube now can fall off the computational grid at internal parallel locations. This differs from the conventional flux tube, in which linear modes can only terminate at the parallel boundaries. However, for the CBC parameters used for this study, we know that the linear mode with zero ballooning angle (i.e. the one passing through the origin) will have the largest growth rate. Looking closely at the example shown in figure \ref{fig:gridLayouts}(b), which uses $N_{x} = 9$, we see that the linear mode with zero ballooning angle has the exact same grid points in the two flux tubes, even though the grids differ for the other two linear modes. This coincidence only occurs at specific values of $N_{x}$ (i.e. $N_{x}$ is an integer multiple of $N_{\text{asp}}$), which makes it a good test case. Studying the growth rates for this test case (shown in red in figure \ref{fig:linearResults}), we see that at $N_{x} = 9$ the two flux tubes produce the same growth rate, as is expected from figure \ref{fig:gridLayouts}(b). Moreover, the two flux tubes agree at every third value of $N_{x}$, while they do not for other values of $N_{x}$. This pattern exactly reflects when the grids of the zero ballooning angle linear mode are identical and when they differ. Additionally, we see that the growth rate at every third value of $N_{x}$ agrees with the growth rate from the $N_{\text{asp}} = 1$ case (shown in black) at $N_{x} / 3$. This is because they also have identical grids for the zero ballooning angle linear mode, which is the case between figures \ref{fig:gridLayouts}(a) and \ref{fig:gridLayouts}(b). The fact that the growth rates exactly inherit all these underlying patterns from the coordinate systems grids confirms that the mode coupling in the parallel derivative is implemented correctly.

Next, we set $N_{\text{pol}} = 3$ to extend the length of the flux tube to three poloidal turns. By also choosing $N_{\text{asp}} = 3$, we model three identical linear modes that are each the same as our first $N_{\text{pol}} = 1$ test case at all values of $N_{x}$. The only thing that has changed is that some of the mode coupling that was occurring at the ends of the flux tube has been moved into the interior. Accordingly, we see that the growth rates in figure \ref{fig:linearResults} (green and black points) agree well at all values of $N_{x}$. This verifies that the non-twisting flux tube has no distinction between mode coupling at the parallel boundary condition and mode coupling within the domain. We note that all of the test cases, in both the non-twisting and conventional flux tubes, converge to the same value at sufficiently high radial resolution. This is important because all are modeling the same physical situation just with different numerical grids.

\begin{figure}
	\centering
	\hspace{-18em} (a) \hspace{17em} (b) \\
	\includegraphics[width=0.49\textwidth]{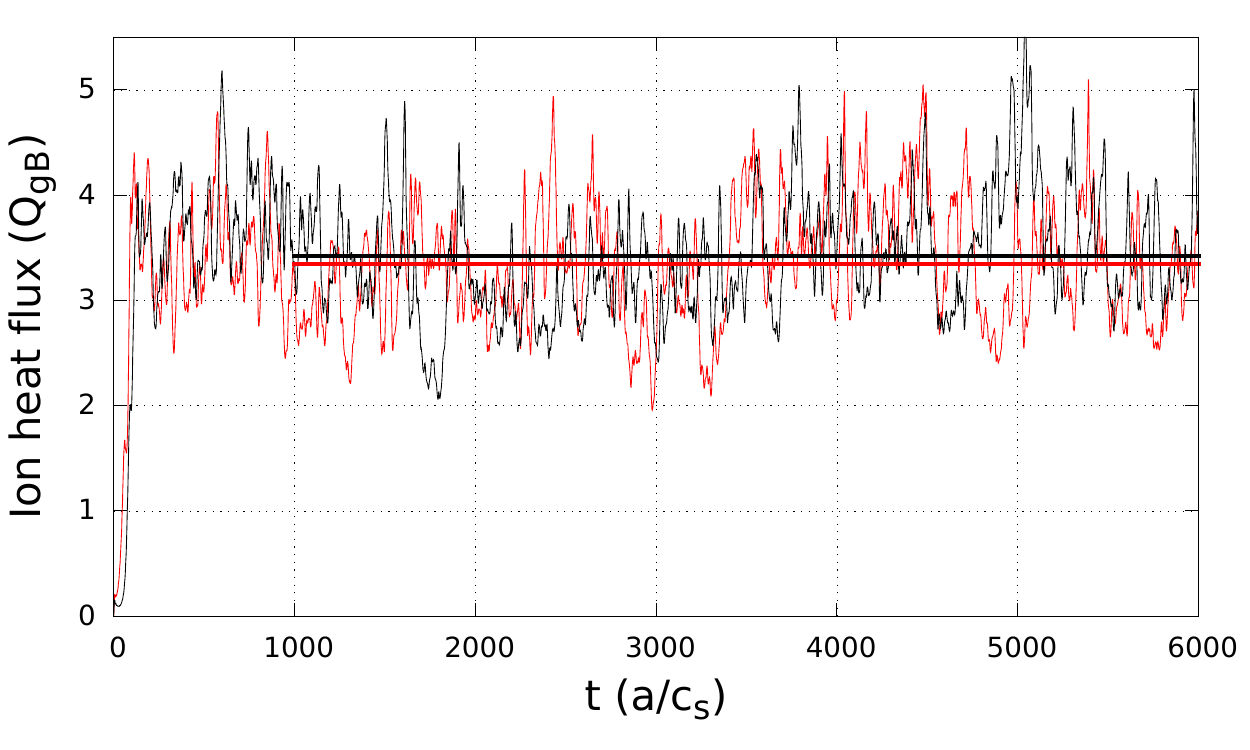}
	\includegraphics[width=0.49\textwidth]{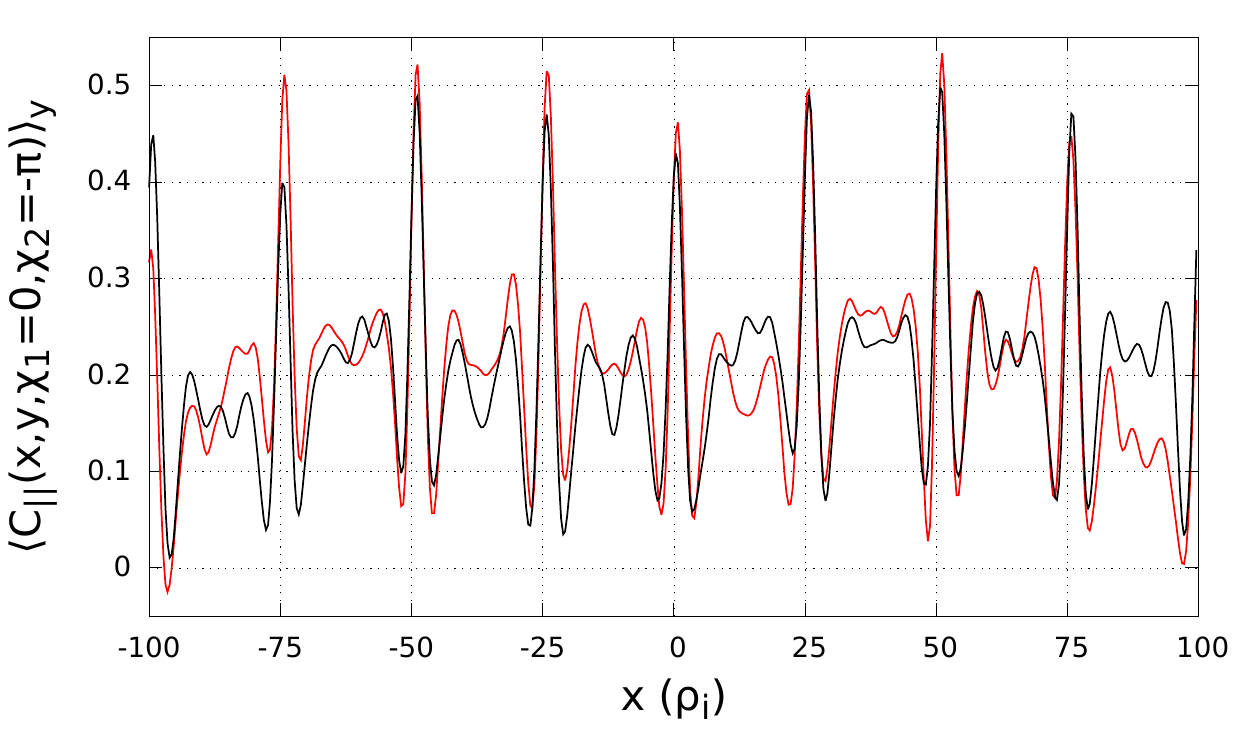}
	\caption{A nonlinear benchmark at the high radial resolution of $N_{x} = 512$ with $N_{\text{pol}} = 1$: (a) the time traces of the turbulent ion heat flux (thin lines) together with their time average over the nonlinearly saturated state (thick lines) and (b) the radial variation of the parallel correlation function for the non-twisting (black) and conventional (red) flux tubes. In (b) the locations of the pseudo-integer magnetic surfaces are indicated by the vertical grid lines.}
	\label{fig:nonlinearBenchmark}
\end{figure}

Lastly, we will perform a nonlinear benchmark. Unless the magnetic shear is zero at all parallel locations, the grids used for nonlinear simulations will always be different. Figure \ref{fig:gridLayouts}(d) shows that the non-twisting flux tube has a grid centered around $K_{x} = 0$, while the conventional grid is centered around $k_{x} = 0$. This is made possible by the fact that linear modes in the non-twisting flux tube can fall off of the computational grid at any parallel location and do not necessarily span an integer number of poloidal turns. On the other hand, figure \ref{fig:gridLayouts}(d) shows that linear modes in the conventional flux tube only ever terminate at the parallel boundary and are, thus, always an integer number of poloidal turns long. Regardless, as with the linear studies, both flux tubes should produce identical results at sufficiently high radial resolution. Accordingly, we performed a nonlinear simulation with the very high radial resolution of $N_{x} = 512$ using the CBC parameters given in table \ref{tab:CBC}, the resolutions given in table \ref{tab:defaultRes}, $N_{\text{pol}} = 1$, and adiabatic electrons. Figure \ref{fig:nonlinearBenchmark}(a) shows that the time-averaged heat flux agrees very well between the two flux tubes and the time traces look qualitatively similar. Note the heat flux is normalized to the gyroBohm value of $Q_{gB} \equiv (\rho_{i} / a)^{2} n_{e} T_{e} c_{S}$.

As an additional check, we calculate the two-point parallel correlation function
\begin{align}
C_{||} \left( x, y, \chi_{1}, \chi_{2} \right) \equiv \frac{\left\langle \bar{\phi}_{NZ} \left( x, y, \chi_{1}, t \right) \bar{\phi}_{NZ} \left( x, y, \chi_{2}, t \right) \right\rangle_{t}}{\sqrt{ \left\langle \bar{\phi}_{NZ}^{2} \left( x, y, \chi_{1}, t \right) \right\rangle_{t} \left\langle \bar{\phi}_{NZ}^{2} \left( x, y, \chi_{2}, t \right) \right\rangle_{t} }} , \label{eq:parCorrFn}
\end{align}
where the subscript $NZ$ signifies the non-zonal portion of the quantity and $\left\langle \ldots \right\rangle_{u}$ indicates an average over any coordinate $u$. The quantity $C_{||}$ indicates the degree of correlation between two points $\chi_{1}$ and $\chi_{2}$ on {\it the same field line}, which is accomplished by holding $x$ and $y$ constant (as opposed to $x$ and $Y$). To transform to the field-aligned coordinate $y$ from the grid used by the non-twisting flux tube, one must use the inverse Fourier transform
\begin{align}
  \bar{\phi} \left( x, y, \chi \right) &= \sum_{K_{x}, k_{y}} \Phi \left( K_{x}, k_{y}, \chi \right) \Exp{- i k_{y} \left( \Nabla x \cdot \Nabla y / \left| \Nabla x \right|^{2} \right) x} \Exp{i K_{x} x + i k_{y} y} , \label{eq:FourierSeriesKxToy}
\end{align}
which can be derived by substituting equations \refEq{eq:Ydef} and \refEq{eq:PhiRealFormY} into equation \refEq{eq:FourierSeriesY}. Figure \ref{fig:nonlinearBenchmark}(b) shows the $y$-averaged correlation between the inboard and outboard midplanes $\left\langle C_{||} ( x, y, \chi_{1} = 0, \chi_{2} = - \pi ) \right\rangle_{y}$ for the two flux tubes. We see that, despite using different grids, the two parallel correlation functions are very similar. As should be the case, we see that both flux tubes display spikes in the parallel correlation function at ``pseudo-integer'' surfaces \cite{BallBoundaryCond2020}. On these surfaces, due to the parallel boundary condition, the magnetic field lines close on themselves after just one poloidal turn, which is artificial unless the flux tube covers the full flux surface. Thus, if turbulent eddies span a full poloidal turn, they will ``bite their own tails.'' This self-interaction alters the statistical properties of the turbulence, resulting in the radially localized spikes shown in figure \ref{fig:nonlinearBenchmark}(b). Interestingly, the extreme radial resolution reveals that each spike is actually shifted slightly outwards (i.e. to the right in figure \ref{fig:nonlinearBenchmark}(b)) from its pseudo-integer surface by a few ion gyroradii. This appears to be a consequence of the magnetic drifts, which prevent the single particle trajectories from precisely following magnetic field lines. Thus, the turbulence may not be {\it exactly} field-aligned, but can have a slight bias one way or the other. A field line that just misses closing on itself can actually compensate for the drifts and enable a larger degree of correlation. While this slight shift is a minor effect, its presence in both flux tube models gives even more confidence that our implementation of the non-twisting flux tube is correct.

\section{Nonlinear performance results}
\label{sec:results}

In this section, we will investigate the performance of the non-twisting flux tube relative to the conventional flux tube using five test cases:
\begin{enumerate}
 \item[(1)] CBC with adiabatic electrons,
 \item[(2)] CBC with a long computational domain,
 \item[(3)] CBC-like with high magnetic shear,
 \item[(4)] a shaped DEMO equilibrium with kinetic electrons, and
 \item[(5)] CBC-like with high magnetic shear and a long computational domain.
\end{enumerate}
We will compare the runtime of the code (i.e. the total computational time needed to reach a given physical time) to see how it is affected by the modifications for the non-twisting flux tube. Additionally, we will study the convergence with radial resolution while holding the radial domain size $L_{x}$ constant. In other words, we will hold the minimum value of the radial wavenumber constant and increase the maximum value by adding more Fourier modes. This will reveal if centering a grid around $K_{x} \approx 0$ enables convergence with fewer radial modes.

\subsection*{Test case 1}

\begin{figure}
	\centering
	\includegraphics[width=0.7\textwidth]{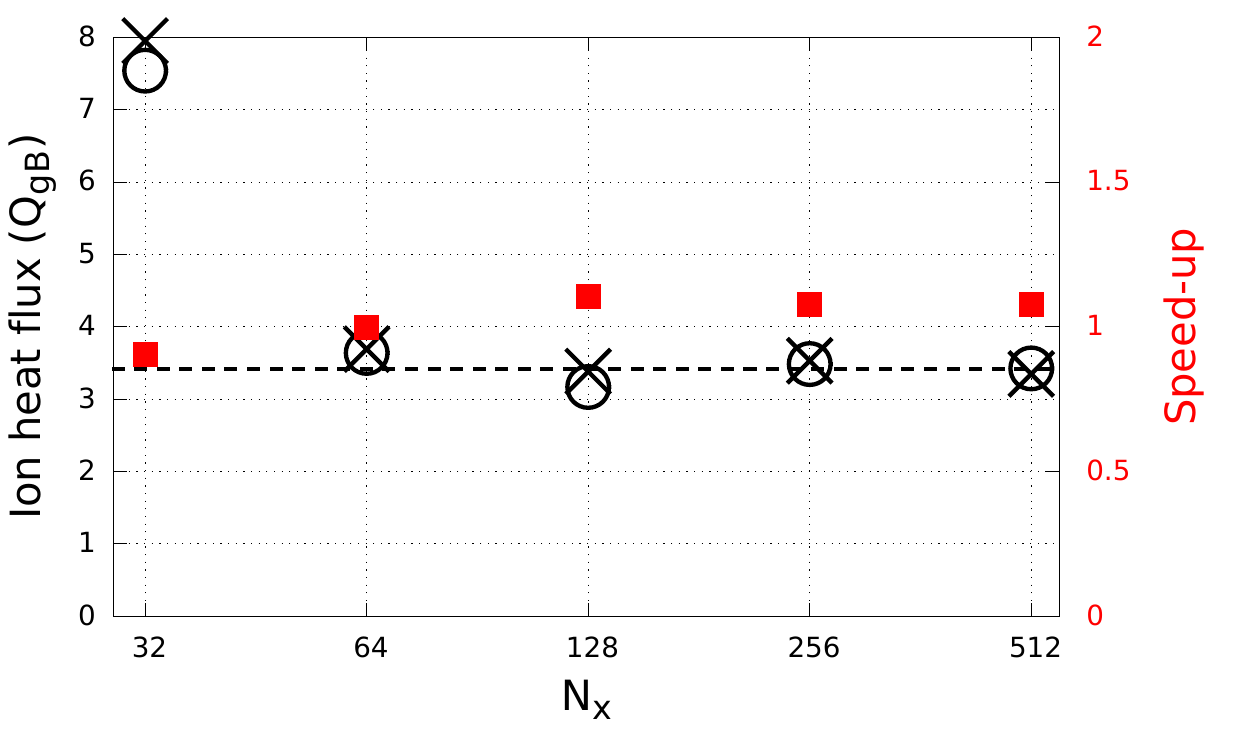}
	\caption{A nonlinear resolution study for the non-twisting (black circles) and conventional (black crosses) flux tube using adiabatic electrons, $N_{\text{pol}} = 1$, CBC parameters with $\hat{s} = 0.8$, and the parameters and resolutions of tables \ref{tab:CBC} and \ref{tab:defaultRes}. The ion heat flux (black) and computational speed-up (red squares) data use the left and right axes respectively and the horizontal dashed line indicates the fully converged heat flux. Each simulation was run on $10 N_{x}$ CPUs (except 2560 CPUs were used for $N_{x} = 512$).}
	\label{fig:nonlinearResStudy}
\end{figure}

For the first test case, we performed standard CBC simulations with the parameters and resolutions given in tables \ref{tab:CBC} and \ref{tab:defaultRes}, the s-$\alpha$ geometry, and adiabatic electrons. These simulations have a modest value of $\hat{s} = 0.8$ and use domains that are just one poloidal turn long. The time averaged turbulent heat fluxes from both flux tubes are shown in figure \ref{fig:nonlinearResStudy}, as is the speed-up (i.e. the total wall clock simulation time of the conventional flux tube divided by that of the non-twisting flux tube). We see that both the heat flux and the computational cost of the two flux tubes are similar at all radial resolutions. Surprisingly, the biggest difference is that the non-twisting flux tube is slightly {\it faster} at high resolution. Looking into the details, while the CPU time required to compute one time step in the non-twisting flux tube is longer by a factor of $\sim 1.2$, the physical time step itself is longer by a factor of $\sim 1.3$. We believe that the slowdown in calculating a time step is because the parallel derivative at fixed $k_{x}$ and $k_{y}$ is no longer taken across continuous sections of memory (see \ref{app:implementation}). However, longer time steps are possible because the local radial wavenumbers of the non-twisting flux tube grid stay closer to $K_{x} = 0$ than in the conventional flux tube, especially at the inboard midplane. This means that, at the same value of $N_{x}$, switching to the non-twisting flux tube simulation will eliminate the finest spatial scales, which allows for a longer time step according to the CFL condition \cite{CourantCFL1967}. Since GENE has a routine for calculating the maximum allowable timestep, it automatically takes advantage of this. In these simulations (as well as all subsequent simulations), no difference was observed in the temporal convergence properties of the turbulent state in the two flux tubes (i.e. all simulations converged to quasi-steady state at roughly the same simulation time).

\subsection*{Test case 2}

\begin{figure}
	\centering
	\hspace{-18em} (a) \hspace{17em} (b) \\
    \includegraphics[width=0.49\textwidth,valign=m]{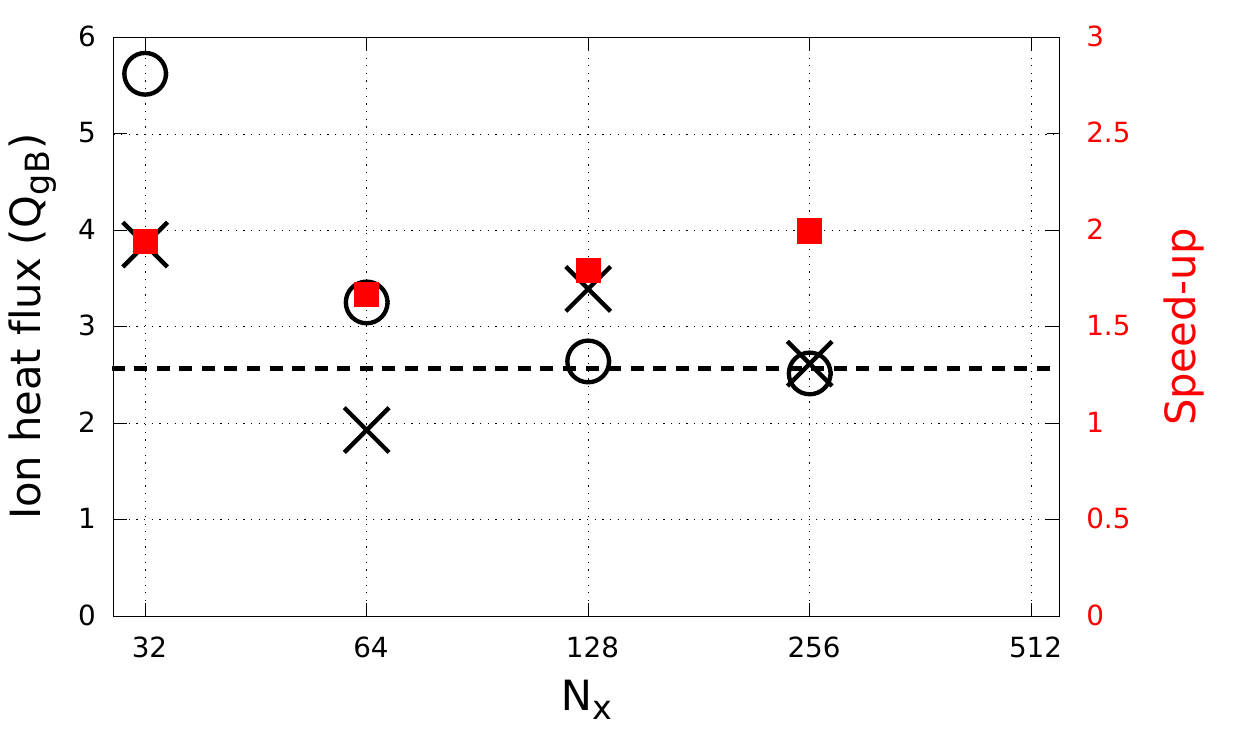}
    \includegraphics[width=0.49\textwidth,valign=m]{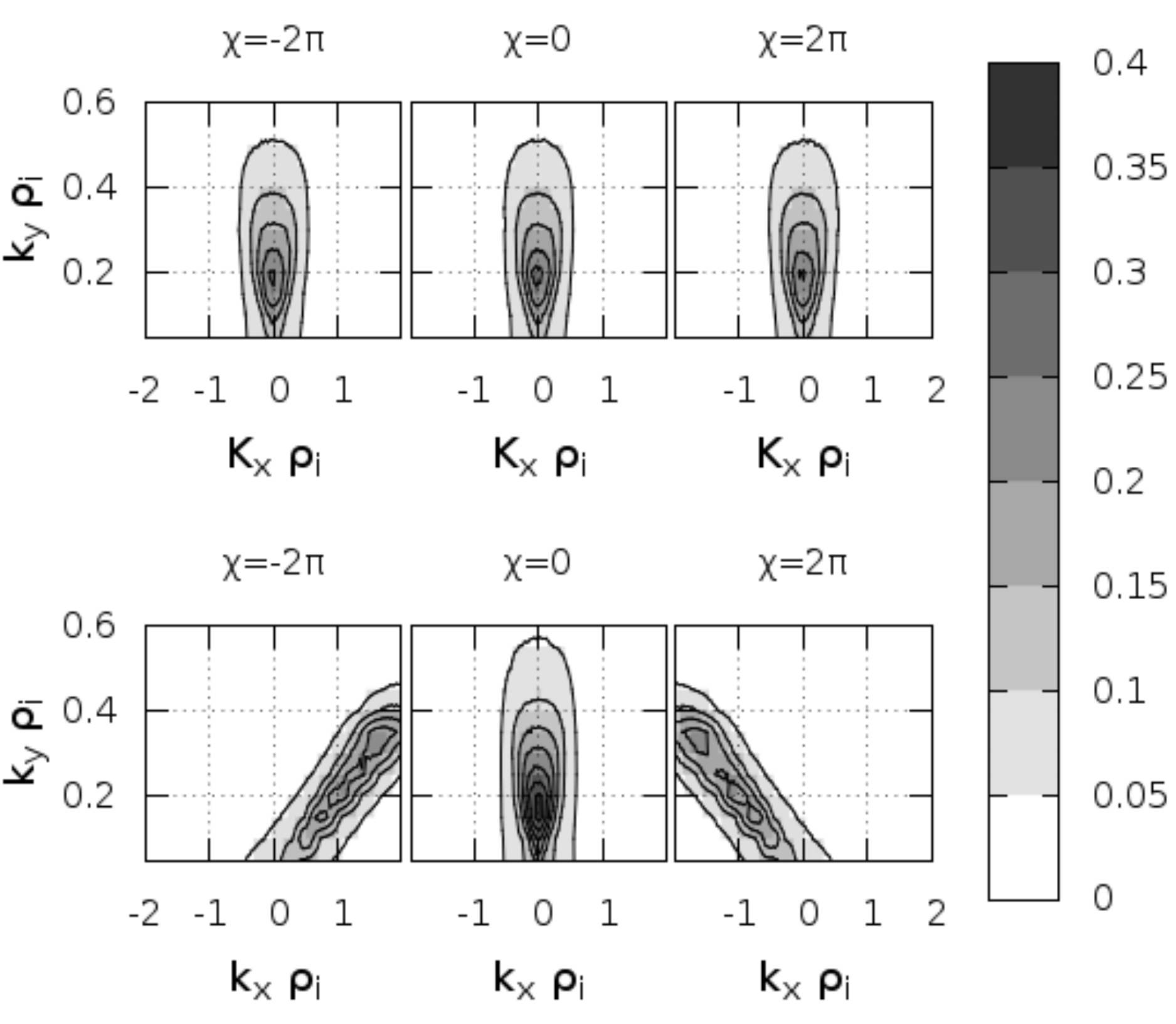}
	\caption{(a) A resolution study for the non-twisting (black circles) and conventional (black crosses) flux tube using adiabatic electrons, $N_{\text{pol}} = 3$, CBC parameters with $\hat{s} = 0.8$, and the parameters and resolutions of tables \ref{tab:CBC} and \ref{tab:defaultRes}. (b) The two-dimensional wavenumber spectra of the electrostatic potential (i.e. $\langle | \Phi | \rangle_{t}$ or $\langle | \phi | \rangle_{t}$) at the three outboard midplanes from the $N_{x} = 128$ simulations. Note the full range in radial wavenumber used by the simulations is shown, while the binormal range is restricted for visual clarity. Each simulation was run on $20 N_{x}$ CPUs.}
	\label{fig:nonlinearNpolResStudy}
\end{figure}

The second test case is identical to the first, except the domain is lengthened from one to three poloidal turns by setting $N_{\text{pol}} = 3$. Though simulation domains with more than one poloidal turn have fallen out of fashion, they were originally recommended to ensure properly converged results \cite{BeerBallooingCoordinates1995}. Without testing for convergence in $N_{\text{pol}}$ (or alternatively $L_{y}$), parallel self-interaction may be artificially large and affect the accuracy of the results \cite{BallBoundaryCond2020, AjaySelfInteraction2020}. Simulations with $N_{\text{pol}} > 1$ are also important for stellarators as each poloidal turn is physically different due to the three-dimensional geometry \cite{MartinParallelBCstell2018, FaberSelfInteractionStellarator2018}. While this test uses a tokamak geometry, it is still expected to be a good indicator for the performance in a stellarator, so long as the magnetic shear is comparable. The important distinguishing feature of this test is that there is now more than one region of turbulent drive. Specifically, turbulence is driven at all three outboard midplanes in the simulation and should be statistically identical because the physical conditions are identical due to axisymmetry.

Figure \ref{fig:nonlinearNpolResStudy}(a) shows the results of the resolution study. First, we note that, for the same value of $N_{x}$, the non-twisting calculation is almost twice as fast. Breaking this down, we find that, while the CPU time required to compute one time step in the non-twisting flux tube is longer by a factor of $\sim 1.7$, the physical time step itself is longer by a factor of $\sim 3.3$. As in the first test, these two effects counteract one another, but the net effect is a computational speed-up for the non-twisting flux tube. Of course, the magnitude of the speed-up is significantly larger for this test case, which is intuitive as the twist from one end of the computational domain to the other is more extreme.

Looking at the convergence in $N_{x}$, we see fairly different behavior. The non-twisting flux tube data converges monotonically to the high resolution limit, while the conventional flux tube data jumps around. To make sense of this, figure \ref{fig:nonlinearNpolResStudy}(b) shows the turbulent spectrum of the electrostatic potential at all three outboard midplanes for both $N_{x} = 128$ simulations. For the non-twisting flux tube, we see the expected results --- since all three outboard midplanes are physically identical, they each have statistically identical turbulence. However, for the conventional flux tube, the turbulence at the central outboard midplane is different from that at the other two outboard midplanes. It has a maximum value of the electrostatic potential that is almost double the maximum value at the other two midplanes. This is because, despite the fact that all three locations are physically identical, they are modeled using different perpendicular grids. The conventional flux tube has a rectangular cross-section at $\chi = 0$, but is twisted into parallelograms at $\chi = -2 \pi$ and $2 \pi$. This means that the turbulent activity is tilted diagonally across the perpendicular wavenumber grid and actually extends off the grid at $\chi = -2 \pi$ and $2 \pi$, even at the high radial resolution of $N_{x} = 128$. As a consequence, the conventional flux tube requires a radial resolution that is more than double that of the non-twisting flux tube. Comparing the runtime of the non-twisting simulation at $N_{x} = 128$ with that of the conventional flux tube simulation at $N_{x} = 256$, we find that the overall computational cost of a properly resolved $N_{\text{pol}} = 3$ CBC simulation is reduced by a factor of 7 by employing the non-twisting flux tube.

\subsection*{Test case 3}

\begin{figure}
	\centering
	\hspace{-18em} (a) \hspace{17em} (b) \\
	\includegraphics[width=0.49\textwidth]{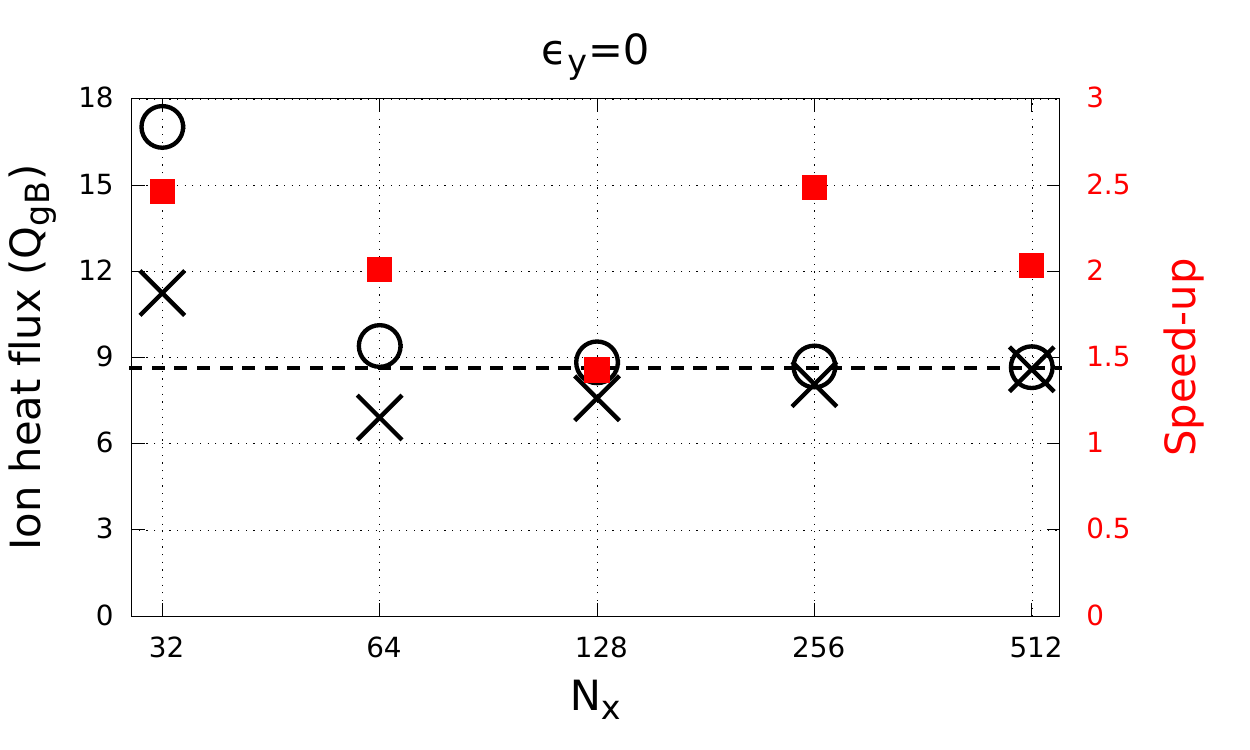}
	\includegraphics[width=0.49\textwidth]{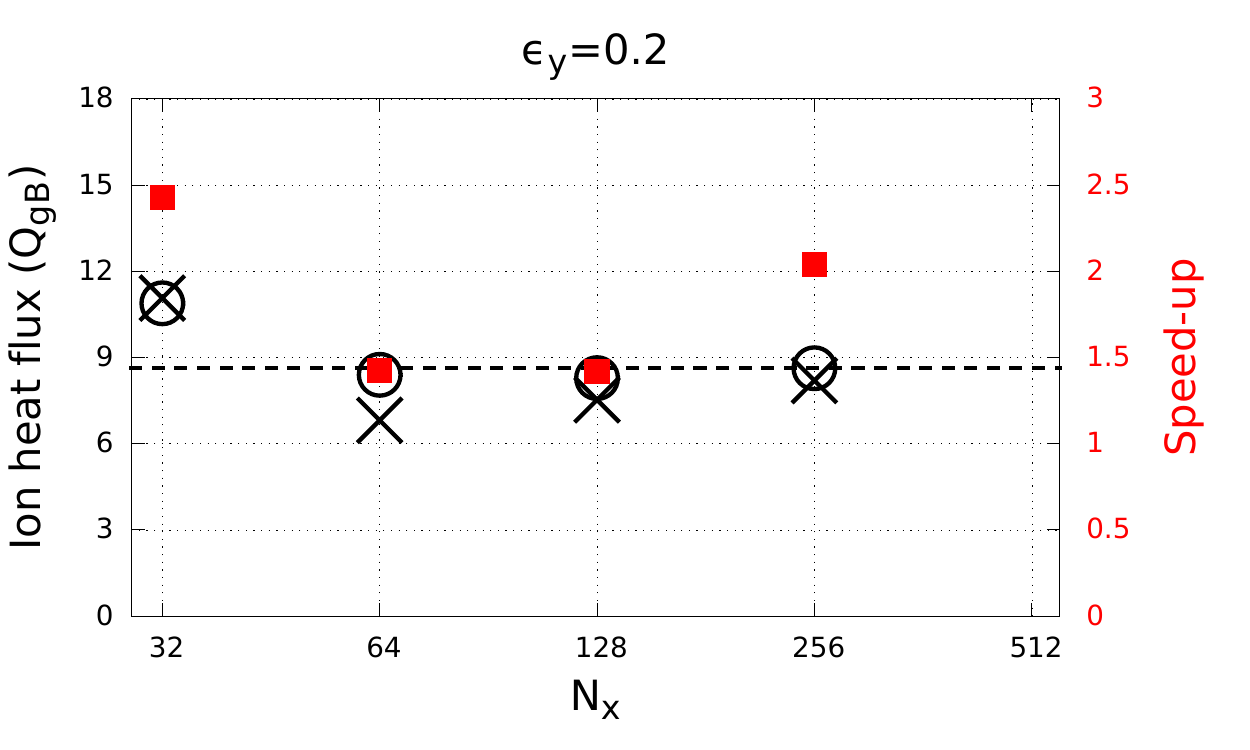} \\
	\hspace{-18em} (c) \hspace{17em} (d) \\
	\includegraphics[width=0.49\textwidth]{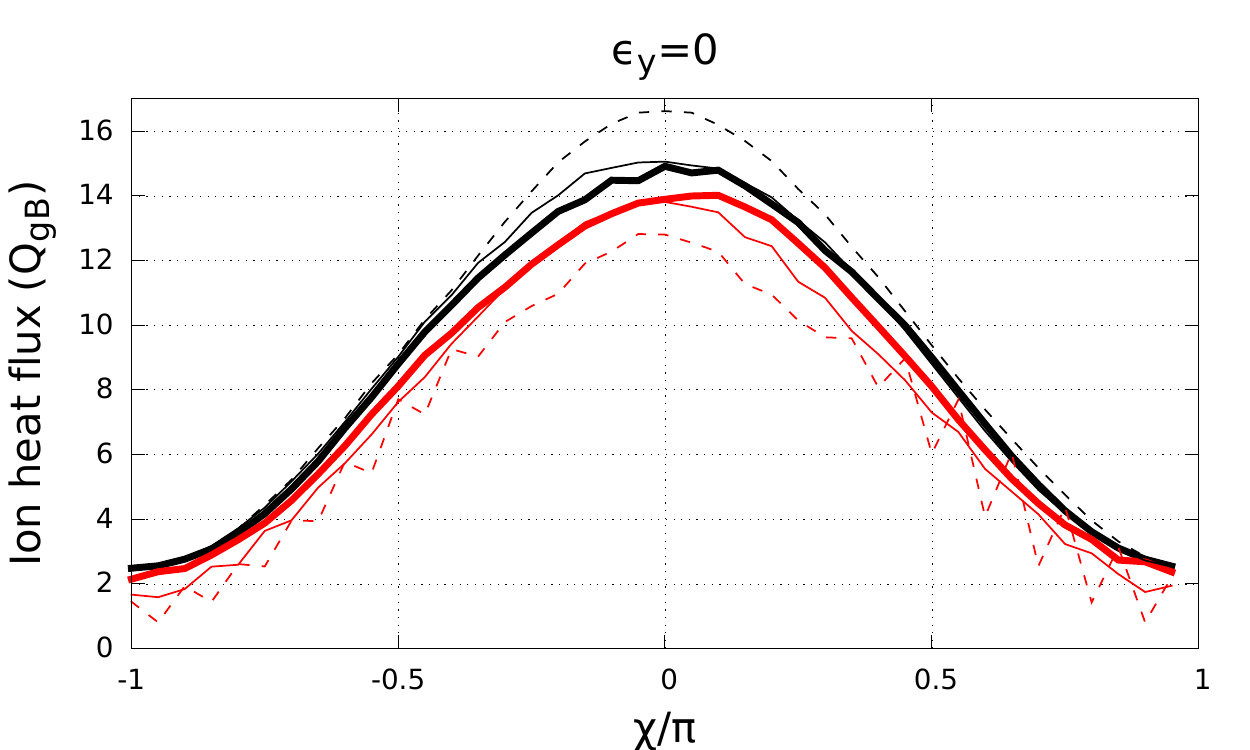}
	\includegraphics[width=0.49\textwidth]{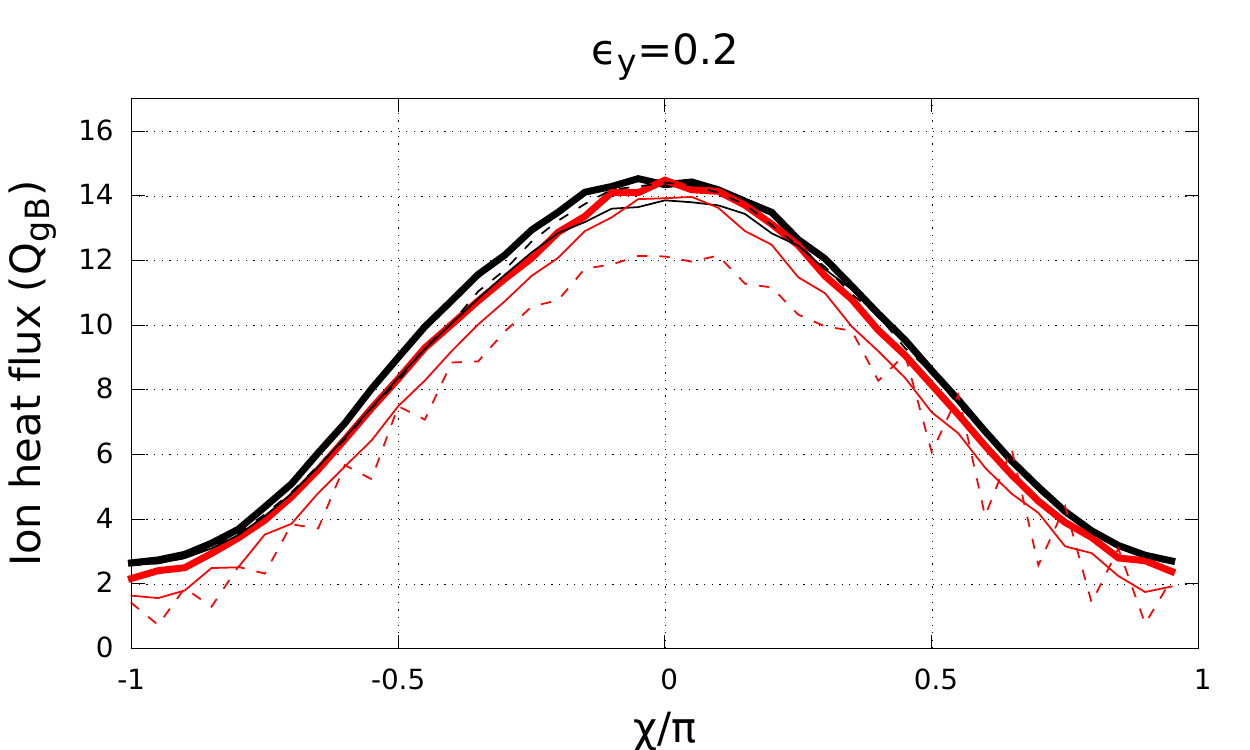}
	\caption{The top row shows a resolution study for the non-twisting (black circles) and conventional (black crosses) flux tube (a) without hyperdiffusion in $y$ or (b) with $\epsilon_{y} = 0.2$. It uses adiabatic electrons, $N_{\text{pol}} = 1$, CBC parameters with $\hat{s} = 4.0$, and the parameters and resolutions of tables \ref{tab:CBC} and \ref{tab:defaultRes} (with the modifications detailed in the text). The bottom row shows the corresponding poloidal distribution of the heat flux (c) without or (d) with hyperdiffusion in $y$ for the non-twisting (black) and conventional (red) flux tube with radial resolutions of $N_{x} = 256$ (thick solid), $N_{x} = 128$ (thin solid), and $N_{x} = 64$ (thin dashed). Each simulation was run on $1.25 N_{x}$ CPUs.}
	\label{fig:nonlinearHighShearResStudy}
\end{figure}

The third test case uses $N_{\text{pol}} = 1$ and is identical to the first, except the strength of the global magnetic shear is quintupled to $\hat{s} = 4.0$. Due to the larger magnetic shear, the background temperature gradient had to be increased to $a / L_{Ti} = 6.5$ to ensure instability. To more efficiently model these new conditions, the radial and binormal domain widths were approximately halved to $L_{x} = 80 \rho_{i}$ and $L_{y} = 63 \rho_{i}$ along with a corresponding decrease in the number of binormal grid points to $N_{y} = 32$. Such a reduced grid still resolves the turbulence well, which is not surprising as stronger magnetic shear tends to reduce the correlation lengths. Additionally, the number of parallel grid points was increased to $N_{\chi} = 40 N_{\text{pol}}$ to better resolve shorter parallel correlation lengths, the parallel hyperdiffusion was lowered to $\epsilon_{\chi} = 0.02$ to match the linear growth rate, and the time average over the nonlinearly saturated state was performed from $t = 500$ to $2000 a/c_{s}$. The resulting resolution study is shown in figure \ref{fig:nonlinearHighShearResStudy}(a). We see that the non-twisting flux tube benefits from a speed-up of around 2 (although it is fairly inconsistent).

Figure \ref{fig:nonlinearHighShearResStudy}(a) also shows that the convergence with $N_{x}$ is fairly similar between the two flux tubes. This may be surprising as it seems like the non-twisting flux tube should be beneficial when the magnetic shear is high because the conventional flux tube cross-section becomes strongly twisted at the inboard midplane. However, figure \ref{fig:nonlinearHighShearResStudy}(c) shows that the turbulence is very localized around the outboard midplane. Thus, the {\it flux surface averaged} fluxes are insensitive to resolution problems around the {\it inboard} midplane because the averages are dominated by behavior at the outboard midplane. Looking carefully at figure \ref{fig:nonlinearHighShearResStudy}(c) we see that, as $N_{x}$ is lowered, an unphysical grid-scale oscillation develops in the parallel direction away from the outboard midplane in the conventional flux tube. Similar behavior was also observed in reference \cite{ToldShiftedMetric2010}. Such oscillations do not occur in the non-twisting flux tube, presumably because it includes modes near $K_{x} = 0$ at all poloidal locations and the turbulent activity does not fall off the grid. To summarize, while the non-twisting flux tube better resolves the turbulence away from $\chi = 0$, this does not appear to have a big impact on the overall fluxes. Thus, for this test case, the non-twisting flux tube is most advantageous if you are interested in local quantities away from $\chi = 0$. Alternatively, the non-twisting flux tube could provide a big advantage if the turbulent drive is not centered around $\chi = 0$ (as is indicated by the second test case). This is true in sheared slab geometry as all parallel locations are statistically identical and have equal turbulent drive \cite{ToldShiftedMetric2010}. It may also be the case in stellarator (see figure 7.12 of reference \cite{MerzThesis2008}) or tokamak pedestal simulations \cite{ToldPedestalGK2008, JenkoNonlinearPedestalGK2009, FultonPedestalGK2014, ParisiPedestal2020}, which often observe that the linear growth rate peaks for a non-zero $k_{x}$ corresponding to the top and bottom of the flux surfaces. In nonlinear simulations, such turbulence would be very expensive to resolve with the conventional flux tube when the magnetic shear is large.

On the other hand, the non-twisting flux tube does have to make a sacrifice in order to better resolve the inboard midplane. Because the radial wavenumber grid always stays centered around $K_{x} \approx 0$, the length of the linear modes is reduced when the binormal mode index $n = k_{y} / k_{y,\text{min}}$ is large. This can be understood using figure \ref{fig:gridLayouts}(d), which has $n=14$. We see that, in the conventional flux tube, all of the linear modes are at least one poloidal turn long. At higher values of $n$, this can only be ensured by sacrificing {\it ballooning angles} (i.e. the poloidal angle $\theta_{0} \equiv k_{x}/ (k_{y} \hat{s})$ at which the linear mode would have $K_{x} = 0$ if the local correction to the shear was zero). 
In the limit of very high $n$, all of the linear modes in the $k_{x}$ grid have ballooning angles that are strongly clustered around the $\chi = 0$, but they remain one poloidal turn long. Hence, the linear modes that are present at high $n$ in the conventional flux are well resolved in the parallel direction, but this comes at the cost of entirely omitting linear modes with ballooning angles away from $\chi=0$. This can lead to problems like the oscillations discussed in the previous paragraph. In contrast, figure \ref{fig:gridLayouts}(d) shows that the non-twisting flux tube maintains the same number of linear modes at all values of $n$, but they can become very short in the parallel direction. In fact, if the $k_{y}$ grid is large enough, the highest values of $n$ will have linear modes that are each comprised of just a single parallel grid point. This can be challenging to implement properly (as discussed in \ref{app:implementation}) and seems like a major concern for the non-twisting flux tube. Fortunately, this behavior appears physically appropriate because the parallel extent of linear modes is indeed reduced at higher values of $k_{y}$. This can be seen clearly in figure 3 of reference \cite{BarnesCriticalBalance2011} and is a consequence of the fact that the magnetic shear acts in proportion to $k_{y} \hat{s}$ (e.g. equations \refEq{eq:bessely} and \refEq{eq:grady}). To summarize, just as the conventional flux tube has difficulty resolving the inboard midplane, the non-twisting flux tube {\it may} have issues at large $n$, i.e. when using grids with many $k_{y}$ modes. The non-twisting simulations shown in figure \ref{fig:nonlinearHighShearResStudy}(a) did exhibit pile-up in the $k_{y}$ spectrum. However, simply adding a small amount of hyperdiffusion in the binormal wavenumber $\epsilon_{y} = 0.2$ resolved this problem and did not affect the overall heat fluxes, as is shown in figure \ref{fig:nonlinearHighShearResStudy}(b,d).

\subsection*{Test case 4}

\begin{figure}
	\centering
	\hspace{-18em} (a) \hspace{17em} (b) \\
	\includegraphics[width=0.49\textwidth]{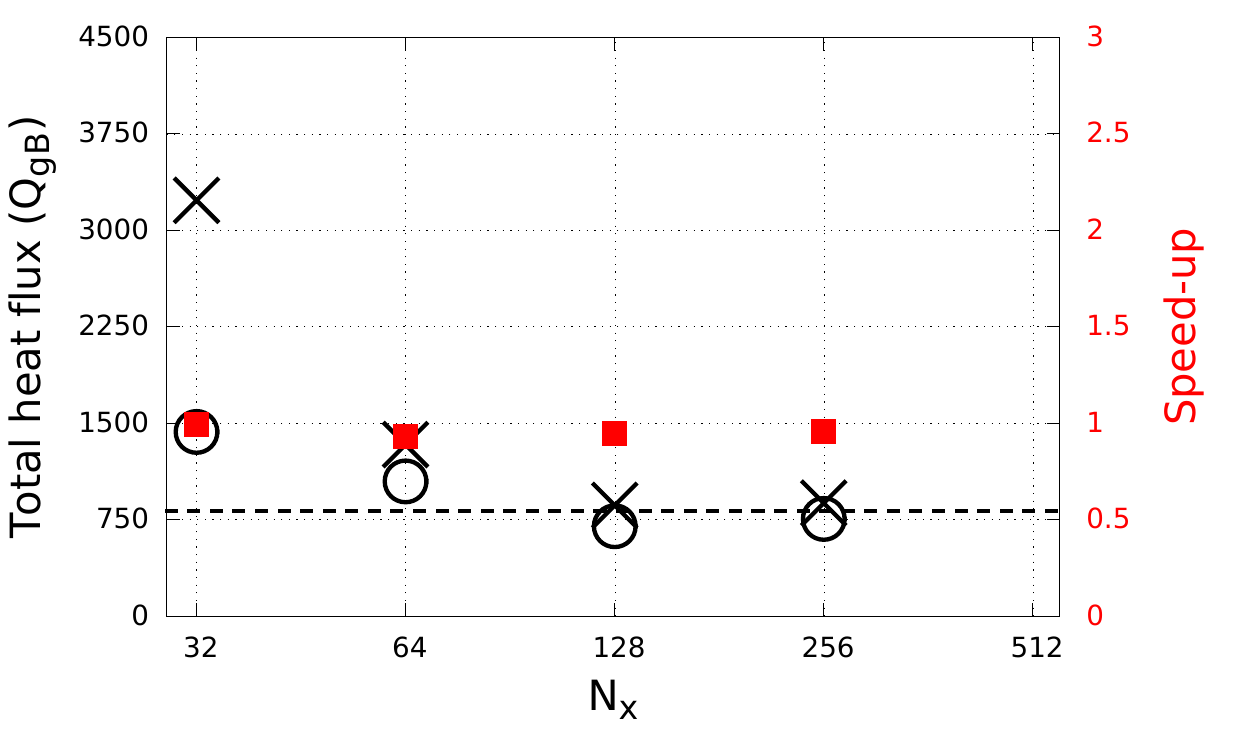}
	\includegraphics[width=0.49\textwidth]{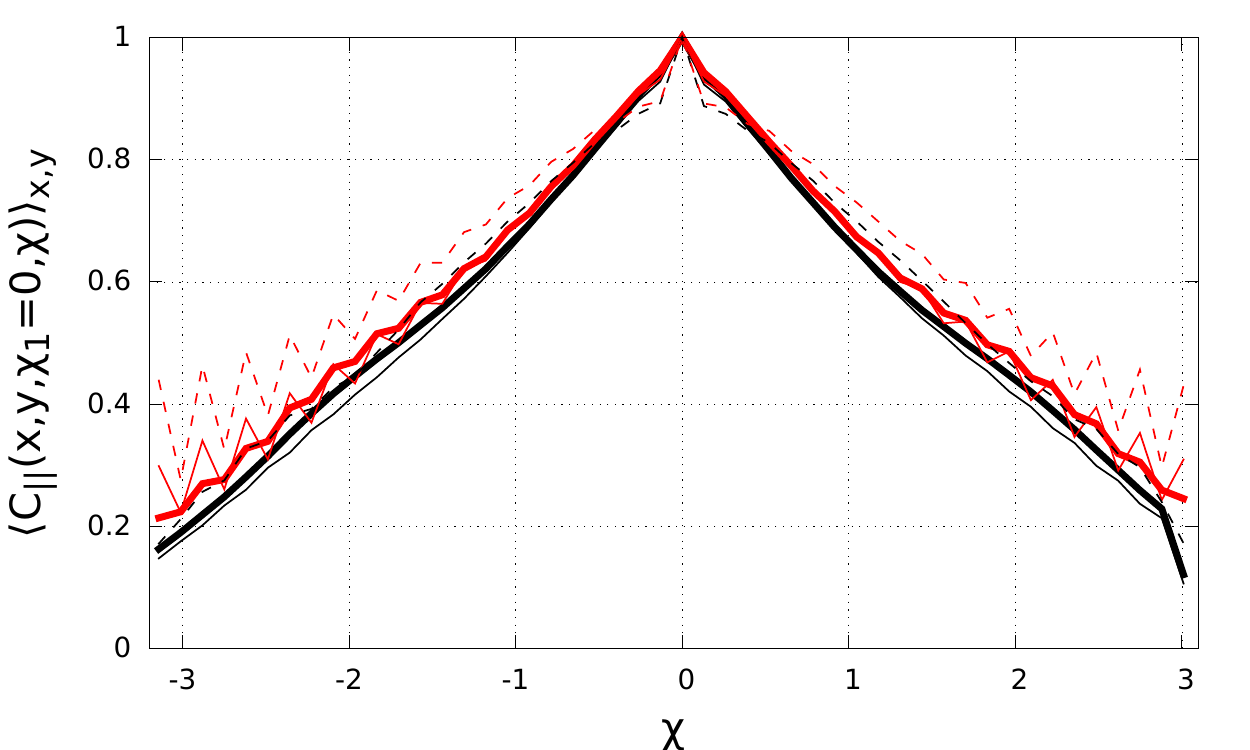}
	\caption{(a) The total heat flux and speed-up for the non-twisting (black circles) and conventional (black crosses) flux tube using kinetic electrons, a realistic shaped DEMO equilibrium with $\hat{s} = 2.4$, $N_{\text{pol}} = 1$, and the resolutions of table \ref{tab:defaultRes}. (b) The corresponding parallel correlation function for the non-twisting (black) and conventional (red) flux tube with radial resolutions of $N_{x} = 256$ (thick solid), $N_{x} = 128$ (thin solid), and $N_{x} = 64$ (thin dashed). All simulations were run on $5120$ CPUs.}
	\label{fig:nonlinearNTTresStudy}
\end{figure}

The fourth test case uses a realistic ``D''-shaped numerical equilibrium based on an inductive EU-DEMO scenario \cite{FablePrivComm2019} with a 50-50 mixture of deuterium-tritium fuel. It has a high magnetic shear of $\hat{s} = 2.4$ and strongly driven Ion Temperature Gradient (ITG) turbulence. Unlike the prior test cases, these simulations include kinetic electrons and local corrections to the magnetic shear. The equilibrium has $x_{0} / R_{0} = 0.43$, $q_{0} = 1.6$, $R_{0} / L_{Ti} = 11.7$, $R_{0} / L_{Te} = 12.0$, and $R_{0} / L_{n} = 0.85$. The parallel hyperdiffusion was increased to $\epsilon_{\chi} = 1.0$ to match the stronger drive, a binormal hyperdiffusion of $\epsilon_{y} = 0.2$ was used, and the gyroBohm heat flux is defined as $Q_{gB} \equiv (\rho_{i} / R_{0})^{2} n_{e} T_{e} c_{S}$ for this test case. In figure \ref{fig:nonlinearNTTresStudy}(a), we see that the computational cost of the two flux tubes is similar at all radial resolutions. Looking into the details, we actually find that the time step does not vary with $N_{x}$ or the type of flux tube, which is very different from the previous test cases with {\it adiabatic} electrons. We believe that this is because including kinetic electrons changes the physics determining the time step. Specifically, as the electron plasma $\beta_{e} \rightarrow 0$, the electrostatic shear Alfv\'{e}n wave becomes very high frequency, thereby limiting the time step (see reference \cite{LeeShearAlfvenWave2001} and section 2.1 of reference \cite{SnyderKinElecFiniteBeta1999}). This can be mitigated by using a small, but finite value of the plasma $\beta$. However, for these simulations the value was limited to $\beta = 10^{-4}$ because higher values were found to modify the heat flux. Unfortunately, at $\beta = 10^{-4}$ it appears that the electrostatic shear Alfv\'{e}n wave still determines the time step. Nevertheless, we would like to note that this issue {\it may} be a particular consequence of the strong turbulent drive. Perhaps for more typical simulations, $\beta$ could be increased further without affecting the fluxes and the smallest radial spatial scale would again determine the time step as in previous adiabatic electron test cases. 

Looking at the convergence properties of figure \ref{fig:nonlinearNTTresStudy}(a), we find that the non-twisting flux tube converges at a slightly lower value of $N_{x}$, but there is not a big difference. However, figure \ref{fig:nonlinearNTTresStudy}(b) shows that there is a more significant difference in the parallel correlation between the outboard and inboard midplanes (i.e. equation \refEq{eq:parCorrFn}). Therefore, although figures \ref{fig:nonlinearHighShearResStudy}(c,d) and \ref{fig:nonlinearNTTresStudy}(b) show different quantities, they reinforce the same conclusion --- the non-twisting flux tube better resolves the inboard midplane at high values of the magnetic shear.

\subsection*{Test case 5}

\begin{figure}
	\centering
    \includegraphics[width=0.7\textwidth]{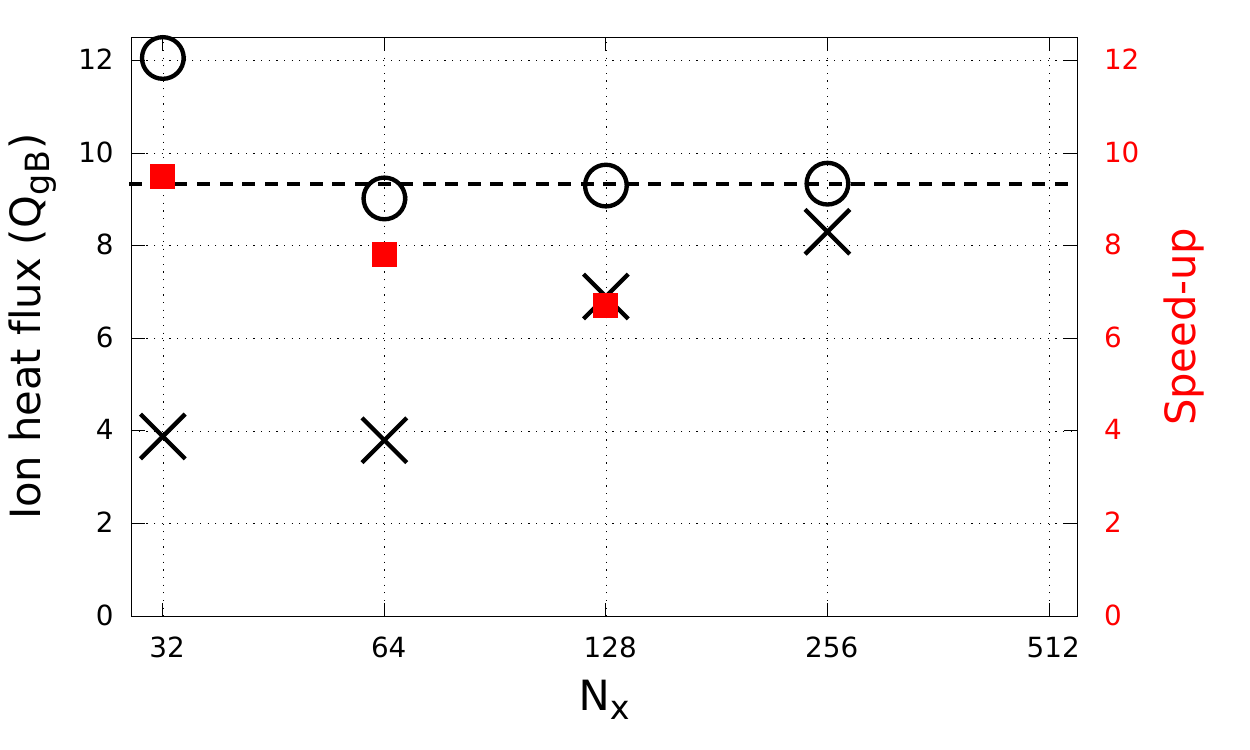}
	\caption{A resolution study for the non-twisting (black circles) and conventional (black crosses) flux tube using adiabatic electrons, $N_{\text{pol}} = 3$, CBC parameters with $\hat{s} = 4.0$, and the parameters and resolutions of tables \ref{tab:CBC} and \ref{tab:defaultRes} (with the modifications detailed in the text). Each simulation was run on $20 N_{x}$ CPUs, but the conventional flux-tube run with $N_{x}=256$ was not fully completed due to its computational cost.}
	\label{fig:nonlinearHighShearNpolResStudy}
\end{figure}

The fifth test case is designed to most clearly demonstrate the advantage of the non-twisting flux tube. It is a repeat of the third test case, except the simulation domain is lengthened to three poloidal turns, so that it has $N_{\text{pol}} = 3$, $\hat{s} = 4.0$, and adiabatic electrons. As before, due to the larger magnetic shear, the background temperature gradient had to be increased to $a / L_{Ti} = 6.5$ to ensure instability. To optimally resolve these new conditions, the number of parallel grid points was increased to $N_{\chi} = 40 N_{\text{pol}}$, the radial domain width was decreased to $L_{x} = 80 \rho_{i}$, the binormal domain width was halved to $L_{y} = 63 \rho_{i}$ along with a corresponding decrease in the number of binormal grid points to $N_{y} = 32$, and the parallel hyperdiffusion was lowered to $\epsilon_{\chi} = 0.02$ to match the linear growth rate. Additionally, the time average over the nonlinearly saturated state was performed from $t = 1000$ to $6000 a/c_{s}$ and the hyperdiffusion in $y$ was set to $\epsilon_{y} = 0.2$.

Figure \ref{fig:nonlinearHighShearNpolResStudy} shows that non-twisting flux tube performs dramatically better than the conventional flux tube. At the same value of $N_{x}$, the non-twisting flux tube is more than six times faster, due to having a much longer time step. Moreover, we see that the non-twisting flux tube is well converged at $N_{x} = 64$, while the conventional flux tube requires many more radial grid points. In fact, we do {\it not} believe that the conventional flux tube is properly converged even at $N_{x} = 256$ because the radial spectrum of $\phi$ still exhibits significant pile-up. Instead we believe the heat flux is oscillating around the converged result in a similar fashion to figure \ref{fig:nonlinearNpolResStudy}(a). All of this means that a fully converged simulation is made at least 30 times less expensive by using the non-twisting flux tube, likely more.

While we deliberately chose the parameters of this test case to increase the computational savings, similar results likely hold in physically motivated simulations. The essential features are high magnetic shear and more than one region of turbulent drive. These could be found in pedestal simulations, which have high magnetic shear due to the proximity to the separatrix and may have two regions of turbulence drive (i.e. the top and bottom) \cite{ToldPedestalGK2008, JenkoNonlinearPedestalGK2009, FultonPedestalGK2014, ParisiPedestal2020}. Alternatively, stellarator simulations generally require more than one poloidal turn \cite{MartinParallelBCstell2018, FaberSelfInteractionStellarator2018}, though they often have small {\it global} magnetic shear. Nevertheless, the non-twisting flux tube could still benefit stellarators with high {\it local} shear. Specifically, if local magnetic shear causes the conventional flux tube cross-section to twist strongly between two regions of turbulent drive, it doesn't matter if it eventually twists back to rectangular. The conventional flux tube would still be strongly sheared for one of the regions of turbulent drive and struggle to resolve it efficiently.

\section{Alternative flux tubes}
\label{sec:altTwist}

In this section we hope to stimulate future progress by discussing what spatial grids are possible in a flux tube and what considerations appear to limit our options. The coordinate system transform of section \ref{sec:derivNonTwist} was deliberately chosen to exactly and entirely eliminate the twist of the flux tube. While this seems like a natural choice to explore, many other possibilities exist, some of which may give better performance. In fact, it may be feasible to optimize the grid for each individual simulation to most efficiently treat the corresponding turbulent conditions. Therefore, we will imagine transforming the gyrokinetic model to a completely general set of coordinates $( \tilde{K}_{x}, \tilde{K}_{y}, \tilde{\chi} )$ and seek constraints on their possible definitions.

The parallel coordinate is treated in real space, due to the variation of the geometric coefficients in this dimension. The $\tilde{\chi}$ grid is already known to be very flexible and the locations of the grid points can be chosen arbitrarily. By default GENE uses a regularly spaced grid in the straight-field line poloidal angle, but it also has the ability to concentrate the grid points around the outboard midplane (see Section 4.3 of reference \cite{ToldThesis2012}). GS2, on the other hand, has a routine that adapts the locations of the parallel grid points based on the geometric coefficients of the equilibrium being simulated \cite{BaumgaertelThesis2012}. For simplicity, in this section we will use a regularly spaced grid in the straight-field line poloidal angle $\tilde{\chi} = \chi$.

\begin{figure}
	\includegraphics[width=0.33\textwidth]{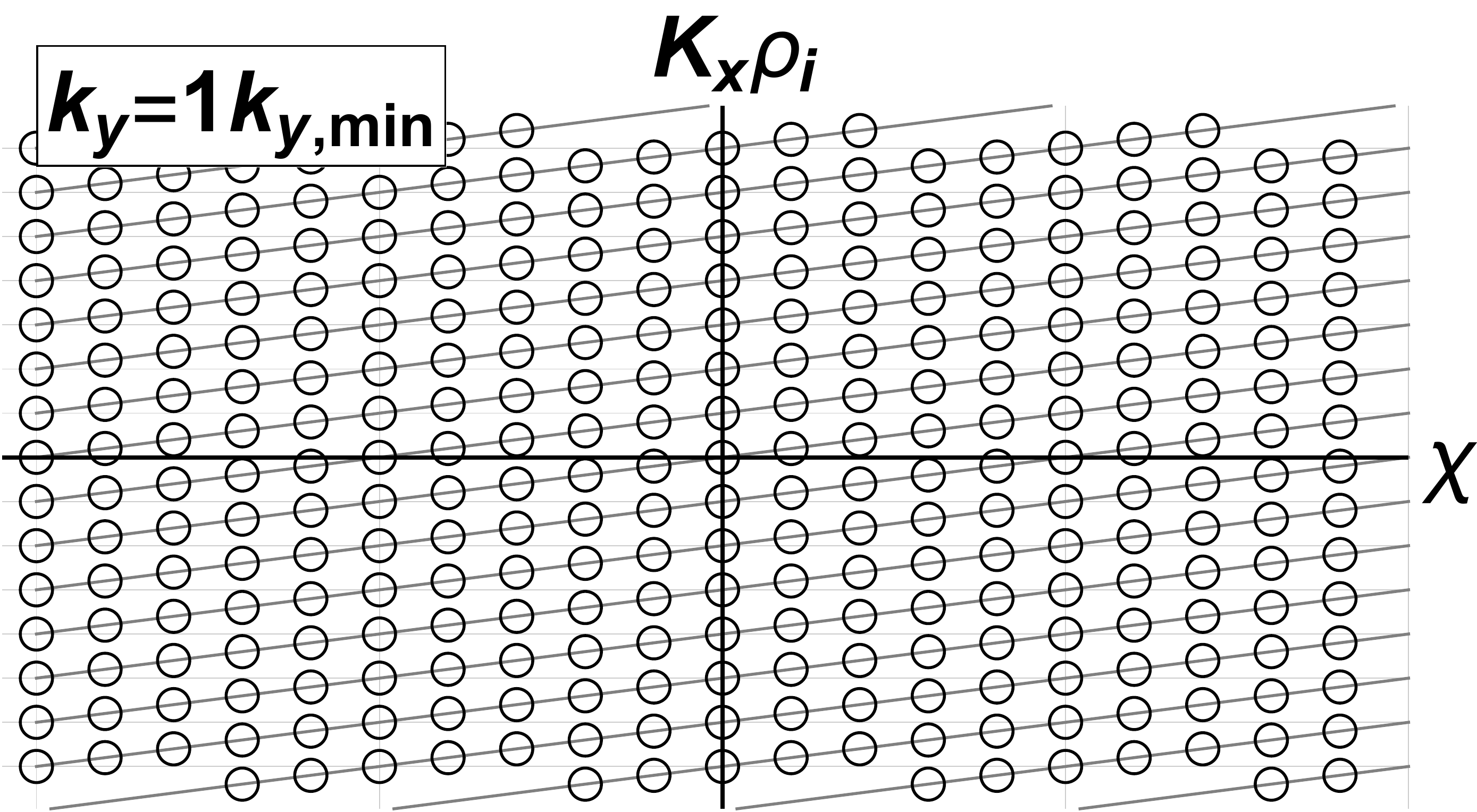}
	\includegraphics[width=0.33\textwidth]{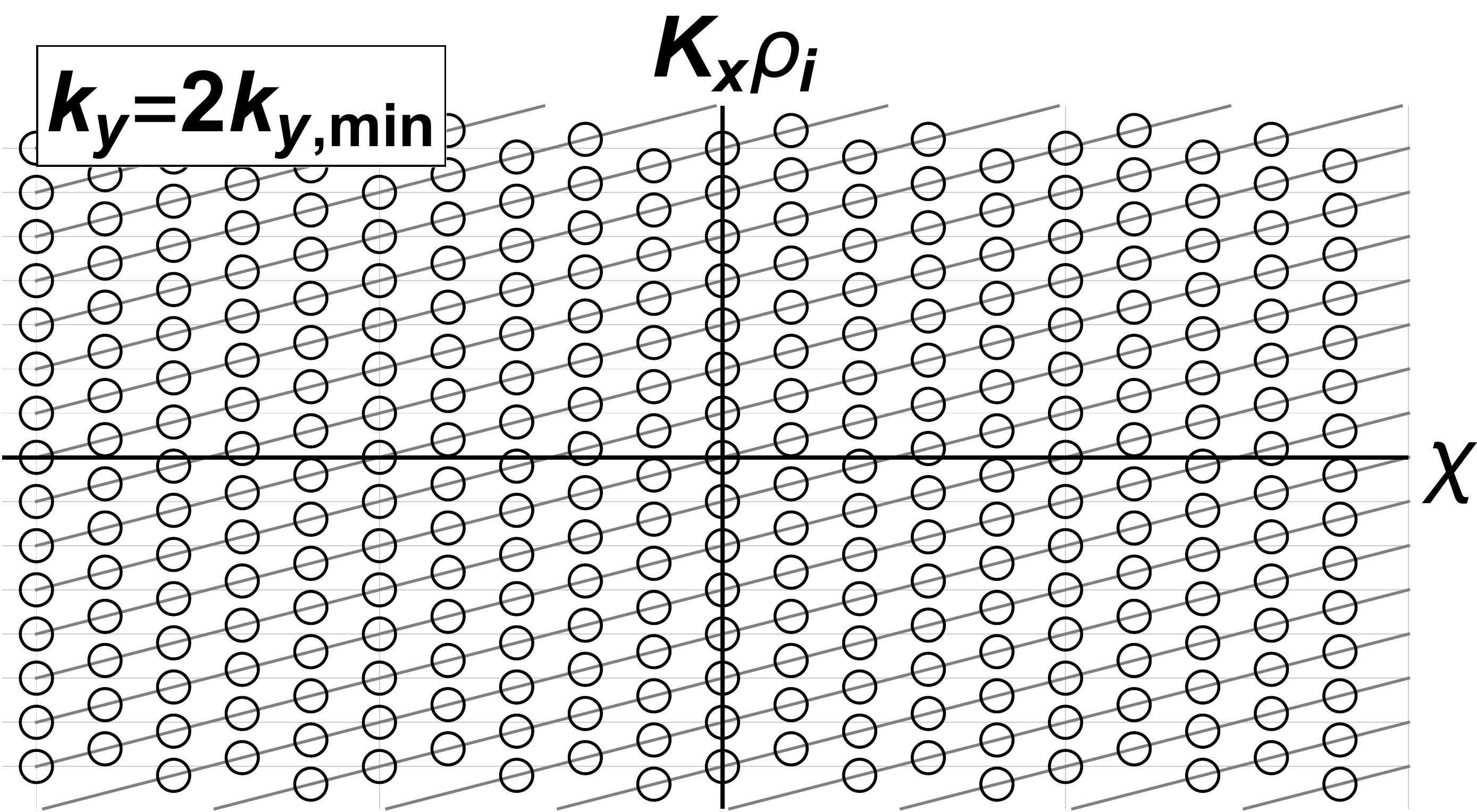}
	\includegraphics[width=0.33\textwidth]{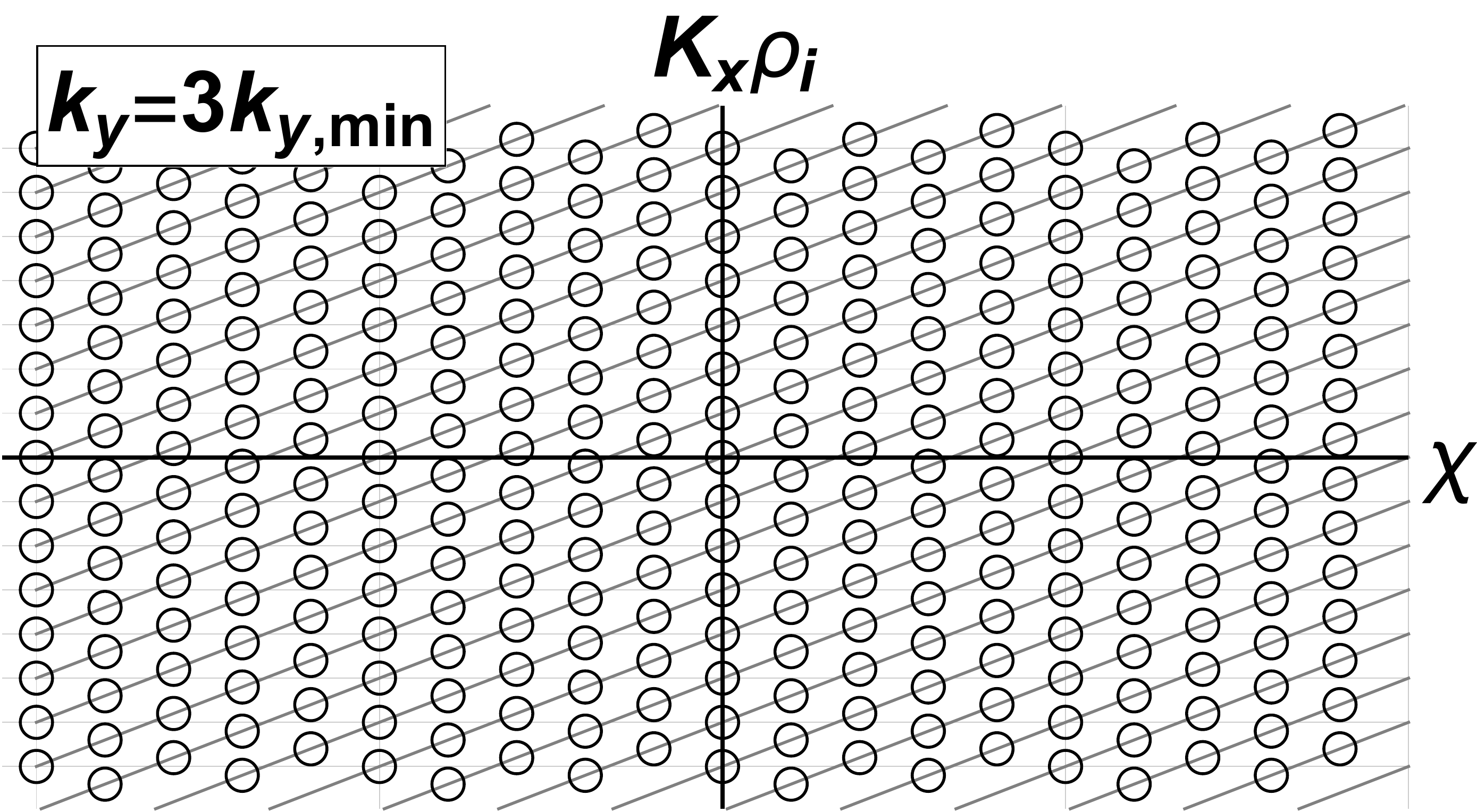}
	\caption{Cuts of the full three-dimensional lattice of allowed spatial grid points (black circles) at the three lowest allowed values of $k_{y}$ (columns). The linear modes (diagonal gray lines) are indicated, $N_{\text{exc}} = 4$, $k_{y, \text{min}} = 2 \pi / L_{y}$ is the minimum value of $k_{y}$ anywhere on the spatial grid, and a regularly spaced $\chi$ grid is used.}
	\label{fig:possibleGridLayouts}
\end{figure}

The other two spatial coordinates, binormal and radial, are represented using Fourier modes. This is well-motivated, given the periodic boundary conditions in these directions and the fact that linear modes are composed of Fourier modes (i.e. a linear mode has a fixed $k_{y}$ and is a chain of linked $K_{x}$ values in $\chi$).  By enforcing the boundary conditions, we discretize the allowed $K_{x}$ and $k_{y}$ Fourier modes into an infinite two-dimensional lattice at each value of $\chi$. Given $L_{x}$, $L_{y}$, and $\Nabla x \cdot \Nabla y / | \Nabla x |^{2}$, we can calculate all of these potential grid points, which are shown in figure \ref{fig:possibleGridLayouts} for an example simulation. Since $L_{x}$ and $L_{y}$ are numerical parameters, we can always increase them to make the grid spacing arbitrarily fine. Moreover, in theory, we have complete freedom to choose at will which of these grid points to include in the simulation. In practice, we will see that there are significant constraints that stem primarily from the physical effects that couple different wavenumbers --- parallel streaming and the quadratic nonlinearity.

\begin{figure}
	(a) \\
	\includegraphics[width=0.33\textwidth]{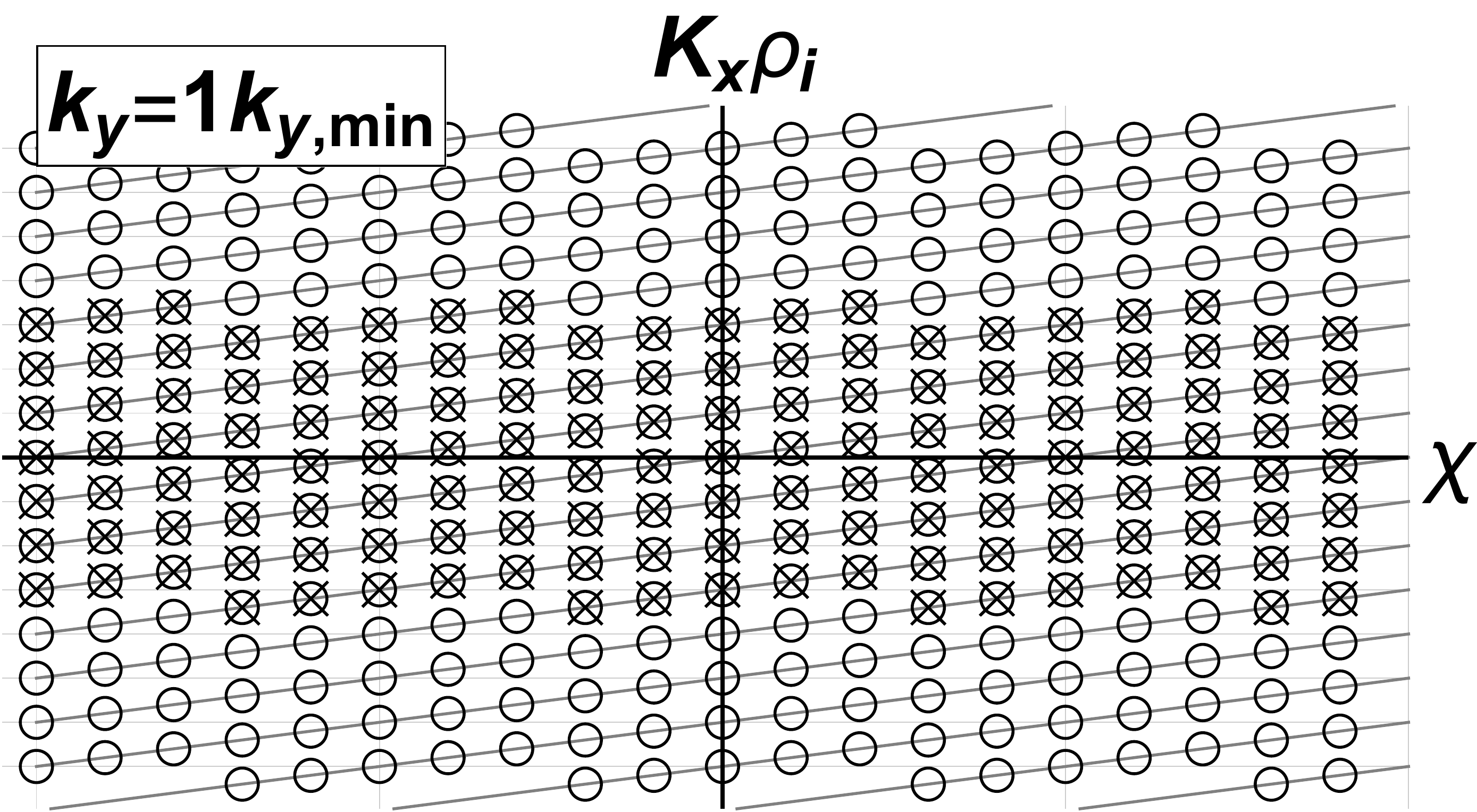}
	\includegraphics[width=0.33\textwidth]{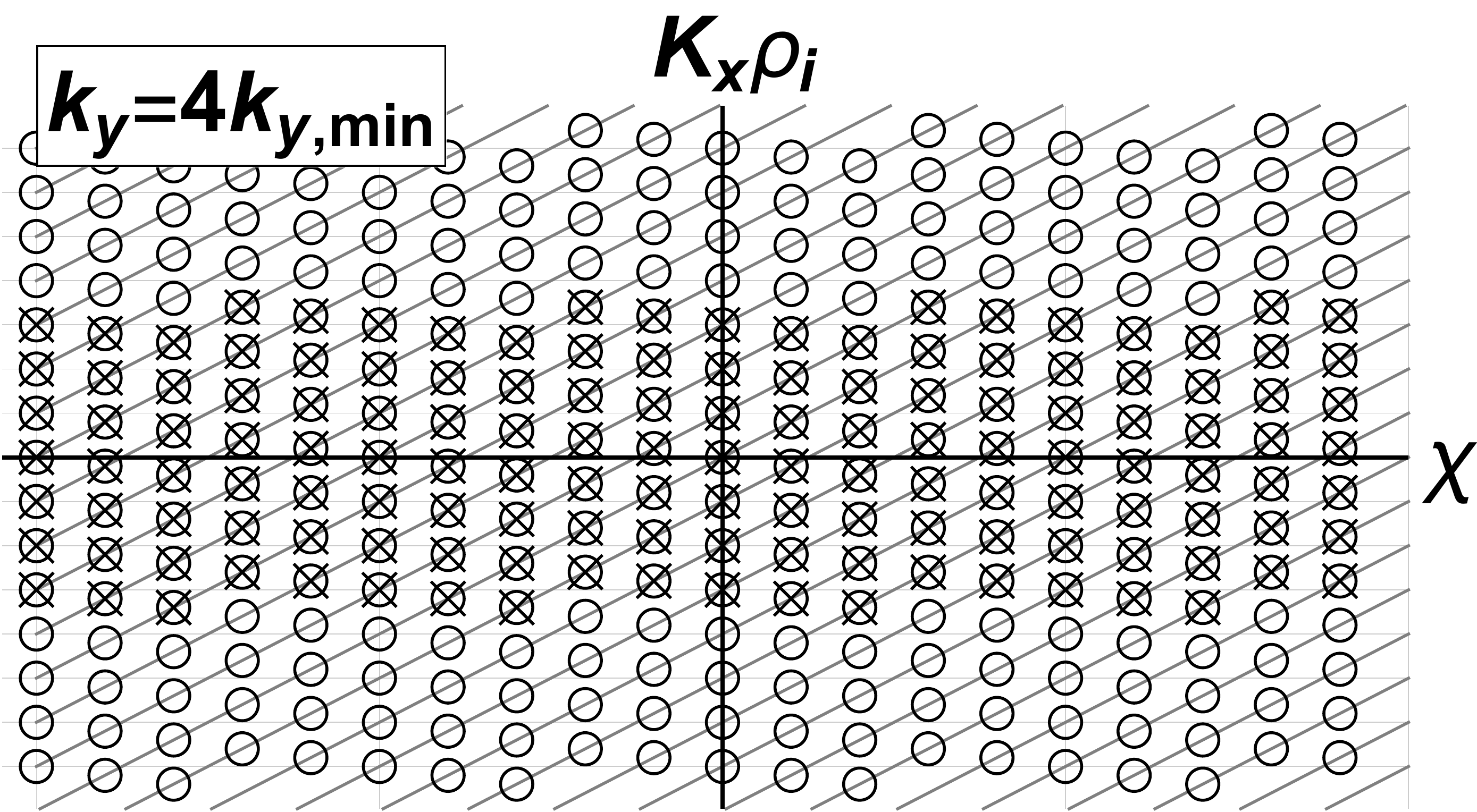}
	\includegraphics[width=0.33\textwidth]{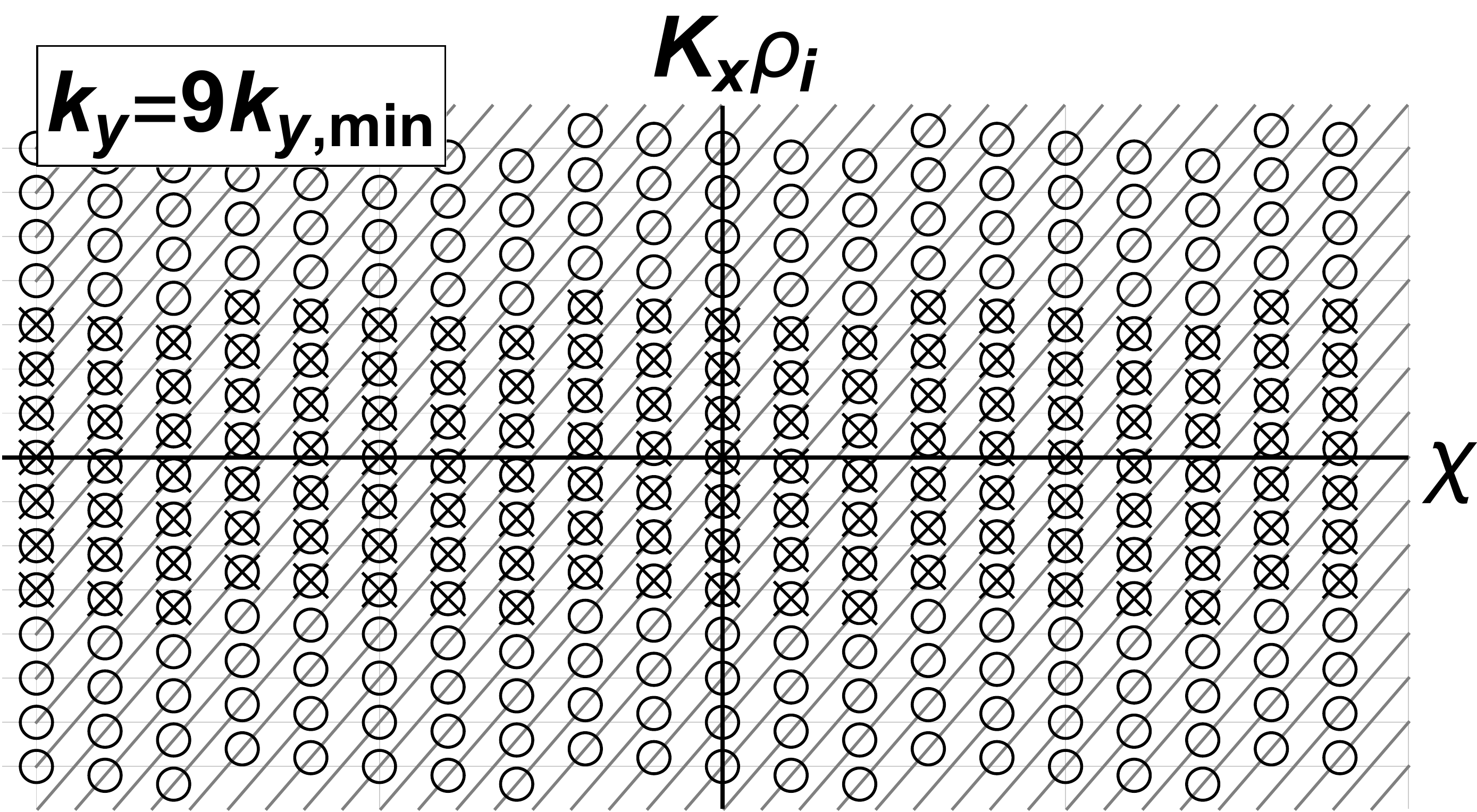}
	(b) \\
	\includegraphics[width=0.33\textwidth]{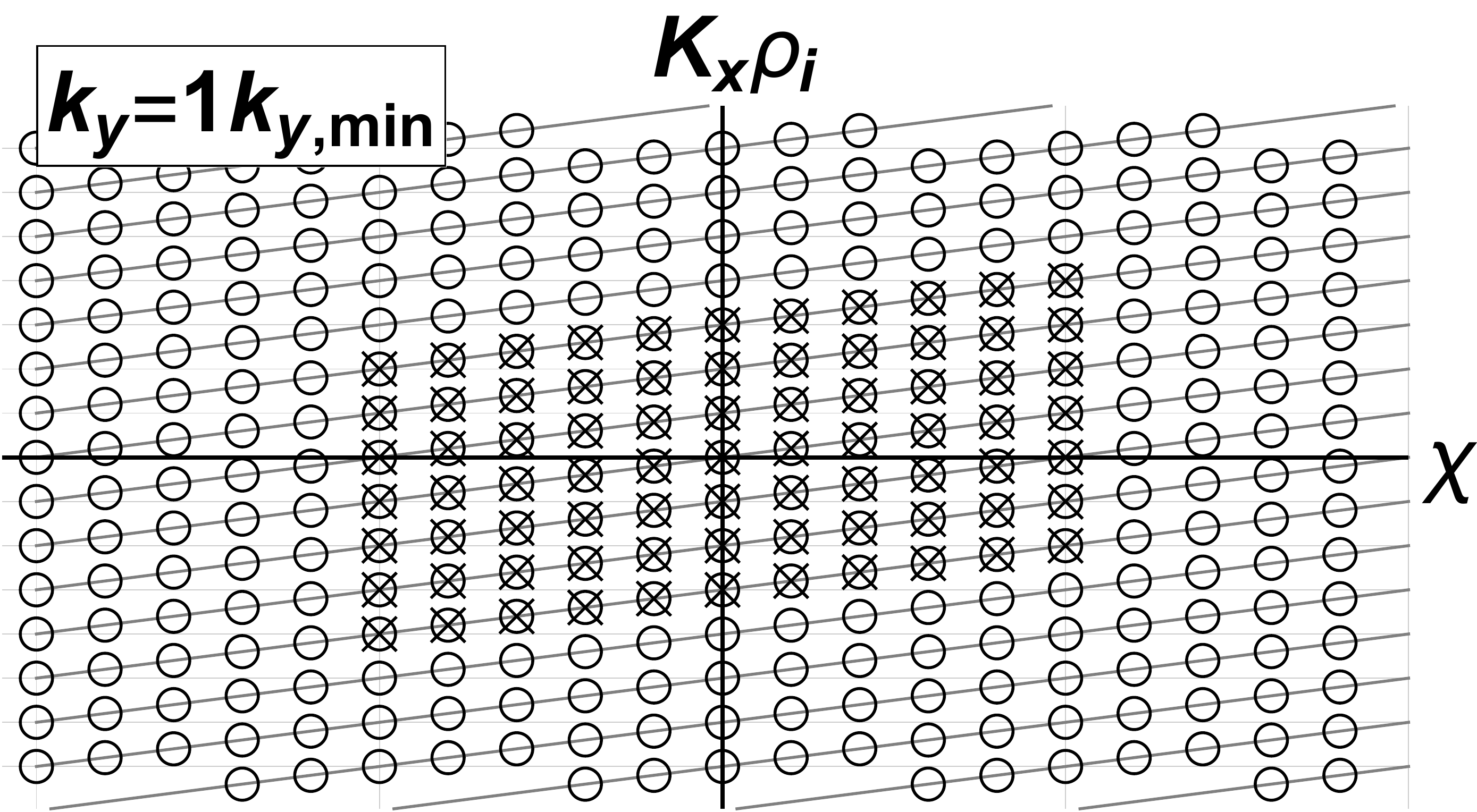}
	\includegraphics[width=0.33\textwidth]{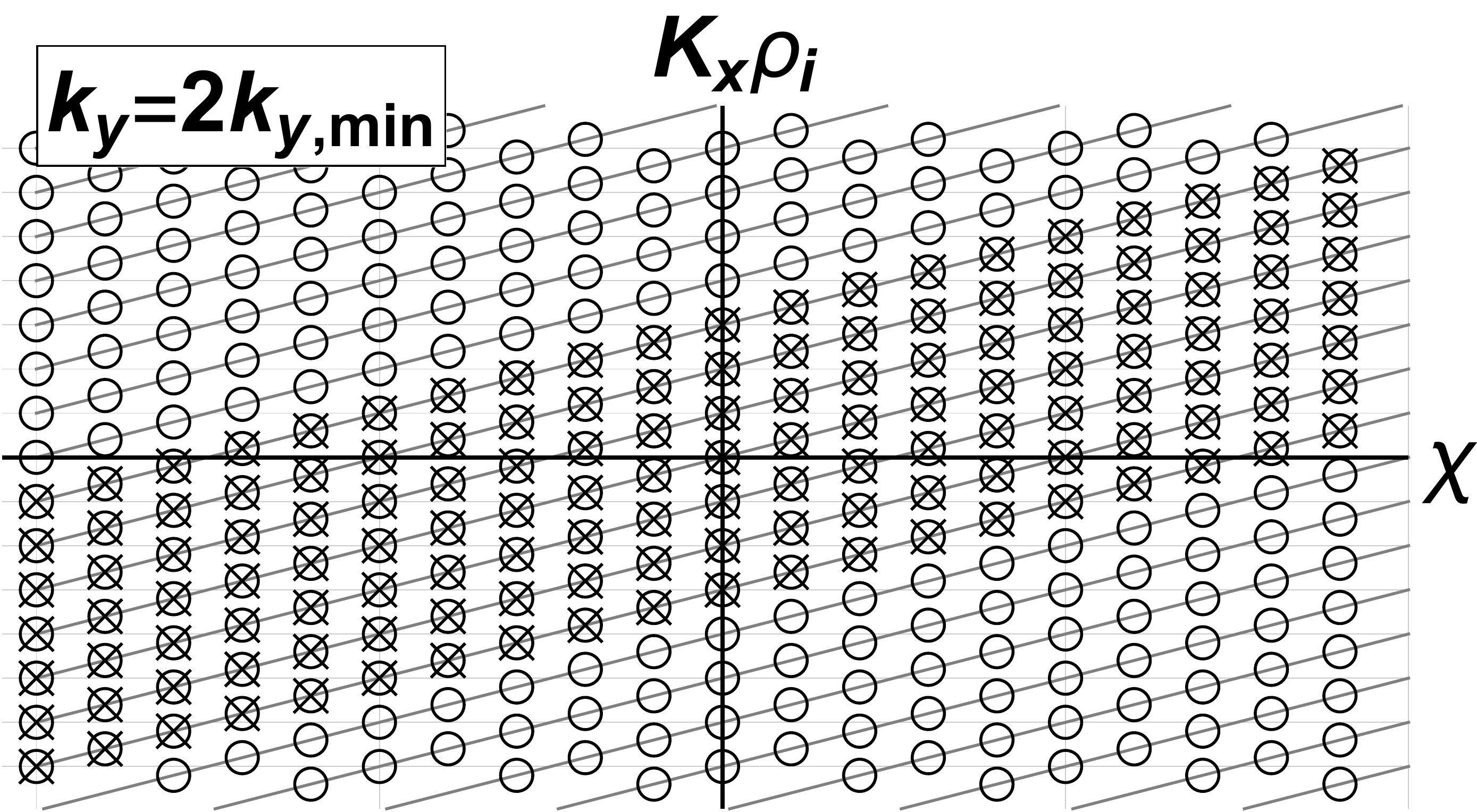}
	\includegraphics[width=0.33\textwidth]{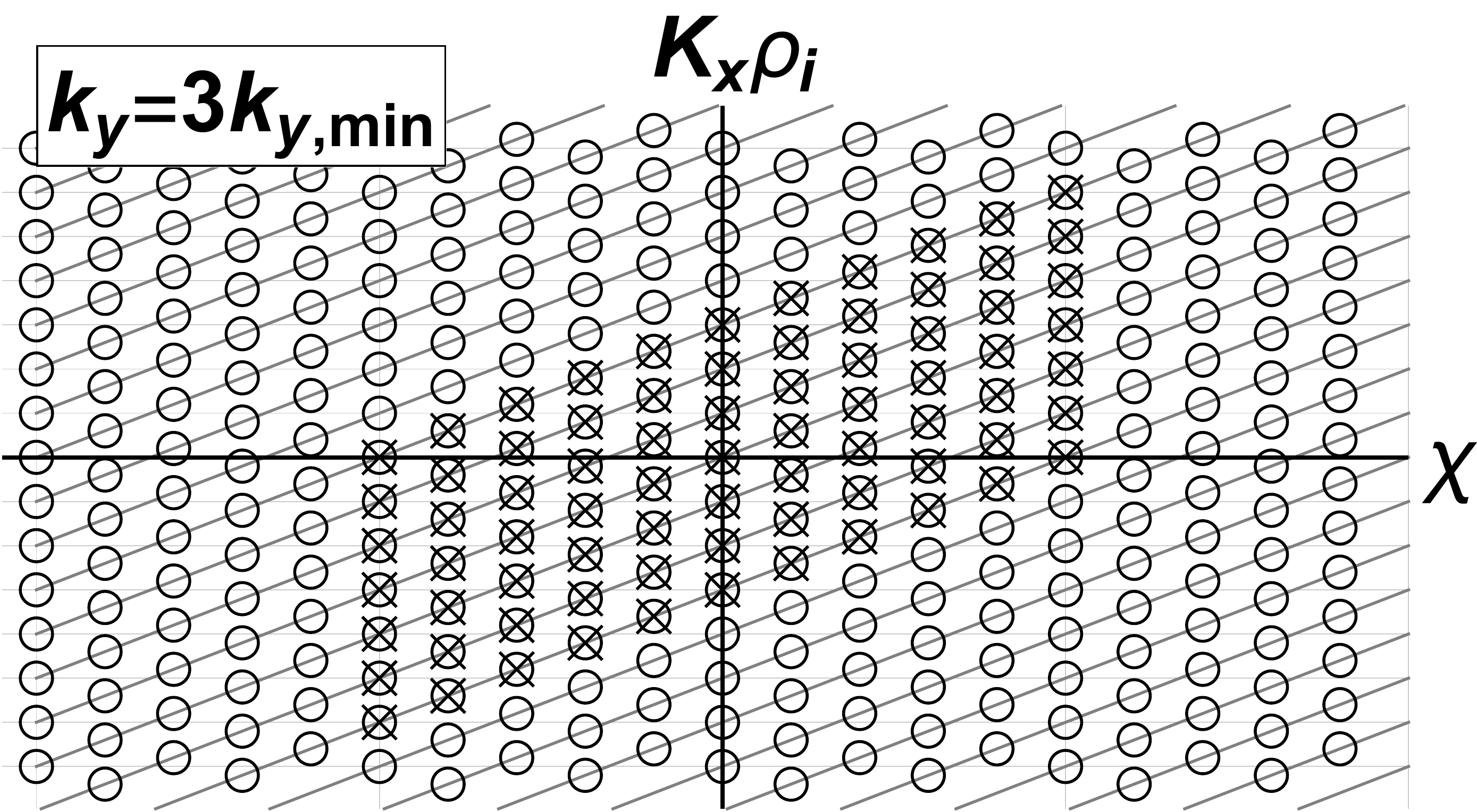}
	(c) \\
	\includegraphics[width=0.33\textwidth]{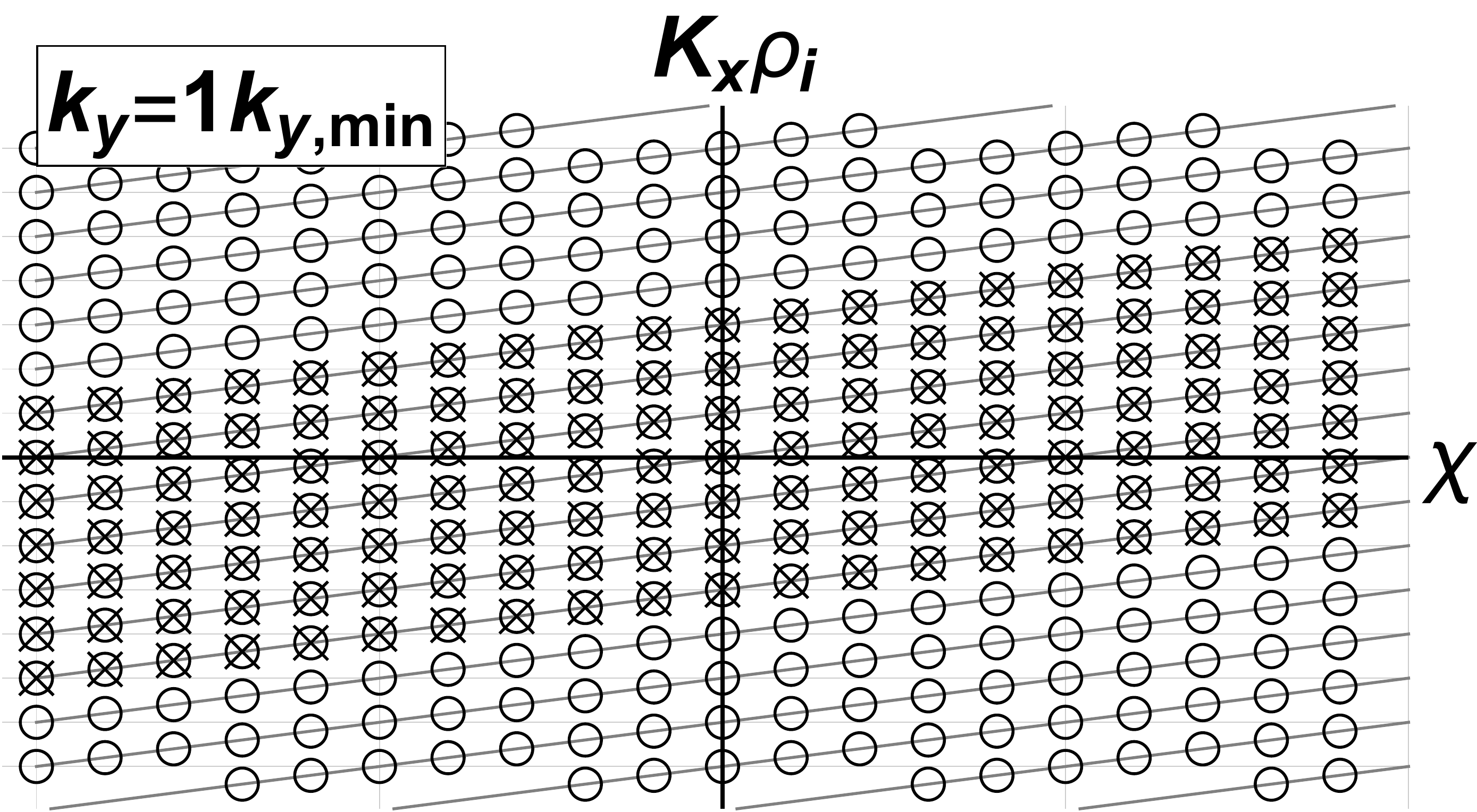}
	\includegraphics[width=0.33\textwidth]{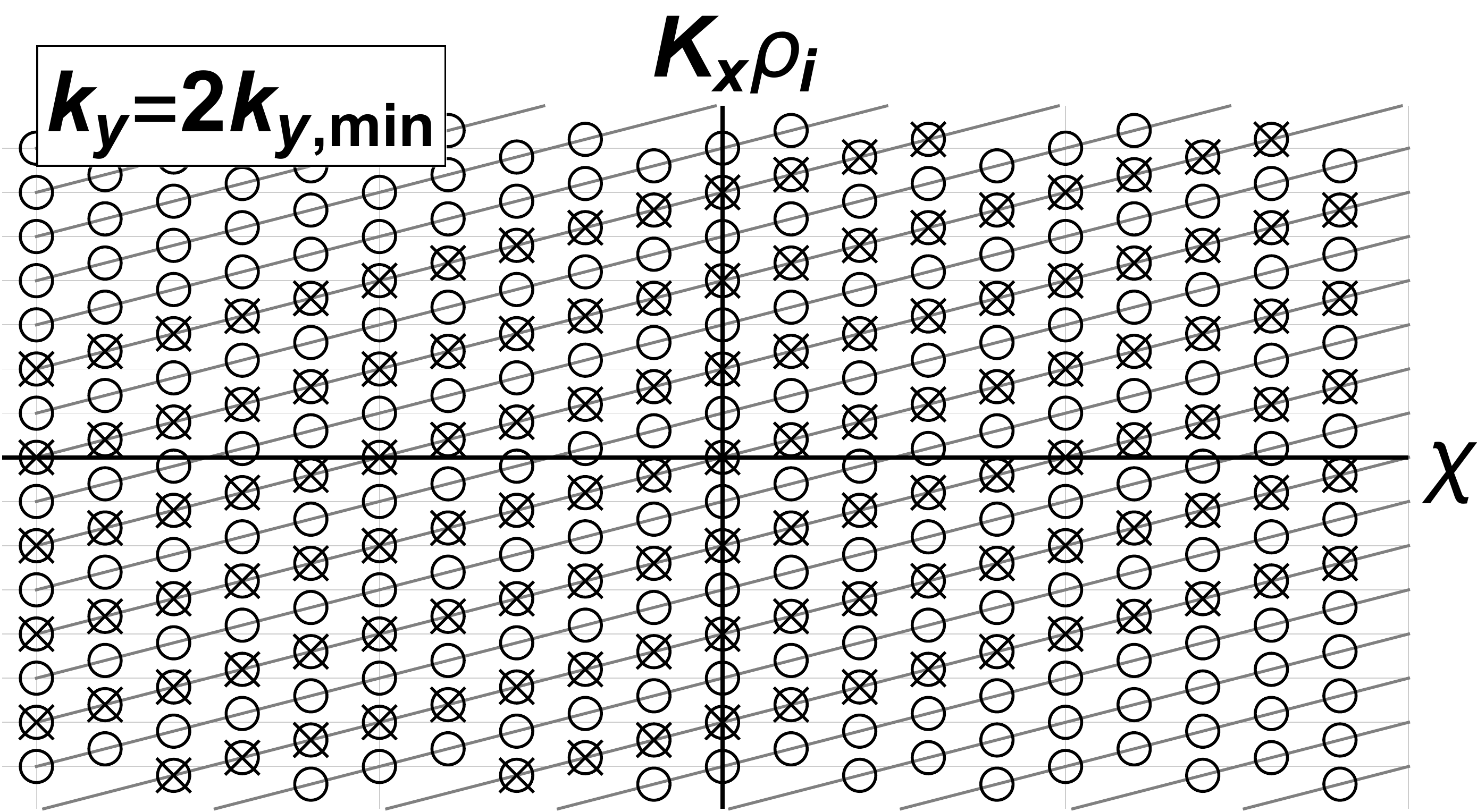}
	\includegraphics[width=0.33\textwidth]{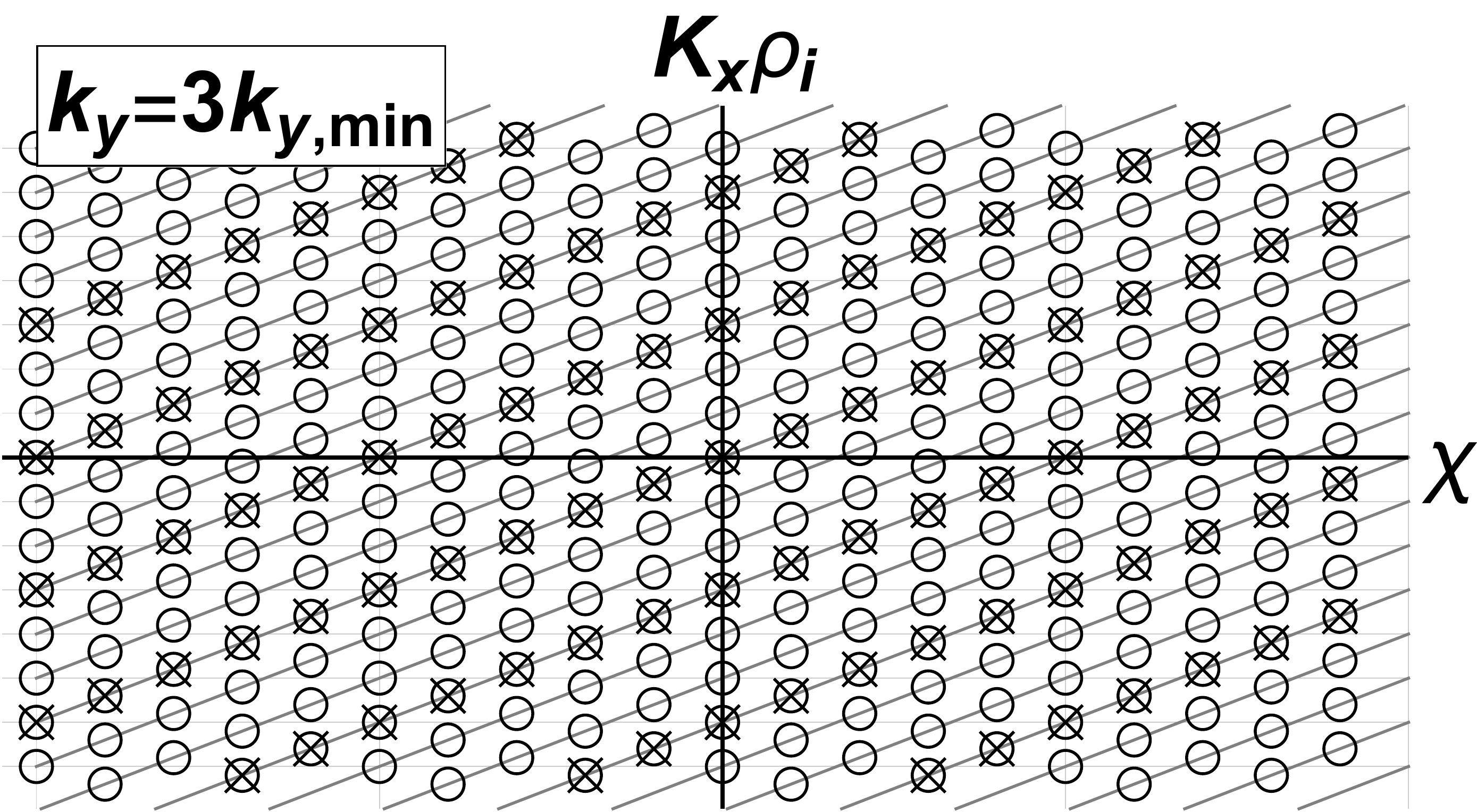}
	(d) \\
	\includegraphics[width=0.33\textwidth]{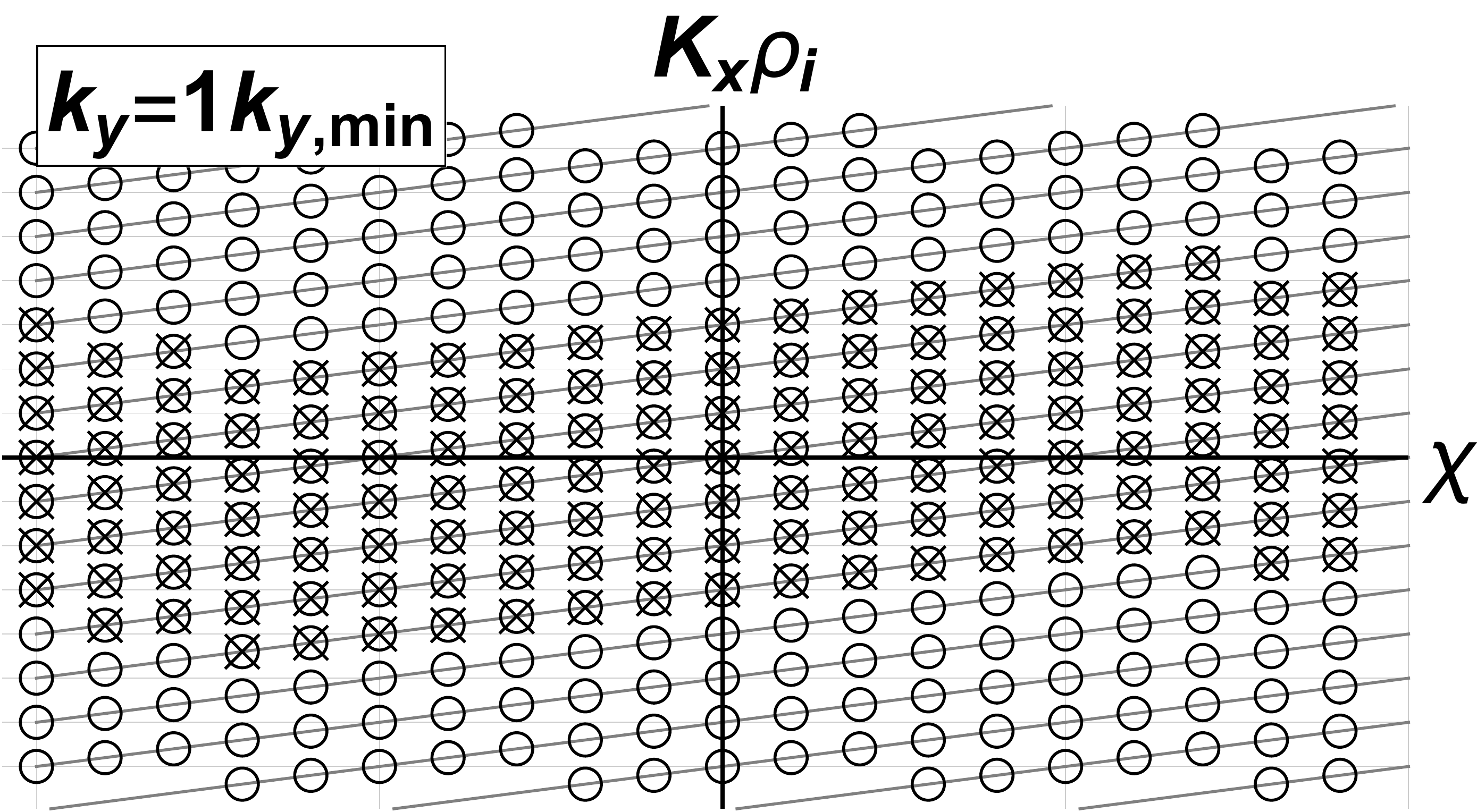}
	\includegraphics[width=0.33\textwidth]{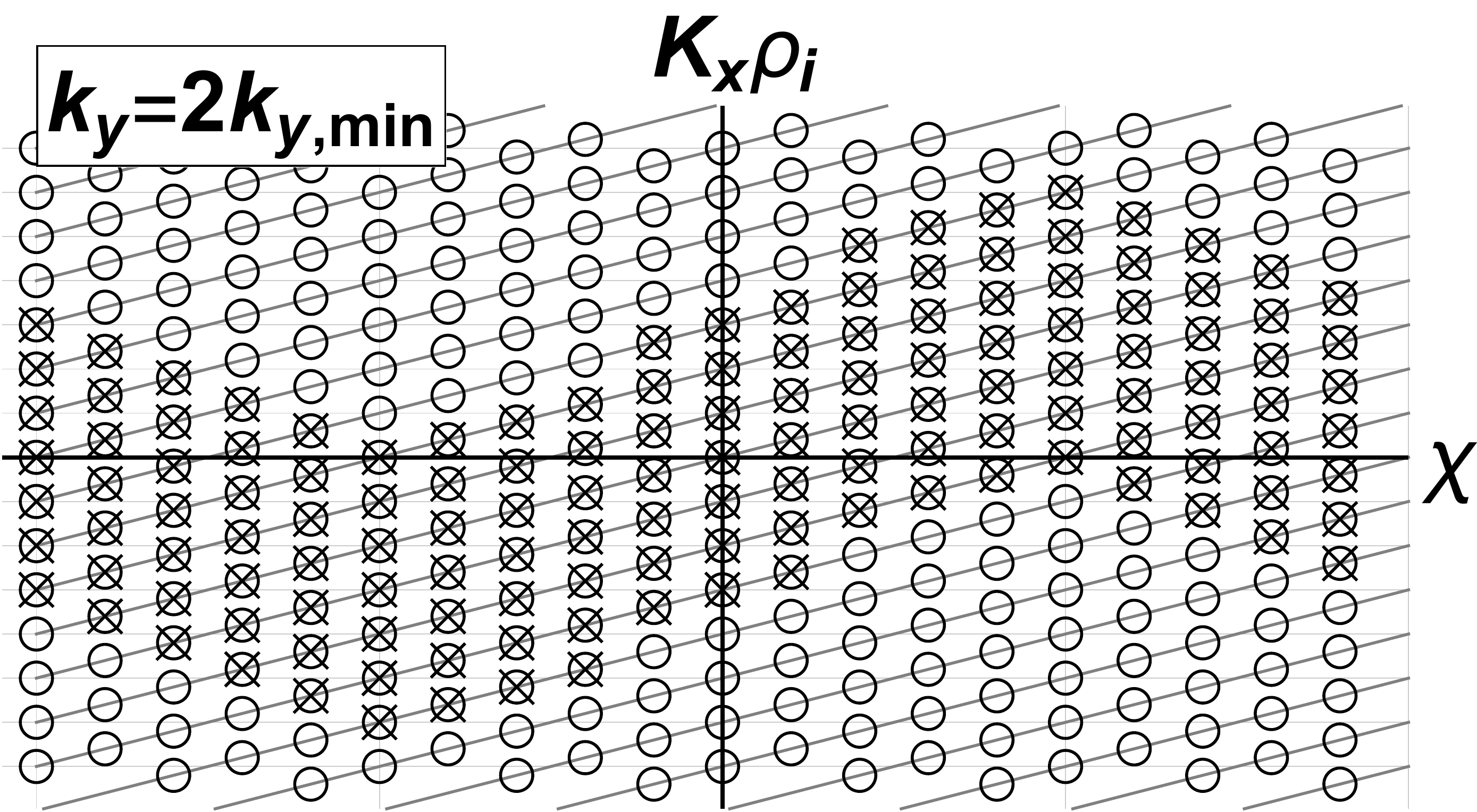}
	\includegraphics[width=0.33\textwidth]{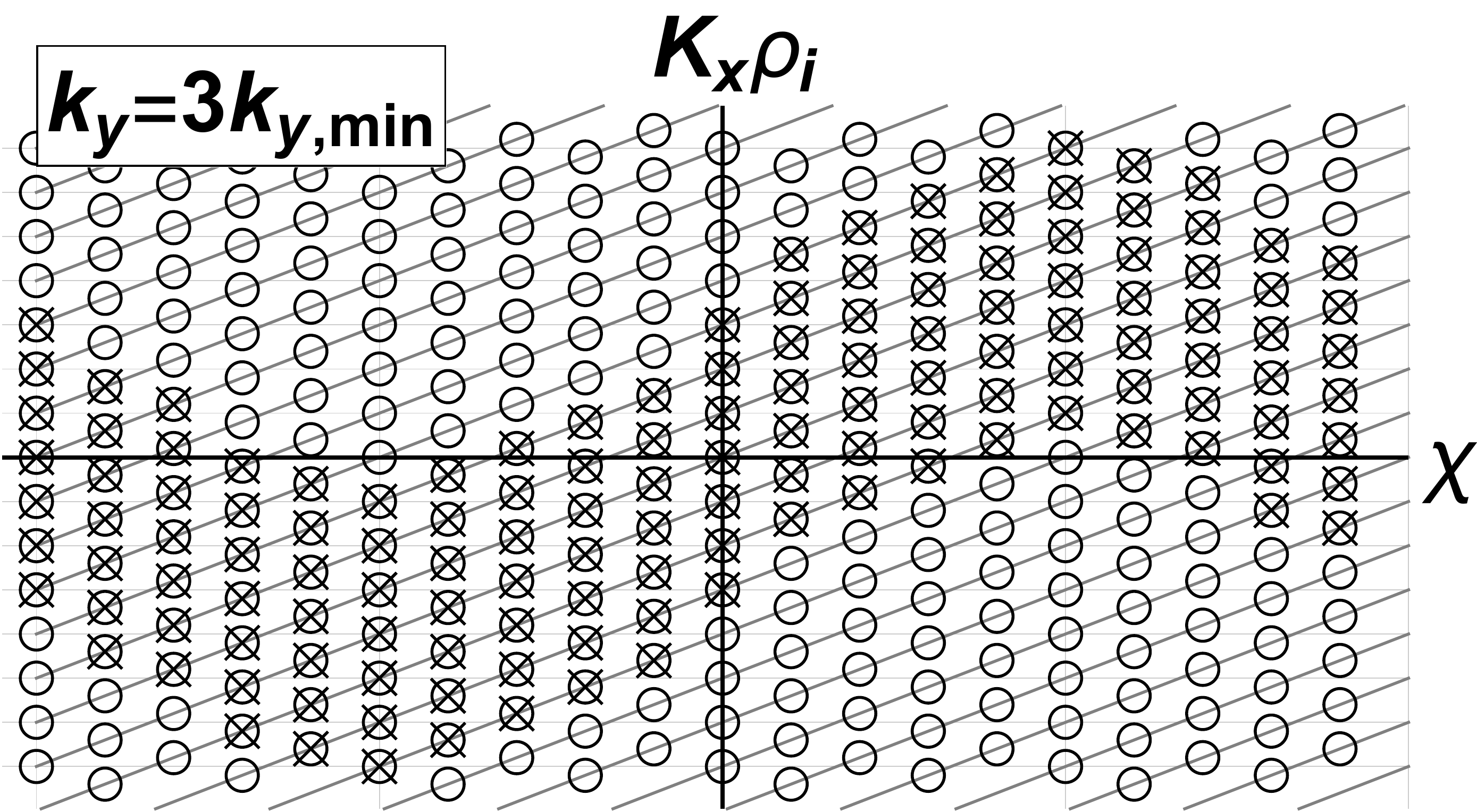}
	\caption{Cuts of the full three-dimensional lattice of the allowed (black circles) and included (black crosses) spatial grid points at the three lowest values of $k_{y}$ included in the simulation (columns) using (a) the local radial wavenumber $\tilde{K}_{x} \equiv K_{x}$ with $\tilde{K}_{y} \equiv \sqrt{ k_{y} 2 \pi / L_{y}}$, (b) the ballooning radial wavenumber $\tilde{K}_{x} \equiv k_{x}$ with $\tilde{K}_{y} \equiv k_{y}/ (1 + \text{NINT} \left[ \left| \chi \right| / \pi \right])$, (c) $\tilde{K}_{x} \equiv k_{x} k_{y,\text{min}} / k_{y}$ with $\tilde{K}_{y} \equiv k_{y}$, and (d) $\tilde{K}_{x} \equiv K_{x} + 2 k_{y} \hat{s} \text{sin} \left( \chi \right)$ with $\tilde{K}_{y} \equiv k_{y}$. The linear modes (diagonal gray lines) are indicated, $k_{y, \text{min}} = 2 \pi / L_{y}$ is the minimum value of $k_{y}$ anywhere on the spatial grid, $N_{\text{pol}} = 1$, and  $N_{x} = 7$.}
	\label{fig:gridLayoutsTilde}
\end{figure}

To understand these constraints, we will start by considering the binormal wavenumber. The most general possible definition of the binormal wavenumber is
\begin{align}
  \tilde{K}_{y} &\equiv \tilde{f}_{1} \left( k_{y}, \chi \right) , \label{eq:KyTildeDef1}
\end{align}
as any dependence on $\tilde{K}_{x}$ can be handled in its definition. Here $\tilde{f}_{j}$ is any arbitrary function of its arguments and the subscript is simply to distinguish the many arbitrary functions that will appear in this section. How does equation \refEq{eq:KyTildeDef1} determine which grid locations a simulation includes from figure \ref{fig:possibleGridLayouts}? As with the definition of $K_{x}$ in the non-twisting flux tube, we lay down a grid in $\tilde{K}_{y}$ that has a uniform spacing of $2 \pi / L_{y}$ and is centered around $\tilde{K}_{y} \approx 0$. Each value of $\tilde{K}_{y}$ that this produces must be rounded to the nearest grid point allowed by the boundary conditions. For the non-twisting flux tube, this was accomplished by including $m_{0} \left( k_{y}, \chi \right)$ in constructing the $K_{x}$ grid of equation \refEq{eq:KxGrid}. As an example, we could define $\tilde{K}_{y} \equiv \sqrt{ k_{y} 2 \pi / L_{y}}$ to produce the grid shown in figure \ref{fig:gridLayoutsTilde}(a), which only includes the $n=\{0,1,4,9,16, \ldots \}$ grid points from the traditional $k_{y}$ grid defined by equation \refEq{eq:kyGridy}.

This makes clear that the form of equation \refEq{eq:KyTildeDef1} does {\it not} necessarily maintain an evenly spaced $k_{y}$ grid, which makes it difficult to implement efficiently and realistically. Specifically, calculating the nonlinear term in real space according to equation \refEq{eq:nonlinearRealY} will become much more expensive because the Fourier transforms rely on computational libraries that assume a uniform grid. Instead, it would likely need to be directly calculated using a convolution in Fourier-space as given by equation \refEq{eq:nonlinearFourierY}, which makes it an $O ( N_{x}^{2} N_{y}^{2} )$ operation rather than $O ( N_{x} N_{y} \log{(N_{x} N_{y})} )$. Additionally, if the $\tilde{K}_{y}$ grid is non-uniform, then Fourier modes can couple via the quadratic nonlinearity to modes that do not exist {\it within} the grid. For example, this occurs for all nonlinear coupling involving different $k_{y}$ modes in the grid of figure \ref{fig:gridLayoutsTilde}(a) because the grid spacing is quadratic. In the conventional flux tube, Fourier modes do nonlinearly couple to non-existing modes, but this only occurs beyond the boundaries of the grid (i.e. above the maximum value of $k_{y}$). Nonlinear coupling involving non-existing Fourier modes internal to the grid can be much more problematic. When a non-existing mode is surrounded by existing modes should it still be taken as having zero amplitude as is done beyond the grid boundaries? Should one somehow interpolate from the existing grid points (even though turbulent mode amplitudes are not continuous in Fourier-space) or use one grid point to represent multiple modes? The ultimate feasibility of non-uniform binormal grids is left for future work. It may be of interest for multi-scale simulations because one could perhaps retain some coupling between ion and electrons scales, but still coarsen the binormal grid at intermediate wavenumbers.

If we want to maintain an evenly spaced binormal grid, $\tilde{K}_{y}$ must have a linear relationship with $k_{y}$ according to
\begin{align}
  \tilde{K}_{y} &\equiv \tilde{f}_{2} \left( \chi \right) k_{y} + \tilde{f}_{3} \left( \chi \right) . \label{eq:KyTildeDef2}
\end{align}
We can immediately set $\tilde{f}_{3} \left( \chi \right) = 0$ because the grid uses the reality condition to handle the negative binormal modes and we want the grid to retain the zonal modes at $k_{y} = 0$. Thus, the only difference from the conventional definition is that $\tilde{f}_{2} \left( \chi \right)$ can be used to vary the binormal domain width with parallel location. For example, figure \ref{fig:gridLayoutsTilde}(b) sets $\tilde{f}_{2} \left( \chi \right) = (1 + \text{NINT} \left[ \left| \chi \right| / \pi \right])^{-1}$ in order to halve the binormal domain width on the inboard side. Such a grid appears feasible to implement and would change the $\tilde{K}_{y}$ grid spacing between neighboring parallel grid locations. Of course, figure \ref{fig:gridLayoutsTilde}(b) shows that doing so can cause many linear modes to end (i.e. couple to zero) in the interior of the $( \tilde{K}_{y}, \chi )$ grid. This is analogous to the issue with nonlinear coupling in non-uniform $k_{y}$ grids, but is perhaps less conceptually concerning as the Fourier coefficients are continuous with $\chi$. Therefore, one could conceivably interpolate individual missing grid points or simply ensure that there is negligible turbulent activity at the parallel locations where a grid transition takes place. Perhaps a domain size that changes in the parallel direction could be appropriate for strongly shaped stellarators that have several independent regions of turbulence divided by regions of strong local magnetic shear. However, if these regions are fully independent, one could just run individual simulations for each. For the remainder of this section we will use the standard binormal wavenumber $\tilde{K}_{y} = k_{y}$.

The radial wavenumber coordinate has even more possibilities. The most general possible form is
\begin{align}
  \tilde{K}_{x} &\equiv \tilde{f}_{4} \left( k_{x}, k_{y}, \chi \right) . \label{eq:KxTildeDef1}
\end{align}
As with $\tilde{K}_{y}$, if we want to maintain an evenly spaced radial grid, $\tilde{K}_{x}$ must have a linear relationship with $k_{x}$ according to
\begin{align}
  \tilde{K}_{x} &\equiv \tilde{f}_{5} \left( k_{y}, \chi \right) k_{x} + \tilde{f}_{6} \left( k_{y}, \chi \right) . \label{eq:KxTildeDef2}
\end{align}
Otherwise, calculating the nonlinear term using Fourier transforms to real space will become much more expensive.

Interestingly, it is unclear if the form of equation \refEq{eq:KxTildeDef2} is sufficient to guarantee an efficient treatment of the nonlinear term. This is because, while the values of $\tilde{K}_{x}$ are evenly spaced for every value of $k_{y}$, the $\tilde{K}_{x}$ grid spacing is {\it not} necessarily the same between different $k_{y}$ values. Nevertheless, it may still be possible to efficiently calculate the nonlinear term in mixed $(x, k_{y})$ space on a non-uniform $x$ grid. Alternatively, one may be able to efficiently treat grids composed of blocks of $k_{y}$ modes with the same $\tilde{K}_{x}$ spacing by Fourier transforming the blocks independently and then combining them in real space. The feasibility of grids with $k_{y}$-dependent $\tilde{K}_{x}$ spacing is left for future work. We point out this category because it includes interesting possibilities such as $\tilde{K}_{x} \equiv k_{x} k_{y,\text{min}} / k_{y}$ (see figure \ref{fig:gridLayoutsTilde}(c)), which is similar to the conventional flux tube except it prevents the linear modes from being closely clustered around zero ballooning angle at high values of $k_{y}$.

However, efficiency aside, if the $\tilde{K}_{x}$ grid does not have the same spacing at all $k_{y}$, then Fourier modes can again nonlinearly couple to modes that do not exist {\it within} the grid. While this may be solvable in some cases, we will set this problem aside in this work. Instead we will ensure a consistent treatment of the nonlinear coupling by using the form
\begin{align}
  \tilde{K}_{x} &\equiv \tilde{f}_{7} \left( \chi \right) k_{x} + \tilde{f}_{8} \left( k_{y}, \chi \right) . \label{eq:KxTildeDef3}
\end{align}
Varying the radial domain width using $\tilde{f}_{7} \left( \chi \right)$ appears feasible and has similar considerations and consequences as varying the binormal domain width in equation \refEq{eq:KyTildeDef2}. Unlike with $\tilde{K}_{y}$, in this case the offset term $\tilde{f}_{8} \left( k_{y}, \chi \right)$ may be useful because the turbulence is not always centered around $\tilde{K}_{x} = 0$. Therefore, we will maintain the offset but choose to have a radial box width that is independent of $\chi$ by assuming the form
\begin{align}
  \tilde{K}_{x} &\equiv k_{x} + \tilde{f}_{8} \left( k_{y}, \chi \right) . \label{eq:KxTildeDef4}
\end{align}

A radial coordinate defined by equation \refEq{eq:KxTildeDef4} has the same grid spacings as the conventional flux tube, but grid points can have a $k_{y}$ and $\chi$ dependent offset. For example, the non-twisting flux tube chooses $\tilde{f}_{8} \left( k_{y}, \chi \right) =  k_{y} \Nabla x \cdot \Nabla y / | \Nabla x |^{2}$. This linear shift in the center of the radial grid with $k_{y}$ can be seen clearly in figure \ref{fig:nonlinearNpolResStudy}(b). It is the reason why the turbulent activity in the non-twisting flux tube remains centered in the perpendicular wavenumber grid at $\chi = 2 \pi$, while it stretches diagonally across the grid in the conventional flux tube. It appears feasible to implement a choice of $\tilde{f}_{8} \left( k_{y}, \chi \right)$ that has a complicated dependence on $k_{y}$, but we do not see any particular physical motivation for this. Instead we expect a relatively simple dependence from the competition of the two physical effects discussed earlier in this paper --- parallel streaming motivates turbulence centered around $k_{x} = 0$ and finite gyroradius damping motivates turbulence centered around $k_{x} = - k_{y} \Nabla x \cdot \Nabla y / | \Nabla x |^{2}$. To accommodate both of these $k_{y}$ dependencies and any combination in between, we assume the form
\begin{align}
  \tilde{K}_{x} &\equiv k_{x} + k_{y} \tilde{f}_{\text{tw}} \left( \chi \right) . \label{eq:KxTildeDef5}
\end{align}
As shown for an example in figure \ref{fig:gridLayoutsTilde}(d), this simply allows a shift in the included $K_{x}$ modes that varies with parallel location. Notably, the form makes it possible to maintain a Fourier series analogous to equations \refEq{eq:FourierSeriesy} and \refEq{eq:FourierSeriesY} by defining a real space binormal coordinate
\begin{align}
  \tilde{Y} \left( x, y, \chi \right) &\equiv y - \tilde{f}_{\text{tw}} \left( \chi \right) x \label{eq:YhatDef}
\end{align}
such that $\tilde{K}_{x} x + k_{y} \tilde{Y} = k_{x} x + k_{y} y$. Thus, we see a clear physical interpretation of the function $\tilde{f}_{\text{tw}} \left( \chi \right)$ --- it controls the twist of the flux tube cross-section as a function of parallel location. This flexibility has the potential to be useful.

To implement this, we must first determine the boundary conditions. Repeating the derivations from sections \ref{sec:derivTwist} and \ref{sec:derivNonTwist} for $\tilde{K}_{x}$ and $\tilde{Y}$, we find
\begin{align}
  \bar{\tilde{\Phi}} \left( x + L_{x}, \tilde{Y} - \tilde{f}_{\text{tw}} \left( \chi \right) L_{x}, \chi \right) &= \bar{\tilde{\Phi}} \left( x, \tilde{Y}, \chi \right) \label{eq:radialBCYtilde} \\
  \bar{\tilde{\Phi}} \left( x, \tilde{Y} + L_{y}, \chi \right) &= \bar{\tilde{\Phi}} \left( x, \tilde{Y}, \chi \right) \label{eq:binormalBCYtilde} \\
  \bar{\tilde{\Phi}} \left( x, \tilde{Y} \mp 2 \pi N_{\text{pol}} \hat{s} x - \left( \tilde{f}_{\text{tw}} \left( \chi + 2 \pi N_{\text{pol}} \right) - \tilde{f}_{\text{tw}} \left( \chi \right) \right) x, \chi + 2 \pi N_{\text{pol}} \right) &= \bar{\tilde{\Phi}} \left( x, \tilde{Y}, \chi \right) \label{eq:parallelBCYtilde}
\end{align}
in real space. In $( \tilde{K}_{x}, k_{y} )$ Fourier-space, we find equation \refEq{eq:kyGridy} and
\begin{align}
  \tilde{\Phi} \left( \tilde{K}_{x} \pm 2 \pi N_{\text{pol}} k_{y} \hat{s} + k_{y} \left( \tilde{f}_{\text{tw}} \left( \chi + 2 \pi N_{\text{pol}} \right) - \tilde{f}_{\text{tw}} \left( \chi \right) \right), k_{y}, \chi + 2 \pi N_{\text{pol}} \right) = \tilde{\Phi} \left( \tilde{K}_{x}, k_{y}, \chi \right) \label{eq:parallelBCKxTilde} \\
  \tilde{K}_{x} \in \frac{2 \pi}{L_{x}} \left( m + \tilde{m}_{0} \left( k_{y}, \chi \right) \right) + k_{y} \tilde{f}_{\text{tw}} \left( \chi \right) ~~~~~~~ \text{for all} ~ m \in \left[ - \frac{N_{x} - 1}{2}, \frac{N_{x} - 1}{2} \right] , \label{eq:KxTildeGrid}
\end{align}
where
\begin{align}
  \tilde{m}_{0} \left( k_{y}, \chi \right) &= - \text{NINT} \left[ \frac{L_{x}}{2 \pi} k_{y} \tilde{f}_{\text{tw}} \left( \chi \right) \right] . \label{eq:selectedModesTilde}
\end{align}
Note that the functions $\bar{\tilde{\Phi}}$ and $\tilde{\Phi}$ for the electrostatic potential are defined analogously to equations \refEq{eq:PhiRealFormY} and \refEq{eq:PhiFourierFormY} as
\begin{align}
  \bar{\tilde{\Phi}} \left( x, \tilde{Y}, \chi \right) &= \bar{\phi} \left( x, \tilde{Y} + \tilde{f}_{\text{tw}} \left( \chi \right) x, \chi \right) = \bar{\phi} \left( x, y \left( x, \tilde{Y}, \chi \right), \chi \right) \label{eq:PhiRealFormYtilde} \\
  \tilde{\Phi} \left( \tilde{K}_{x}, k_{y}, \chi \right) &= \phi \left( \tilde{K}_{x} - k_{y} \tilde{f}_{\text{tw}} \left( \chi \right), k_{y}, \chi \right) = \phi \left( k_{x} \left( \tilde{K}_{x}, k_{y}, \chi \right), k_{y}, \chi \right) . \label{eq:PhiFourierFormYtilde}
\end{align}
Like before, we can combine the real space radial and parallel boundary conditions of equations \refEq{eq:radialBCYtilde} and \refEq{eq:parallelBCYtilde} (as well as equation \refEq{eq:binormalBCYtilde}) to find constraints on the domain. However, here we actually find two conditions. First, we find the standard discretization of the domain aspect ratio (i.e. equation \refEq{eq:aspRatioDiscrete}) by evaluating the parallel boundary condition at $x \rightarrow x + L_{x}$, applying the radial boundary condition to both sides, and then applying the parallel boundary condition to the left side. Second, by evaluating the radial boundary condition at $\chi \rightarrow \chi + 2 \pi N_{\text{pol}}$, applying the parallel boundary condition to both sides, applying the radial boundary condition to the left side, and then using equation \refEq{eq:aspRatioDiscrete}, we find
\begin{align}
  \frac{\tilde{f}_{\text{tw}} \left( \chi + 2 \pi N_{\text{pol}} \right) - \tilde{f}_{\text{tw}} \left( \chi \right)}{2 \pi N_{\text{pol}} \left| \hat{s} \right|} &=  \frac{N_{\mathbb{Z}}}{N_{\text{asp}}} , \label{eq:twistConstraint}
\end{align}
where $N_{\mathbb{Z}} \in \mathbb{Z}$ is some integer. Equation \refEq{eq:twistConstraint} is a novel result and is interesting. We see that any $2 \pi N_{\text{pol}}$-periodic function can be added to $\tilde{f}_{\text{tw}} \left( \chi \right)$ without causing concern. However, the global twist of the simulation domain is discretized by the magnetic shear just like the domain aspect ratio. Finishing the calculation, we find that, despite the generality of this transformation, the geometric coefficients and equilibrium quantities retain simple forms of
\begin{align}
  \Nabla x \cdot \Nabla \tilde{Y} &= \Nabla x \cdot \Nabla y - \tilde{f}_{\text{tw}} \left( \chi \right) \left| \Nabla x \right|^{2} \label{eq:gradXdotGradYhat} \\
  \left| \Nabla \tilde{Y} \right|^{2} &= \left| \Nabla y \right|^{2} - 2 \tilde{f}_{\text{tw}} \left( \chi \right) \Nabla x \cdot \Nabla y + \tilde{f}_{\text{tw}}^{2} \left( \chi \right) \left| \Nabla x \right|^{2} \\
  \Nabla \tilde{Y} \cdot \Nabla \chi &=  \Nabla y \cdot \Nabla \chi - \tilde{f}_{\text{tw}} \left( \chi \right) \Nabla x \cdot \Nabla \chi \\
  J^{-1} &\equiv \mp \left( \Nabla x \times \Nabla \tilde{Y} \right) \cdot \hat{b} = \mp \left( \Nabla x \times \Nabla y \right) \cdot \hat{b} \label{eq:jacobHat} \\
  \vec{B} &= \frac{1}{C_{y}} \frac{d \psi}{d x} \Nabla x \times \Nabla y = \frac{1}{C_{y}} \frac{d \psi}{d x} \Nabla x \times \Nabla \tilde{Y} \\
  \Nabla B &= \left( \frac{\partial B}{\partial x} + \tilde{f}_{\text{tw}} \left( \chi \right) \frac{\partial B}{\partial y} \right) \Nabla x + \frac{\partial B}{\partial y} \Nabla \tilde{Y} + \frac{\partial B}{\partial \chi} \Nabla \chi .
\end{align}

In summary, we see that it is possible to implement a wide variety of flux tubes apart from just the conventional and non-twisting versions. While it is currently unclear if certain exotic options are computationally feasible and useful, it does appear straightforward to achieve complete control of the twist of the flux tube along its length by defining the radial wavenumber according to equation \refEq{eq:KxTildeDef5}. In fact, all local computational domains considered so far in the literature can be expressed through different choices of the function $\tilde{f}_{\text{tw}} \left( \chi \right)$. Specifically, the conventional flux tube, the non-twisting flux tube, the globally non-twisting flux tube (which only removes the twist from the {\it global} magnetic shear), and the flux tube train \cite{WatanabeFluxTubeTrain2015} are given by
\begin{align}
  \tilde{f}_{\text{tw}} \left( \chi \right) &= 0 \\
  \tilde{f}_{\text{tw}} \left( \chi \right) &= \frac{\Nabla x \cdot \Nabla y}{\left| \Nabla x \right|^{2}} \\
  \tilde{f}_{\text{tw}} \left( \chi \right) &= \mp \hat{s} \chi \\
  \tilde{f}_{\text{tw}} \left( \chi \right) &= \mp 2 \pi \hat{s} \text{NINT} \left[ \frac{\chi}{2 \pi} \right] = \left. \frac{\Nabla x \cdot \Nabla y}{\left| \Nabla x \right|^{2}} \right|_{\chi' = 2 \pi \text{NINT} \left[ \chi/(2 \pi) \right]}
\end{align}
respectively. The first three have already been implemented in GENE and any choice of $\tilde{f}_{\text{tw}} \left( \chi \right)$ looks simple to add. For unusual turbulence, finding the best choice of $\tilde{f}_{\text{tw}} \left( \chi \right)$ may offer significant computational savings. More broadly, it may be possible to formulate an automated routine that performs linear tests to optimize $\tilde{f}_{\text{tw}} \left( \chi \right)$ before beginning a nonlinear simulation or adapts the choice of $\tilde{f}_{\text{tw}} \left( \chi \right)$ during a nonlinear simulation.

\section{Conclusions}
\label{sec:conclusions}

In this paper, we have presented a novel derivation of the ``non-twisting flux tube'' --- a local simulation domain that does not twist due to the effects of magnetic shear. It maintains a field-aligned spatial grid, but abandons having a field-aligned domain. In other words, grid points are placed in rows along magnetic field lines, but the boundaries of the domain do {\it not} follow field lines. This is made possible by allowing field lines to pass through the binormal boundary of the domain, but, whenever they exit, binormal periodicity is invoked so that they immediately enter back in on the opposite side. Importantly, this derivation shows that such a flux tube is consistent with a periodic radial boundary condition and the Fourier representation that is typically used in the perpendicular directions of local domains.

By comparing the spatial coordinate grids of the non-twisting and conventional flux tubes, we found that they are optimized to model different physical effects. The conventional flux tube grid prioritizes parallel streaming and is constructed to follow linear modes along field lines from the outboard midplane. On the other hand, the non-twisting flux tube grid prioritizes the Fourier modes that are least damped by finite gyroradius effects (i.e. those with a small {\it local} radial wavenumber $K_{x}$). Thus, although both simulation domains should give the same result in the limit of infinite radial resolution, one may be more computationally efficient and allow accurate simulations at lower grid resolution.

To investigate this, we successfully implemented and benchmarked the non-twisting flux tube in the gyrokinetic code GENE. Then, using five different test cases, we analyzed the runtime and convergence properties. We found {\it no} case where the non-twisting flux tube performed worse than the conventional flux tube. Moreover, we found cases where the non-twisting flux tube was seven, or even thirty times less expensive. This computational savings came partially from a longer time step and partially from enabling a lower radial resolution. The non-twisting flux tube was found to perform best when the magnetic shear was high and when there was more than one region of turbulent drive in the parallel direction. This makes it potentially useful for stellarator simulations, pedestal simulations, and tokamak simulations with several poloidal turns.

Lastly, we showed that the non-twisting flux tube is just one example of many possible flux tube formulations. By generalizing the non-twisting flux tube derivation, we explored the feasibility of some exotic flux tube grids. Moreover, we demonstrated that it is straightforward to fully control how the flux tube twists as it extends along the field line. Thus, it may be possible to tailor the flux tube cross-section to the particular turbulence being modeled. This could enable even more efficient spatial grids and, ultimately, significant computational savings.

\ack

The authors would like to thank Tobias Goerler, Daniel Told, and Ajay C. J. for useful discussions pertaining to this work. 
This work was supported by the EUROfusion - Theory and Advanced Simulation Coordination (E-TASC). This work has been carried out within the framework of the EUROfusion Consortium and has received funding from the Euratom research and training programme 2014-2018 and 2019-2020 under grant agreement No 633053. The views and opinions expressed herein do not necessarily reflect those of the European Commission. 
We acknowledge the CINECA award under the ISCRA initiative, for the availability of high performance computing resources and support. 
This work was supported by a grant from the Swiss National Supercomputing Centre (CSCS) under project ID s863, s956, and 1050. 
This work was carried out using the JFRS-1 supercomputer system at Computational Simulation Centre of International Fusion Energy Research Centre (IFERC-CSC) in Rokkasho Fusion Institute of QST (Aomori, Japan).

\appendix

\section{The straight-field line and non-twisting poloidal angles}
\label{app:straightCoord}

We are seeking a field line label $y$ to construct a field-aligned coordinate system. We already have a radial coordinate $x$ that selects the flux surface. The coordinate $y$ will be used to select the field line and any arbitrary poloidal angle $\theta$ will determine the parallel position. Since $y$ must be constant along a magnetic field line, we know it must respect the field line trajectory relation
\begin{align}
  \left. \frac{\partial \zeta}{\partial \theta} \right|_{x,y} = \frac{\vec{B} \cdot \Nabla \zeta}{\vec{B} \cdot \Nabla \theta} ,
\end{align}
where the vertical bar indicates the quantities that are held constant in the operation. This says that moving poloidally while staying on a field line (i.e. staying at constant $x$ and $y$) involves moving toroidally an amount proportional to the field line pitch. Integrating this equation gives
\begin{align}
  \zeta = \left. \int_{0}^{\theta} \right|_{x,y} d \theta' \frac{\vec{B} \cdot \Nabla \zeta}{\vec{B} \cdot \Nabla \theta'} + f(x,y) ,
\end{align}
where we see that $y$ only appears in the integration constant. Choosing a form for $f(x,y)$ is unconstrained and reflects the flexibility we have in defining $y$. The conventional choice is $f(x,y) = \pm y / C_{y}$, which after rearranging gives the typical definition \cite{GoerlerGENE2011} of
\begin{align}
  y = \pm C_{y} \left( \zeta - \left. \int_{0}^{\theta} \right|_{x} d \theta' \frac{\vec{B} \cdot \Nabla \zeta}{\vec{B} \cdot \Nabla \theta'} \right) . \label{eq:yDefGen}
\end{align}
Here we have indicated that the integral is performed at constant $x$, but do not need to specify a second constant because of toroidal symmetry. Now this definition of $y$ holds for any poloidal angle $\theta$ (including the geometric poloidal angle). However, if we chose to define the poloidal angle according to
\begin{align}
  \chi ( x, \theta ) \equiv \frac{1}{q(x)} \left. \int_{0}^{\theta} \right|_{x} d \theta' \frac{\vec{B} \cdot \Nabla \zeta}{\vec{B} \cdot \Nabla \theta'} , \label{eq:chiDef}
\end{align}
we can substitute it into equation \refEq{eq:yDefGen} and arrive at the simpler definition of $y$ given by equation \refEq{eq:yDef}. This poloidal angle $\chi$ is called a ``straight-field line'' poloidal angle because we can invert equation \refEq{eq:yDef} to directly calculate that
\begin{align}
\left. \frac{\partial \zeta}{\partial \chi} \right|_{x,y} = q(x) .
\end{align}
In other words, the magnetic field lines appears as straight lines when plotted in the $(\chi, \zeta)$ coordinate plane.

In this paper, we use the straight-field line poloidal angle $\chi$ because it is what is used by the GENE code. Moreover, it makes it straightforward to calculate the geometric coefficients for the non-twisting flux tube from the geometric coefficients that were already calculated for the conventional flux tube (i.e. according to equations \refEq{eq:gradXdotGradY} through \refEq{eq:gradB}). However, if one was coding a non-twisting flux tube from scratch, there is another poloidal coordinate that may be more convenient. Analogously to the new non-twisting binormal coordinate $Y$, one can define a new ``non-twisting poloidal angle'' that has no minor radial variation and is a straight-field line poloidal angle only at $x=0$. Such a non-twisting poloidal angle is defined by
\begin{align}
  \vartheta ( \theta ) \equiv \chi ( x = 0, \theta ) = \frac{1}{q_{0}} \left. \int_{0}^{\theta} \right|_{x=0} d \theta' \frac{\vec{B} \cdot \Nabla \zeta}{\vec{B} \cdot \Nabla \theta'} . \label{eq:varthetaDef}
\end{align}

This is useful primarily because $\Nabla x \cdot \Nabla \vartheta = 0$, which will allow it to cleanly separate the effects of global magnetic shear from the local corrections. To see how, substitute the definition of $\vartheta$ and the radial derivative of the definition of
\begin{align}
  q \left( x \right) \equiv \frac{1}{2 \pi} \left. \oint_{0}^{2 \pi} \right|_{x} d \theta' \frac{\vec{B} \cdot \Nabla \zeta}{\vec{B} \cdot \Nabla \theta'} ,
\end{align}
into a Taylor expansion of equation \refEq{eq:yDefGen} about $x=0$. This gives
\begin{align}
   y = \pm C_{y} \zeta \mp C_{y} q_{0} \vartheta \mp \hat{s} \vartheta x \mp x \int_{0}^{\vartheta} d \vartheta' \hat{l} \left( \vartheta' \right) , \label{eq:yVartheta}
\end{align}
where
\begin{align}
  \hat{l} \left( \vartheta \right) \equiv C_{y} \frac{d \theta}{d \vartheta} \left. \frac{\partial}{\partial x} \right|_{\theta} \frac{\vec{B} \cdot \Nabla \zeta}{\vec{B} \cdot \Nabla \theta} - \frac{C_{y}}{2 \pi} \left. \oint_{0}^{2 \pi} \right|_{x=0} d \theta \left( \left. \frac{\partial}{\partial x} \right|_{\theta} \frac{\vec{B} \cdot \Nabla \zeta}{\vec{B} \cdot \Nabla \theta} \right) \label{eq:localShearDef}
\end{align}
is the local correction to the magnetic shear (which is defined to be the total shear minus the global shear) and $\theta \left( \vartheta \right)$ can be computed by inverting equation \refEq{eq:varthetaDef}.

Comparing equation \refEq{eq:yVartheta} with equation \refEq{eq:yForm}, we see that a new term has appeared, which makes explicit the local correction to the magnetic shear. This is the advantage of the new poloidal angle. Previously, the local correction was hidden away within the definition of the poloidal angle $\chi$, specifically in the second term of equation \refEq{eq:yForm}. In this new non-twisting poloidal angle, one can use equation \refEq{eq:yVartheta} to show that the geometric coefficients of the conventional flux tube are
\begin{align}
  \Nabla y &= \pm C_{y} \Nabla \zeta \mp C_{y} q_{0} \Nabla \vartheta \mp \hat{s} \vartheta \Nabla x \mp \Nabla x \int_{0}^{\vartheta} d \vartheta' \hat{l} \left( \vartheta' \right) \\
  \frac{\Nabla x \cdot \Nabla y}{\left| \Nabla x \right|^{2}} &= \mp \hat{s} \vartheta \mp \int_{0}^{\vartheta} d \vartheta' \hat{l} \left( \vartheta' \right) \\
  \left| \Nabla y \right|^{2} &= C_{y}^{2} \left| \Nabla \zeta \right|^{2} + C_{y}^{2} q_{0}^{2} \left| \Nabla \vartheta \right|^{2} + \left( \hat{s} \vartheta + \int_{0}^{\vartheta} d \vartheta' \hat{l} \left( \vartheta' \right) \right)^{2} \left| \Nabla x \right|^{2} ,
\end{align}
which look pretty similar to what they were using the typical straight-field line poloidal angle $\chi$ (e.g. equations \refEq{eq:grady} and \refEq{eq:Yfactor}). However, using equation \refEq{eq:Ydef} we can calculate the geometric coefficients of the non-twisting flux tube to be
\begin{align}
  Y &= \pm C_{y} \zeta \mp C_{y} q_{0} \vartheta \\
  \Nabla Y &= \pm C_{y} \Nabla \zeta \mp C_{y} q_{0} \Nabla \vartheta \\
  \Nabla x \cdot \Nabla Y &= 0 \\
  \left| \Nabla Y \right|^{2} &= C_{y}^{2} \left| \Nabla \zeta \right|^{2} + C_{y}^{2} q_{0}^{2} \left| \Nabla \vartheta \right|^{2} \\
  \Nabla Y \cdot \Nabla \vartheta &= \mp C_{y} q_{0} \left| \Nabla \vartheta \right|^{2} , \\
  \Nabla x \cdot \Nabla \vartheta &= 0 ,
\end{align}
which are considerably simpler than equations \refEq{eq:Yform}, \refEq{eq:gradY}, \refEq{eq:gradYSq}, and \refEq{eq:gradYGradChi}.

\section{Computational implementation}
\label{app:implementation}

In this appendix we will outline the practicalities of implementing the non-twisting flux tube in the gyrokinetic code GENE \cite{JenkoGENE2000, GoerlerGENE2011}. If we compare the equations for the non-twisting flux tube (i.e. equations \refEq{eq:kyGridy}, \refEq{eq:quasineuty}, \refEq{eq:parallelBCfourierKx}, \refEq{eq:KxGrid}, \refEq{eq:kperpDefY}, \refEq{eq:GKeqY}, and \refEq{eq:nonlinearRealY}) with the equations for the conventional flux tube (i.e. equations \refEq{eq:kxGridy}, \refEq{eq:kyGridy}, \refEq{eq:parallelBCfourierkx}, \refEq{eq:GKeqy}, \refEq{eq:quasineuty}, \refEq{eq:kperpDefy}, and \refEq{eq:nonlinearRealy}) we see that relatively few changes are required. The binormal $k_{y}$ grid requires no changes. The geometric quantities must be transformed, but this is simple to accomplish using equations \refEq{eq:gradXdotGradY} through \refEq{eq:gradB}.

However, the modification of the radial grid is more substantial. It should be constructed according to equation \refEq{eq:KxGrid}, which has different values of $K_{x}$ at each parallel location $\chi$ and for each binormal wavenumber $k_{y}$. Changing the radial grid to a three dimensional array would have ramifications all throughout the code, so we will make some adjustments to the analytic derivation to avoid it. To do this, we notice that, even though the physical meanings of the ballooning radial wavenumber $k_{x}$ and the local radial wavenumber $K_{x}$ are quite different, the numerical values of the grid are very similar. Equation \refEq{eq:kxGridy} is a rectangular grid centered around $k_{x} = 0$ with a regular grid spacing of $2 \pi / L_{x}$ and equation \refEq{eq:KxGrid} is a rectangular grid centered around $K_{x} \approx 0$ with a regular spacing of $2 \pi / L_{x}$. We see that the only difference in the numerical values is a small shift of less than a grid point that comes from being centered at $K_{x} \approx 0$ instead of $K_{x} = 0$. Thus, we will write the local radial wavenumber as a sum of two contributions
\begin{align}
  K_{x} &= \mathscr{K}_{x} + \Delta K_{x} , \label{eq:scriptDecomp}
\end{align}
where
\begin{align}
  \Delta K_{x}  \equiv k_{y} \frac{\Nabla x \cdot \Nabla y}{\left| \Nabla x \right|^{2}} - \frac{2 \pi}{L_{x}} \text{NINT} \left[ \frac{L_{x}}{2 \pi} k_{y} \frac{\Nabla x \cdot \Nabla y}{\left| \Nabla x \right|^{2}} \right] \label{eq:smallShift}
\end{align}
is the small shift that is the only difference between the $K_{x}$ and $k_{x}$ modes in the non-twisting and conventional flux tubes. By substituting equations \refEq{eq:scriptDecomp} and \refEq{eq:smallShift} into equation \refEq{eq:KxGrid}, we find that the grid for $\mathscr{K}_{x}$ is
\begin{align}
\mathscr{K}_{x} &\in \frac{2 \pi}{L_{x}} m ~~~~~~~ \text{for all} ~ m \in \left[ - \frac{N_{x} - 1}{2}, \frac{N_{x} - 1}{2} \right] ,  \label{eq:scriptKxGrid}
\end{align}
which no longer depends on $k_{y}$ nor $\chi$ and is numerically identical to the $k_{x}$ grid of the conventional flux tube (i.e. equation \refEq{eq:kxGridy}). Thus, if we used $\mathscr{K}_{x}$ as the radial wavenumber, the radial grid would not require any modification. To do so, we substitute equation \refEq{eq:scriptDecomp} into the gyrokinetic model for the non-twisting flux tube (i.e. equation \refEq{eq:GKeqY}) to find
\begin{align}
  \frac{\partial \mathscr{H}_{s}}{\partial t} &+ v_{||} \hat{b} \cdot \vec{\nabla} \chi \left. \frac{\partial \mathscr{H}_{s}}{\partial \chi} \right|_{\mathscr{K}_{x} + \Delta K_{x} - k_{y} \frac{\Nabla x \cdot \Nabla y}{\left| \Nabla x \right|^{2}}, k_{y}} + i \vec{v}_{d s} \cdot \left( \mathscr{K}_{x} \Nabla x + \Delta K_{x} \Nabla x + k_{y} \Nabla Y \right) \mathscr{H}_{s} + a_{s ||} \frac{\partial \mathscr{H}_{s}}{\partial v_{||}} \nonumber \\
  &\mp \frac{1}{J B} \left\{ \mathscr{H}_{s}, \varphi J_{0} \left( \mathscr{K}_{\perp} \rho_{s} \right) \right\} = \frac{Z_{s} e F_{M s}}{T_{s}} \frac{\partial \varphi}{\partial t} J_{0} \left( \mathscr{K}_{\perp} \rho_{s} \right) \mp i \frac{k_{y}}{J B} \varphi J_{0} \left( \mathscr{K}_{\perp} \rho_{s} \right) \frac{\partial F_{Ms}}{\partial x} , \label{eq:GKeqYwithSmallShift}
\end{align}
where the perpendicular wavenumber is
\begin{align}
  \mathscr{K}_{\perp} = \sqrt{ \left( \mathscr{K}_{x} + \Delta K_{x} \right)^{2} \left| \Nabla x \right|^{2} + k_{y}^{2} \left| \Nabla Y \right|^{2}} \label{eq:kPerpDefScriptKx}
\end{align}
and the quasineutrality equation remains unchanged except for the new form of the perpendicular wavenumber, the electrostatic potential becomes a function of $\mathscr{K}_{x}$ according to
\begin{align}
  \varphi \left( \mathscr{K}_{x}, k_{y}, \chi \right) = \phi \left( \mathscr{K}_{x} + \Delta K_{x} - k_{y} \frac{\Nabla x \cdot \Nabla y}{| \Nabla x |^{2}}, k_{y}, \chi \right) = \phi \left( k_{x} ( \mathscr{K}_{x}, k_{y}, \chi ), k_{y}, \chi \right) , \label{eq:PhiFourierFormScript}
\end{align}
and the distribution function $\mathscr{H}_{s}$ is defined analogously. The nonlinear term is computed in real space according to
\begin{align}
\left\{ \mathscr{H}_{s}, \varphi J_{0} \left( \mathscr{K}_{\perp} \rho_{s} \right) \right\} &= \frac{1}{N_{x} \left( 2 N_{y} - 1 \right)} \sum_{x, Y} \left[ \left( \sum_{\mathscr{K}_{x}', k'_{y}} \left( \mathscr{K}_{x}' + \Delta K_{x}' \right) \mathscr{H}_{s}' \Exp{i \Delta K_{x}' x} \Exp{i \mathscr{K}_{x}' x + i k'_{y} Y} \right) \right. \nonumber \\
  &\times \left( \sum_{\mathscr{K}_{x}'', k''_{y}} k_{y}'' \varphi'' J_{0} \left( \mathscr{K}_{\perp}'' \rho_{s} \right) \Exp{i \Delta K_{x}'' x} \Exp{i \mathscr{K}_{x}'' x + i k''_{y} Y} \right) \label{eq:nonlinearRealYwithSmallShift} \\
  &- \left( \sum_{\mathscr{K}_{x}', k'_{y}} k_{y}' \mathscr{H}_{s}' \Exp{i \Delta K_{x}' x} \Exp{i \mathscr{K}_{x}' x + i k'_{y} Y} \right) \nonumber \\
  &\times \left. \left( \sum_{\mathscr{K}_{x}'', k''_{y}} \left( \mathscr{K}_{x}'' + \Delta K_{x}'' \right) \varphi'' J_{0} \left( \mathscr{K}_{\perp}'' \rho_{s} \right) \Exp{i \Delta K_{x}'' x} \Exp{i \mathscr{K}_{x}'' x + i k''_{y} Y} \right) \right] \Exp{-i \Delta K_{x} x} \Exp{-i \mathscr{K}_{x} x - i k_{y} Y} \nonumber
\end{align}
using the three-halves rule as before and the modified Fourier transforms
\begin{align}
  \bar{\varphi} \left( x, Y, \chi \right) &= \sum_{\mathscr{K}_{x}, k_{y}} \varphi \left( \mathscr{K}_{x}, k_{y}, \chi \right) \Exp{i \Delta K_{x} x} \Exp{i \mathscr{K}_{x} x + i k_{y} Y} \label{eq:FourierSeriesYwithSmallShift} \\
  \varphi \left( \mathscr{K}_{x}, k_{y}, \chi \right) &= \frac{1}{N_{x} \left( 2 N_{y} - 1 \right)} \sum_{x, Y} \bar{\varphi} \left( x, Y, \chi \right) \Exp{- i \Delta K_{x} x} \Exp{- i \mathscr{K}_{x} x - i k_{y} Y} . \label{eq:invFourierSeriesYwithSmallShift}
\end{align}
This eliminates the need to modify the radial wavenumber grid because the preexisting $k_{x}$ grid is identical to the new $\mathscr{K}_{x}$ grid. Instead all we must do is add the small $\Delta K_{x}$ correction terms in the perpendicular wavenumber within the argument to the Bessel functions, the magnetic drift term, and the nonlinear term (including the exponential phase factors in the Fourier transforms).

Additionally, there is a numerical subtlety related to the nearest integer rounding used in constructing the new radial grid. The $\text{NINT}$ function that appears in equations \refEq{eq:selectedModes}, \refEq{eq:smallShift}, etc. behaves unpredictably when its argument evaluates to half-integer values. This is because any incidental round-off errors will then determine whether the function rounds up or down. This is particularly important at the $\chi = - \pi N_{\text{pol}}$ grid point, as can be seen from equation \refEq{eq:linearProofNoLocalShear}. Note that GENE does {\it not} put a grid point at $\chi = \pi N_{\text{pol}}$, so we do not need to worry about it. To ensure well-defined and consistent rounding, instead of calculating the argument of $\text{NINT}$ at a given grid location $\chi$, it is actually approximated at a slightly larger value using the linear interpolation
\begin{align}
  \text{NINT} \left[ \frac{L_{x}}{2 \pi} k_{y} \left. \frac{\Nabla x \cdot \Nabla y}{\left| \Nabla x \right|^{2}} \right|_{\chi} \right] \rightarrow \text{NINT} \left[ \frac{L_{x}}{2 \pi} k_{y} \left( (1-\delta) \left. \frac{\Nabla x \cdot \Nabla y}{\left| \Nabla x \right|^{2}} \right|_{\chi} + \delta \left. \frac{\Nabla x \cdot \Nabla y}{\left| \Nabla x \right|^{2}} \right|_{\chi + \Delta \chi} \right) \right] ,
\end{align}
where $\Delta \chi$ is the parallel grid spacing. This means that, for typical simulations, the $\chi = - \pi N_{\text{pol}}$ grid point will be rounded in the same way as the $\chi = - \pi N_{\text{pol}} + \Delta \chi$ grid point. The constant $\delta = 0.01313$ was chosen fairly arbitrarily to be small enough to be negligible from a physical perspective, but large enough to overcome the significant numerical errors that can be present when using geometric coefficients calculated by external MHD equilibrium codes. It is also important that $\delta$ is not a rational number with a small denominator, otherwise such a round-off problem may inadvertently occur at another parallel grid location.

The last and most significant challenge to implement the non-twisting flux tube is the parallel derivative because it is crucial to maintain an exactly field-aligned grid. In a conventional flux tube this is straightforward to accomplish because taking a derivative exactly along the field line corresponds to taking the derivative at constant Fourier mode {\it index}. To see this, note that the parallel derivative in equation \refEq{eq:GKeqy} is taken at constant $k_{x}$ and $k_{y}$, which is the same as holding the Fourier mode indexes $m$ and $n$ constant because of equations \refEq{eq:kxGridy} and \refEq{eq:kyGridy}. In the code, these mode numbers directly correspond to the array indexes of the variables for $h_{s}$ and $\phi$, so standard finite difference schemes along $\chi$ naturally accomplish what we want. For example, the fourth-order centered finite difference scheme typically used by GENE calculates the parallel derivative at a point using the value of $h_{s}$ from the nearest two parallel locations on each side. Writing this at constant $m$ and $n$ is simply
\begin{align}
\left. \frac{\partial h_{s}}{\partial \chi} \right|_{k_{x}, k_{y}} \left[ m, n, l \right] \approx \frac{1}{12 \Delta \chi} &\left( - h_{s} \left[ m, n, l-2 \right] + 8 h_{s} \left[ m, n, l-1 \right] \right. \nonumber \\
&\left. - 8 h_{s} \left[ m, n, l+1 \right] + h_{s} \left[ m, n, l+2 \right] \right) ,  \label{eq:finiteDiffy}
\end{align}
where we are treating $h_{s}$ as an array with the index $m$ in the $k_{x}$ direction, $n$ in the $k_{y}$ direction, and $l$ in the $\chi$ direction (assuming uniform grid spacing in $\chi$). The primary challenge for the conventional flux tube is taking the parallel derivative across the parallel boundary because it requires correctly coupling $k_{x}$ modes according to equation \refEq{eq:parallelBCfourierkx}. In the non-twisting flux tube, this complicated mode coupling gets spread out over the interior of the domain.

Thus, for the non-twisting flux tube, we must be careful to select the proper radial wavenumber index at the various parallel locations in our parallel finite difference. This is accomplished, in accordance with equation \refEq{eq:GKeqYwithSmallShift}, by selecting the indexes that have the same value of $\mathscr{K}_{x} + \Delta K_{x} - k_{y} \Nabla x \cdot \Nabla y / | \Nabla x |^{2}$. By substituting equations \refEq{eq:smallShift} and \refEq{eq:scriptKxGrid}, we see that
\begin{align}
  \mathscr{K}_{x} + \Delta K_{x} - k_{y} \frac{\Nabla x \cdot \Nabla y}{\left| \Nabla x \right|^{2}} = \frac{2 \pi}{L_{x}} \left( m + m_{0} \left( k_{y}, \chi \right) \right) ,
\end{align}
where $m_{0} \left( k_{y}, \chi \right)$ is given by equation \refEq{eq:selectedModes}. Thus, instead of holding $m$ and $n$ constant as in the conventional flux tube, we hold $m + m_{0} \left( k_{y}, \chi \right)$ and $n$ constant in the non-twisting flux tube. While this may seem like a minor change, it is fairly significant because $m_{0}$ has a piecewise dependence on $k_{y}$ and $\chi$. In other words, we are now taking the parallel derivative diagonally across the $( \mathscr{K}_{x}, k_{y} )$ grid. However, because the grid is composed of discrete points, it is really a diagonal line that is everywhere rounded to the closest $\mathscr{K}_{x}$ grid location. Thus, the fourth-order centered finite difference scheme becomes
\begin{align}
  \left. \frac{\partial \mathscr{H}_{s}}{\partial \chi} \right|_{\mathscr{K}_{x} + \Delta K_{x} - k_{y} \frac{\Nabla x \cdot \Nabla y}{\left| \Nabla x \right|^{2}}, k_{y}} \left[ m, n, l \right] \approx \frac{1}{12 \Delta \chi} &\left( - \mathscr{H}_{s} \left[ m + m_{0} \left[ n, l \right] - m_{0} \left[ n , l-2 \right], n, l-2 \right] \right. \nonumber \\
  &+ 8 \mathscr{H}_{s} \left[ m + m_{0} \left[ n, l \right] - m_{0} \left[ n , l-1 \right], n, l-1 \right] \nonumber \\
  &- 8 \mathscr{H}_{s} \left[ m + m_{0} \left[ n, l \right] - m_{0} \left[ n , l+1 \right], n, l+1 \right] \label{eq:finiteDiffY} \\
  &\left.+ \mathscr{H}_{s} \left[ m + m_{0} \left[ n, l \right] - m_{0} \left[ n , l+2 \right], n, l+2 \right] \right) \nonumber
\end{align}
because the radial wavenumber index $m'$ at a different location $l'$ must satisfy the relationship $m' + m_{0}  \left[ n, l' \right] = m + m_{0}  \left[ n, l \right]$. 

This is implemented together with the parallel boundary condition, which the derivation of section \ref{sec:derivNonTwist} indicates is simpler than for the conventional flux tube. However, we still must derive its precise form for the slightly modified local radial wavenumber $\mathscr{K}_{x}$. Equation \refEq{eq:parallelBCfourierKx} shows it should be taken holding $K_{x}$ and $k_{y}$ constant according to
\begin{align}
  \varphi \left( \mathscr{K}_{x} \left( K_{x}, k_{y}, \chi + 2 \pi N_{\text{pol}} \right), k_{y}, \chi + 2 \pi N_{\text{pol}} \right) = \varphi \left( \mathscr{K}_{x} \left( K_{x}, k_{y}, \chi \right), k_{y}, \chi \right) .
\end{align}
Using equations \refEq{eq:aspRatioDiscrete}, \refEq{eq:geoCoeffPeriodicity}, \refEq{eq:scriptDecomp}, and \refEq{eq:smallShift}, we see that $\mathscr{K}_{x} \left( K_{x}, k_{y}, \chi + 2 \pi N_{\text{pol}} \right) = \mathscr{K}_{x} \left( K_{x}, k_{y}, \chi \right)$. Therefore, the parallel boundary condition is
\begin{align}
  \varphi \left( \mathscr{K}_{x}, k_{y}, \chi + 2 \pi N_{\text{pol}} \right) = \varphi \left( \mathscr{K}_{x}, k_{y}, \chi \right) , \label{eq:parallelBCfourierScriptKx}
\end{align}
which can be shown to be consistent with the conventional parallel boundary condition of equation \refEq{eq:parallelBCfourierkx} by using equations \refEq{eq:geoCoeffPeriodicity}, \refEq{eq:KxDef}, \refEq{eq:scriptDecomp}, and \refEq{eq:PhiFourierFormScript}. Thus, we can see that the parallel boundary condition for coding the non-twisting flux tube is actually a subset of its prior form. This makes it easy to implement --- we simply treat all modes in the same way that zonal modes were treated in the conventional flux tube.

This completely specifies the GENE implementation of the non-twisting flux tube, which is defined by equations \refEq{eq:kyGridy}, \refEq{eq:quasineuty} (using $\varphi$, $\mathscr{H}_{s}$, and $\mathscr{K}_{\perp}$), \refEq{eq:smallShift}, \refEq{eq:scriptKxGrid}, \refEq{eq:GKeqYwithSmallShift}, \refEq{eq:nonlinearRealYwithSmallShift}, \refEq{eq:kPerpDefScriptKx}, \refEq{eq:finiteDiffY}, and \refEq{eq:parallelBCfourierScriptKx}. However, there are still a few important numerical details. As a result of the numerical scheme of equation \refEq{eq:finiteDiffy} or \refEq{eq:finiteDiffY}, grid points with odd parallel indexes are closely tied together, as are grid points with even parallel indexes, but the connection between these two subsets is weak (see section 3.1.2 of reference \cite{MerzThesis2008}). This can cause problems when the total number of parallel grid points in a linear mode is odd, because the two subsets will not have an equal number of points. Thus, unless the linear mode is very long, the smaller subset will be more affected by the Dirichlet boundary condition at the ends of the linear mode (i.e. $\varphi = 0$). This can lead to grid-scale oscillations in the parallel direction that significantly affect accuracy. In the conventional flux tube, this is typically addressed by simply choosing $N_{\chi}$ to be even. However, in the non-twisting flux tube, linear modes are not constrained to span an integer number of poloidal turns. In fact, even at the same value of $k_{y}$, the number of $\chi$ grid points in the individual linear modes (with different ballooning angles) can differ by one. This makes it practically impossible to ensure that all linear modes are composed of an even number of $\chi$ grid points. To solve this, we first tried changing the boundary condition at the ends of the linear modes to be periodic, instead of Dirichlet. Specifically, whenever you would like to use the mode $m=N_{x}/2$, which is outside the grid given by equation \refEq{eq:scriptKxGrid}, you use the $m = -(N_{x}-1)/2$ mode at the same $\chi$ location instead. This works, but is observed to cause an overestimate of the linear growth rates when the linear modes are short. Instead, a better solution was to retain the Dirichlet boundary conditions, but to calculate the number of grid points in each linear mode. If it is odd, you artificially remove a grid point from that mode (i.e. the point with the lowest value of $\mathscr{K}_{x}$). This is especially important for the highest $k_{y}$ modes in nonlinear simulations because the linear modes can have a very short extent in $\chi$. In fact, when using large $k_{y}$ grids, small $\mathscr{K}_{x}$ grids, and a coarse parallel resolution, the highest values of $k_{y}$ can have linear modes with just one parallel grid point. If such linear modes are not omitted from the simulation, they can cause fictitiously large heat fluxes. To test that the code treats such modes properly without running full nonlinear simulations, it is helpful to use linear simulations with large $N_{\text{asp}}$ compared to $N_{x}$.

While implementing the parallel derivative seems complicated in the non-twisting flux tube, in practice it can be done at little computational cost. This is because the coordinate system grids stay fixed throughout the simulation, so all related work can be done when the simulation is initialized and does not need repeating. In practice, it was implemented in GENE by calculating a four-dimensional coupling matrix that is a function of $\mathscr{K}_{x}$, $k_{y}$, $\chi$, and the points in the finite difference stencil (e.g. four in equation \refEq{eq:finiteDiffY}). This coupling matrix simply holds the radial mode indexes that should be used in every finite difference computation possible in the simulation. Note that it is necessary for the coupling matrix to be a function of $\mathscr{K}_{x}$ because, near the edges of the grid, the finite difference will extend off the radial grid. Thus, the entries of the coupling matrix near the boundaries in $\mathscr{K}_{x}$ are unique as they implement the Dirichlet boundary condition.

Because coupling between different radial modes now occurs all throughout the domain (instead of just at the parallel boundaries), one might worry about the computational cost of data communication between different processors when running on a supercomputer. Fortunately, while GENE has a very flexible parallelization scheme to best distribute the five dimensional gyrokinetic equation across many processors, it does not allow any parallelization in the $k_{x}$ dimension. This was because the conventional flux tube still requires a lot of data communication between $k_{x}$ modes for the parallel boundary condition and computing the Fourier transforms in the nonlinear term. Thus, the changes to the parallel derivative needed for the non-twisting flux tube are not expected to significantly increase the data communication between processors. However, the fact that the parallel derivative uses array values that are no longer contiguous in memory likely increases the number of cache misses within each processor. We believe that this carries the most significant computational cost of all the changes required to implement the non-twisting flux tube.

\section*{References}
\bibliographystyle{unsrt}
\bibliography{references.bib}

\end{document}